\documentclass[preprint,10.5pt,3p]{elsarticle}
\usepackage[utf8]{inputenc}
\usepackage{epsfig} 
\usepackage[dvipsnames]{xcolor}
\usepackage{graphicx}
\usepackage{amsmath}
\usepackage{mathrsfs} 
\usepackage{amssymb}
\usepackage{xcolor}
\usepackage{color, colortbl}
\usepackage{caption}
\usepackage{subcaption}
\usepackage{hyperref}
\usepackage{optidef}
\usepackage{ragged2e}
\usepackage{float}
\usepackage{array}
\usepackage{adjustbox}
\usepackage{dblfloatfix}
\usepackage{setspace}
\usepackage{multirow}
\usepackage{booktabs}
\usepackage{boldline}
\usepackage[font=small]{caption}
\usepackage[hang,flushmargin]{footmisc}
\usepackage{url}

\usepackage{etoolbox,lipsum}
\usepackage{cleveref}

\newcommand{\upchi}{\text{\usefont{U}{psy}{m}{n}\symbol{'143}}}

\usepackage{mathtools}
\newcounter{tableeqn}[equation]
\renewcommand{\thetableeqn}{\theequation}
\newcounter{tablesubeqn}[tableeqn]
\renewcommand{\thetablesubeqn}{\thetableeqn\alph{tablesubeqn}}

\hypersetup{
  colorlinks   = true, 
  urlcolor     = blue, 
  linkcolor    = blue, 
  citecolor   = red 
}

\usepackage{enumitem}

\usepackage{tikz}

\newrobustcmd*{\mytriangle}[1]{\tikz{\filldraw[draw=#1,fill=#1] (0,0) -- (0.2cm,0) -- (0.1cm,0.2cm);}}

\makeatletter
\newcommand{\mathleft}{\@fleqntrue\@mathmargin0pt}
\newcommand{\mathcenter}{\@fleqnfalse}
\makeatother

\newcommand{\eg}{{\em e.g.}}
\newcommand{\etal}{{\em et~al.}}
\newcommand{\ie}{{\em i.e.}}

\newcommand{\RNum}[1]{\uppercase\expandafter{\romannumeral #1\relax}}

\journal{Computer Methods in Applied Mechanics and Engineering}

\begin{document}

\begin{frontmatter}

\title{IH-GAN: A Conditional Generative Model for Implicit Surface-Based Inverse Design of Cellular Structures}

\author[label1]{Jun Wang\corref{cor1}}
\address[label1]{Department of Mechanical Engineering, University of Maryland, College Park, MD 20742, USA}

\cortext[cor1]{Corresponding author.}

\ead{jwang38@umd.edu}

\author[label2]{Wei (Wayne) Chen}
\address[label2]{Department of Mechanical Engineering, Northwestern University, Evanston, IL 60208, USA}

\author[label2]{Daicong Da}

\author[label1]{Mark Fuge}

\author[label3]{Rahul Rai}
\address[label3]{International Center for Automotive Research, Clemson University, Clemson, SC 29607, USA}


\begin{abstract}
Variable-density cellular structures can overcome connectivity and manufacturability issues of topologically optimized structures, particularly those represented as discrete density maps. 
However, the optimization of such cellular structures is challenging due to the multiscale design problem. Past work addressing this problem generally either only optimizes the volume fraction of single-type unit cells but ignores the effects of unit cell geometry on properties, or considers the geometry-property relation but builds this relation via heuristics. In contrast, we propose a simple yet more principled way to accurately model the property to geometry mapping using a conditional deep generative model, named \textit{Inverse Homogenization Generative Adversarial Network (IH-GAN)}. It learns the conditional distribution of unit cell geometries given properties and can realize the one-to-many mapping from properties to geometries. We further reduce the complexity of IH-GAN by using the implicit function parameterization to represent unit cell geometries. Results show that our method can 1)~generate various unit cells that satisfy given material properties with high accuracy ({$R^2$-scores between target properties and properties of generated unit cells $>98\%$}) and 2)~improve the optimized structural performance over the conventional variable-density single-type structure. In the minimum compliance example, our IH-GAN generated structure achieves a $79.7\%$ reduction in concentrated stress and an extra $3.03\%$ reduction in displacement. In the target deformation examples, our IH-GAN generated structure reduces the target matching error by 86.4\% and 79.6\% for two test cases, respectively. 
We also demonstrated that the connectivity issue for multi-type unit cells can be solved by transition layer blending.

\end{abstract}



\end{frontmatter}

\section{Introduction}

Complexity-free manufacturing techniques such as additive manufacturing (AM) have enabled the fabrication of intricate geometric features. This permits the design of complex structures that fulfill specific functional criteria while possessing lower weight. Access to this new design space makes complex structural designs (\eg, cellular structures) coveted in various engineering applications \cite{fleck2010micro, gibson2005biomechanics, callanan2018hierarchical, evans2001topological, brennan2012design, cheng2018coupling}. In this context, topology optimization (TO) plays a major role in designing light-weight structures that satisfy functional goals~\cite{bendsoe2013topology}. However, the most widely used TO algorithms (\eg, solid isotropic material with penalization or SIMP) \cite{bendsoe1989optimal} deliver discrete density maps as the outcome, leading to poor manufacturability due to 1)~connectivity issues within variable-density bulk materials and 2)~the limited resolution of the density map. 

Conventional bulk material/density-based TO methods alleviated the connectivity issues (\eg, the checkerboard problem and intermediate density) by applying filters (\eg, sensitivity and density filters) \cite{sigmund1998numerical, andreassen2011efficient, lazarov2011filters}. To enable the manufacturability of the optimized density-based structure, researchers enhanced the resolution of the density map by employing the super-computing power \cite{aage2017giga, liu2018narrow} and post-processing techniques \cite{groen2018homogenization, zegard2016bridging}, or incorporated manufacturing constraints directly into the TO methods \cite{sutradhar2017incorporating, carstensen2018projection}. However, these conventional approaches cannot fundamentally solve the intermediate density (multiple materials) and smoothness (continuity) issues due to their inherent vice relying on discrete density-based structures.

\begin{figure*}[hbt!]
\begin{subfigure}[b]{0.5\linewidth}
\centering
\includegraphics[width=0.8\linewidth]{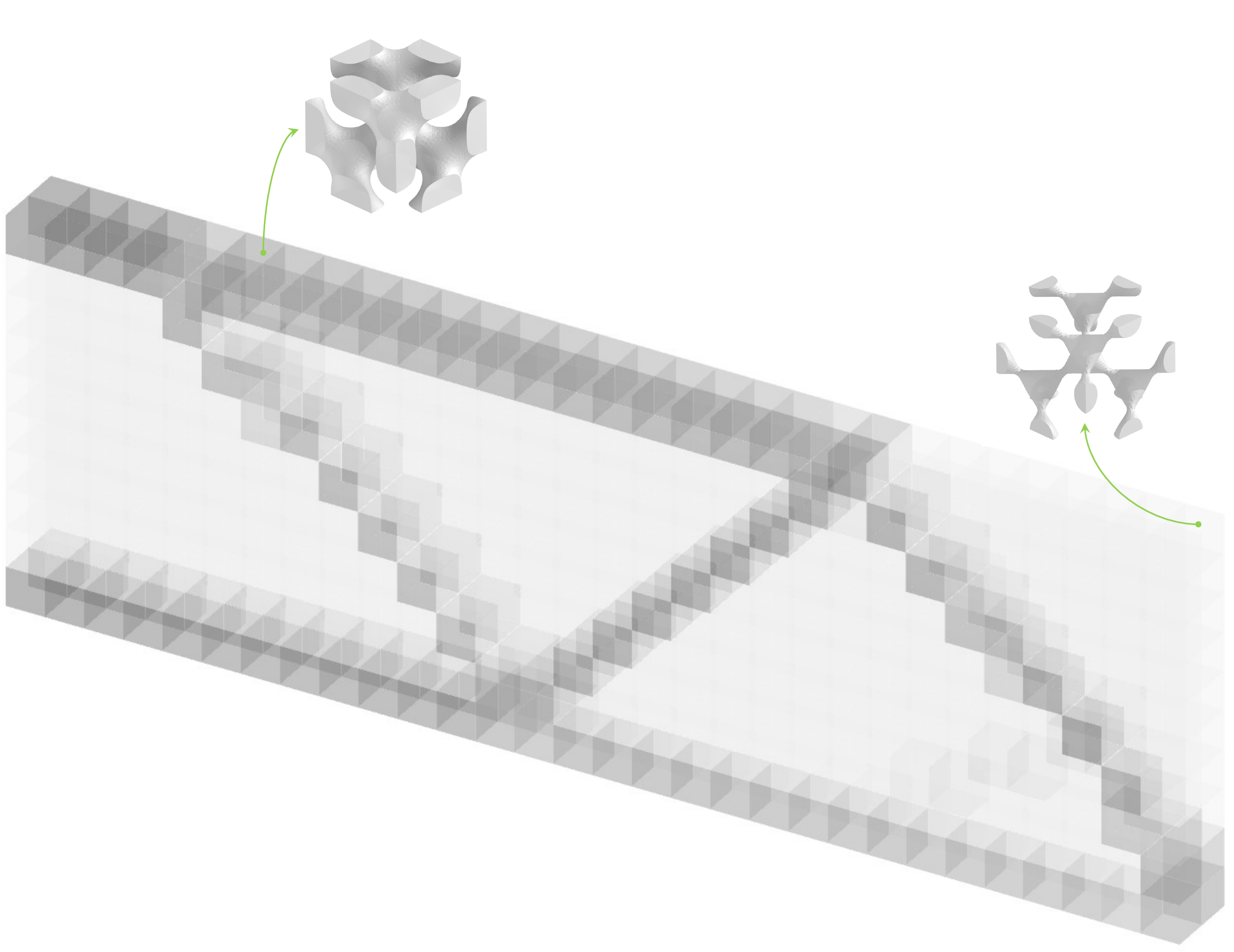}
\caption{}
\end{subfigure}%
\begin{subfigure}[b]{0.5\linewidth}
\centering
\includegraphics[width=0.85\linewidth]{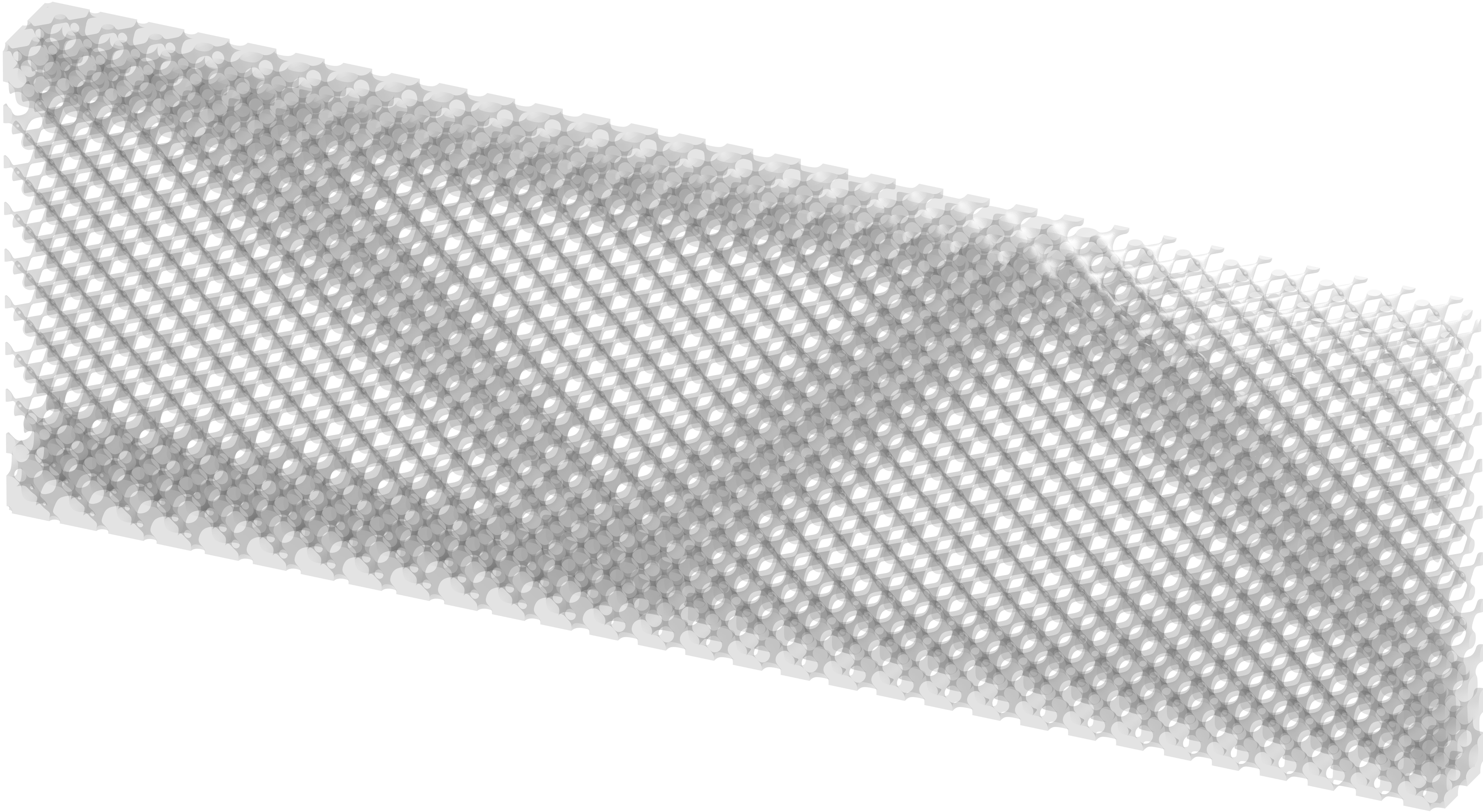}
\caption{}
\end{subfigure}

\caption{Variable-density cellular structure. (a) Density map obtained from TO. (b) Replacement with variable-density unit cells.}
\label{fig:varDen}
\end{figure*}

In recent years, researchers typically replace the density map with variable-density cellular unit cells (Figure~\ref{fig:varDen}), enabling high-resolution functionally graded structure design and manufacturing with a single material~\cite{montoya2019density}. However, a na\"ive one-to-one replacement with unit cells having corresponding densities/volume fractions (without considering the unit cell shape) can break the optimized behavior because, unlike continuum solids, equivalent density alone does not guarantee equivalent mechanical properties for unit cells. The unit cell shape also affects its mechanical properties. To address this issue, researchers commonly \textit{homogenize} each unit cell by pre-computing their effective properties~\cite{bensoussan2011asymptotic, andreassen2014determine} such that they can approximate the continuum solid domain with the homogenized cellular structure, hopefully retaining equivalent properties. By establishing a mapping between densities and homogenized properties (\eg, elastic tensors) following a scaling law\footnote{The scaling law hypothesizes the mechanical properties of cellular structures have polynomial law relationships with their relative densities.} \cite{ashby2006properties}, TO can generate an optimized density map for a specific type/shape of cellular unit cells.

However, there are several issues in building such a mapping. First, the scaling law creates a mapping between the material property space and the unit cell density space. This means the unit cell's material property is controlled only by its density (volume fraction), while variable unit cell types are not allowed. Second, being subject to that setting, the scaling law only allows bijective mapping, whereas the mapping from a material property space to a unit cell shape space can be one-to-many. Finally, the mapping itself is hard-coded (polynomial) and has limited flexibility/accuracy under different scenarios.

\begin{figure}[hbt!]
\begin{center}
\includegraphics[width=1\linewidth]{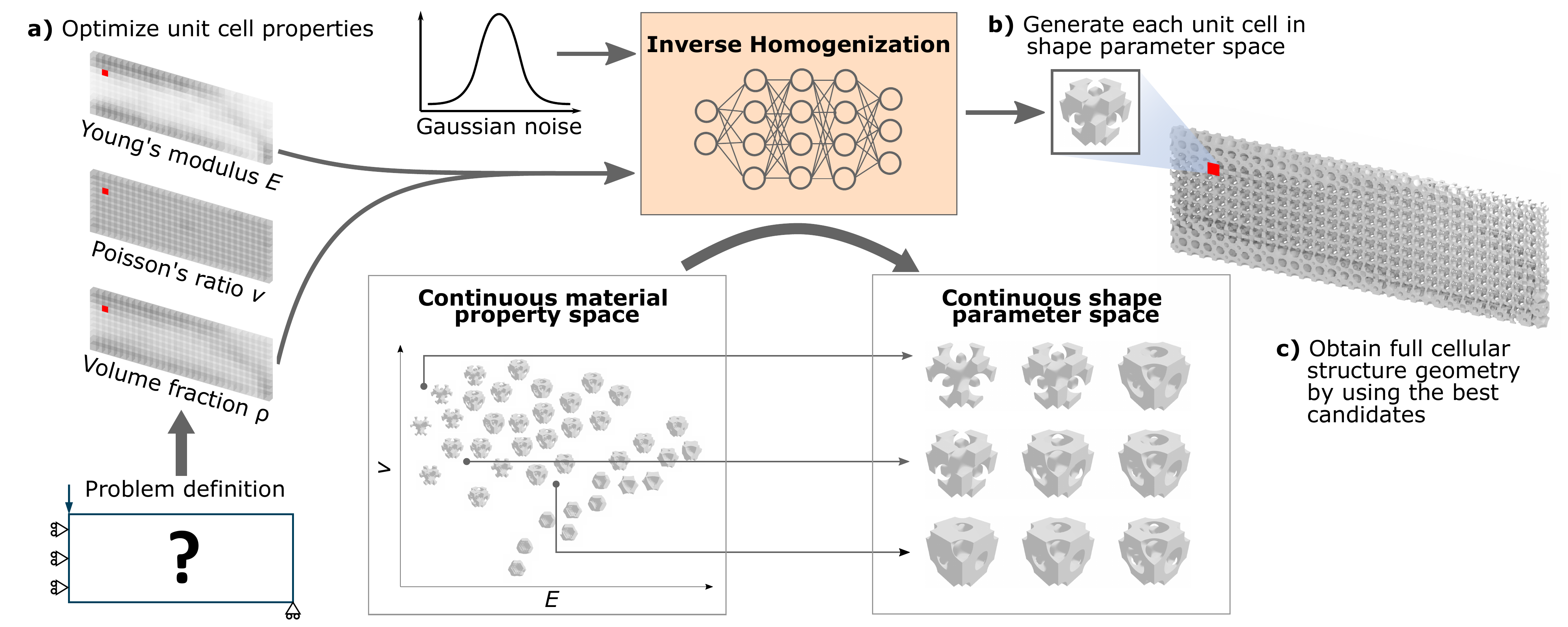}
\caption{Multi-scale design synthesis of functionally graded cellular structures.}
\label{fig:flowchart}
\end{center}
\end{figure}

We solve these issues by constructing an \emph{inverse homogenization (IH) mapping}\textemdash a generalized, direct, and accurate mapping from properties to multi-type unit cell shapes. 
The IH mapping allows designers to efficiently retrieve correct unit cell shapes given the optimized properties. Rather than the previous approaches that were limited to polynomial scaling and enforced bijectivity
(\ie, using the scaling law and level set field \cite{zhu2017two, li2019design}), we propose an end-to-end generative model that automatically learns a data-driven one-to-many IH mapping and suggests unit cell shapes conditioned on the input properties. Figure~\ref{fig:flowchart} illustrates the role of our work in assisting the multi-scale design synthesis for functionally graded cellular structures.

Our generative model is developed by training a conditional generative adversarial network (GAN) upon a family of implicit function-based cellular structures\textemdash \ie, triply periodic minimal surfaces (TPMS) \cite{gandy2001nodal}. The implicit function representation (FRep) modeling method represents cellular structures as iso-surfaces that enable control of unit cell shapes by manipulating the functions' coefficients and level set constants. Compared to expensive volumetric representations \cite{schumacher2015microstructures}, the FRep minimizes the number of design variables representing unit cell geometries. We propose a deep generative model, named \textit{Inverse Homogenization GAN (IH-GAN)}, to learn a mapping from properties to the implicit function describing the cellular unit cell geometry. We demonstrate our method's efficacy by rapidly generating various TPMS unit cells with accurate elastic properties (\ie, Young's modulus and Poisson's ratio) and relative density (\ie, volume fraction). To further demonstrate IH-GAN's practical usage in designing functionally graded cellular structures, we show a cantilever beam design example where the structure is assembled by variable types of unit cells generated from IH-GAN.
This paper's primary contributions are as follows:
\begin{enumerate}
    \item We describe a conditional generative model, IH-GAN, to enable an accurate ($R^2$-scores $>98\%$) and efficient IH mapping. After 38 seconds of training, IH-GAN can generate corresponding unit cell shapes in less than one second given feasible ranges of Young's modulus, Poisson's ratio, and relative density.
    \item We improve the IH mapping learned by the conditional GAN by adding an auxiliary property regressor. We show through ablation experiments that this additional regressor reduces about half of the average mapping error.
    \item We demonstrate the feasibility and the extremely low computational complexity when using the implicit function parameterization in deep generative modeling of unit cells while enabling downstream cellular structure inverse design tasks.
    \item We ensure high-quality connections between different types of adjacent unit cells in designing functionally graded cellular structures by taking advantage of the filtering kernel used in structural optimization, the continuous shape variation of IH-GAN's generated unit cells in the property space, and the transition layer blending technique.
    \item We demonstrate that our method can improve the structural performance over the conventional variable-density single-type structure in both minimum compliance problem ($79.7\%$ reduction in concentrated stress and extra $3.03\%$ reduction in displacement) and target deformation problem (86.4\% and 79.6\% reductions of the target matching error for the two case studies).
    \item We create a unit cell shape database for future studies on data-driven cellular structure design (available at \url{https://github.com/IDEALLab/IH-GAN_CMAME_2022}).
\end{enumerate}


\section{Related work}

Our proposed method constructs the IH mapping from material properties to cellular unit cells using a GAN-based model. In this section, we first review prior work in cellular structure modeling. We then review generative models pertinent to the inverse design of cellular structures. Finally, we introduce conditional GANs (cGANs)~\cite{mirza2014conditional}, which our proposed IH-GAN builds upon.

\subsection{Cellular structure modeling}

Main approaches to representing cellular structures include boundary (\eg, NURBS surfaces) representations (BRep) \cite{gorguluarslan2017improved, chen20073d, naing2005fabrication}, volume (\eg, voxels) representations (VRep) \cite{schumacher2015microstructures, verges20083d, panetta2015elastic}, and FRep (\eg, trigonometric periodic functions) \cite{maldovan2009periodic, yoo2011porous, wang2020generative}. Given the high complexity of cellular structures, BRep and VRep incur large data size and processing costs (\eg, in the number of voxels or polygons), limited geometry precision (\eg, cracks in surfaces and self-intersections of polygons), limited or no support for parameterization (\eg, VRep models need to be regenerated using a separate, high-level procedure or method), and poor manufacturability (\eg, voxels have aliasing problems unless they are given at high resolutions requiring large amounts of memory) \cite{pasko2011procedural}. FRep is a shape parameterization that maps a set of parameters to points along the iso-surface of an implicit function. It offers a compact but precise representation of cellular structures that solves most of the above issues. To enable an accurate and efficient IH mapping using a deep neural network, we benefit from the compactness of FRep that allows representing cellular unit cells by a small set of parameters. 

\subsubsection{FRep of TPMS-based cellular structures}
Surfaces whose mean curvature is everywhere zero are minimal surfaces. A \textit{triply periodic} minimal surface is infinite and periodic in three independent directions. A TPMS is of special interest because it has no self-intersections and partitions space into two labyrinthine regions. These regions commonly appear in various natural and human-made structures, providing a concise description of a wide variety of cellular structures \cite{wohlgemuth2001triply, wang2007periodic, jung2007variational}. Compared to the other families of periodic surfaces \cite{maldovan2009periodic, von1991nodal, halse1969fermi, pasko1995function}, TPMS are particularly fascinating because a TPMS derives one of the crystallographic space groups as its symmetry group \cite{lord2003periodic}. Those with cubic symmetry simplify the homogenization process by having the same properties along three orthogonal axes \cite{li2019design}. Additionally, TPMS-based cellular structures afford high specific surface area, high porosity, and low relative density while maintaining outstanding mechanical properties \cite{wu2016insect, khaderi2014stiffness}. 

TPMS can be approximated by the periodic implicit surfaces of a sum defined in terms of Fourier series \cite{mackay1993crystallographic, klinowski1996curved, gandy2001nodal}:
\begin{equation}
f(x, y, z) = \displaystyle\sum_{hkl} |F(hkl)|\cos[2\pi(hx + ky +lz) - \alpha(hkl)] = 0
\label{eqTPMS}
\end{equation}
where ($h$, $k$, $l$) are the reciprocal lattice vectors for a given lattice, $\alpha(hkl)$ is a phase shift, and the structure factor $F(hkl)$ is an amplitude associated with a given vector ($h$, $k$, and $l$).

Other than the concise representation and cubic symmetry, TPMS-based cellular unit cells can also be diversified by combining several functions and adjusting corresponding coefficients \cite{maldovan2009periodic, wang2019investigation}.

\subsubsection{Variable-density cellular structures and inverse homogenization mapping}
Topology optimization is a computational process that optimizes the material distribution in a design space by literally removing material within it, intending to reveal the most efficient design given a set of constraints. Conventional TO methods (\eg, SIMP, ESO/BESO\footnote{Evolutionary structural optimization (ESO) and its later version bi-directional ESO (BESO).}, and level set) \cite{bendsoe1989optimal, huang2007convergent, wang2003level} often output the optimal structural design as a discrete field (\eg, SIMP produces discrete densities and BESO generates discrete mesh elements).
Thus, the early-stage TO outputs failed to consider aesthetics, manufacturability, or any other design constraints that one would normally need in a design process. To fill the gap between the discrete representation and those practical considerations, researchers have leveraged high-performing computational tools \cite{aage2017giga, liu2018narrow} and post-processing \cite{groen2018homogenization, zegard2016bridging} to increase the representation's resolution and incorporated the manufacturability constraints \cite{sutradhar2017incorporating, carstensen2018projection} into the conventional TO directly. However, these methods still rely on the discrete field, which cannot fully take care of the smoothness and aesthetics considerations with high efficiency and precision. To further overcome these limitations, one needs to convert a discrete field into concrete geometries that can be physically realized while maintaining the same performance. 

Variable-density cellular structures are a promising candidate to replace the density map (generated by SIMP). They resolve the geometric connectivity and manufacturability issues and possess unique combinations of physical properties (\eg, high strength-to-weight ratio, high energy absorption, and high thermal conductivity) \cite{hou2018design, maloney2012multifunctional, yin2014damping}. In addition, variable-density cellular structures make the density map manufacturable with a single material to avoid using costly multi-material techniques \cite{miyamoto2013functionally, bandyopadhyay2018additive, xu2021stress, ghasemi2018multi}. First-generation variable-density cellular structures were designed using simple repeating elements such as cubic trusses with variable strut thickness \cite{cheng2017efficient} or hexagonal cells with variable hole diameter \cite{zhang2015efficient, zhang2020topological}. Since then, the three main approaches, namely BReps (\eg, B-splines) \cite{wang2019cellular}, VReps (\eg, voxels) \cite{li2021topology}, and FReps \cite{li2019design}, have been investigated for topology optimization of more complex cellular structures. FReps have become a more versatile representation that can produce miscellaneous cellular structures as implicit surfaces whose volume fraction can be conveniently controlled by the level set constant to resemble the given density map. By forming specific implicit functions like TPMS, one can even mimic complicated real-world structures, such as silicates (\eg, Schwartz Diamond structures in diamonds) \cite{gandy1999exact} and biomorphic formations (\eg, Gyroid structures in butterfly wing scales) \cite{michielsen2008gyroid}. These implicit surfaces also take care of geometric continuity and coherence at the connections or interfaces between variable-density unit cells.

Besides the geometric modeling challenge, we need to ensure that generated variable-density cellular structures still preserve desired property performance to complete the density map conversion. To do that, one needs to homogenize the variable-density cellular unit cells to acquire their effective material properties and construct an IH mapping from the homogenized properties to their geometries. Most existing work builds the IH mapping by following the scaling law\textemdash \ie, the mechanical properties of cellular structures have polynomial law relationships with their relative densities (Gibson-Ashby model \cite{gibson_ashby_1997, ashby2006properties}). These relationships have been successfully fitted for low dimensional properties like elastic constants and thermal conductivity constants using polynomial and exponential functions \cite{li2019design}. For material properties with higher dimensionality, a level set field has been proposed to construct a continuous gamut representation of the material properties \cite{zhu2017two}. 

As mentioned previously, the scaling law performs a preliminary and simplified version of IH mapping with limited data to be fitted. However, it can only be applicable to approximating the low dimensional (\eg, one-dimensional) property space with simple relationships (\eg, one-to-one mapping). Every time the unit cell type changes, one needs to repeat that fitting process; this forces the mapping to be bijective and difficult to generalize across cell types. The level set method supports the IH mapping for a higher dimensional property space but suffers from a costly manual process requiring extensive computation. For example, Zhu \etal~\cite{zhu2017two} needed to perform a series of procedures, including a discrete sampling of the microstructures, a continuous optimization of the microstructures, and a continuous representation of the material gamut by computing a signed distance field (up to 93 hours computational cost in total) to construct an accurate mapping in their application. Unlike these methods, our work eliminates these limitations by training an end-to-end deep generative model (\ie, conditional GAN) to automatically learn the IH mapping from two-dimensional elastic properties (\ie, Young’s modulus and Poisson’s ratio) to multiple types of unit cells. 

\subsubsection{Structures assembled by multiple types of cellular unit cells}
Rather than using single-type variable-density unit cells that can limit the spectrum of physical properties, researchers have explored assembling structures with different types of unit cells to expand the range of physical behaviors. The major bottleneck in the assembly is the lack of sufficient interface connection area due to significantly different geometries at the intersection between two adjacent unit cells. Some existing works have focused on solving this bottleneck by optimizing geometric compatibility between adjacent unit cells. Zhou and Li \cite{zhou2008design} have summarized three methods (\ie, connective constraint, pseudo load, and unified formulation with nonlinear diffusion) to ensure the connectivity between adjacent 2D unit cells. Li \etal~\cite{li2018topology} used a similar kinematic approach to solve the compatibility issue. Radman \etal~\cite{radman2013topology} performed topology optimization of three adjacent base cells by considering the connectivity constraints simultaneously: the base unit cells were designed for the target stiffness while maintaining smooth connectivity. Garner \etal~\cite{garner2019compatibility} focused on finding the optimal connectivity between more diverse topology optimized unit cells while maximizing bulk moduli of the graded structures. Schumacher \etal~\cite{schumacher2015microstructures} precomputed a database of tiled unit cells indexed by the elastic properties and applied a global optimization to select the optimal tiling that can best connect adjacent tile. Another approach was to use geometric interpolation to obtain intermediate structures that have no geometric frustration \cite{cramer2016microstructure}. A similar interpolation strategy was commonly used to design variable-density cellular structures to generate a smooth connection between unit cells with different volumes \cite{li2019design}. Wang \etal~\cite{wang2021hierarchical} tackled the geometric compatibility by proposing non-periodic implicit functions that can generate two compatible unit cells with good overlap at the intersection face. In this paper, our approach takes advantage of the integrated TO and IH-GAN framework to generate functionally graded cellular structures with multiple types of unit cells and address the connectivity bottleneck simultaneously without a need for compatibility optimization. The generated structures are also precise and manufacturable for AM techniques.

\subsubsection{Multiscale topology optimization}
Due to the constant increase of computing power and the availability of big data in recent years, the simultaneous topological design of macroscale structures (\ie, structure) and their underlying microscale structures (\ie, material) have allowed researchers to optimize structures with hierarchically fine features for improved performance and manufacturability \cite{robbins2016efficient}. Equipped with data-driven methods, the multiscale TO becomes another important tool for efficiently overcoming the limitations of mono-scale or homogeneous structures. To more accurately map from the macroscale to the microscale structures, White \etal~\cite{white2019multiscale} developed single layer feedforward Gaussian basis function networks as a surrogate model. While Wu \etal~\cite{wu2019topology} constructed the map from the density to super-element stiffness matrix using a surrogate model built with the assistance of proper orthogonal decomposition and diffuse approximation. 
Patel \etal~\cite{patel2022improving} employed two deep neural networks (\ie, ConnectivityNet and SILONet) to enable multiscale TO, with ConnectivityNet improving the connectivity while SILONet~\cite{bielecki2021multi} accelerating the multiscale TO computations.
Wang \etal~\cite{wang2020deep} proposed systematic data-driven methods for the design of metamaterial microstructure, graded family, and multiscale system using a variational autoencoder (VAE).
Rather than using a single type of underlying microstructures, the multiscale TO can also deal with multiple classes of truss structures to accommodate spatially varying desired behaviors with the aid of data-driven models (\eg, interpolation model and multi-response latent-variable Gaussian process model) \cite{liu2020data, wang2021data}. Those multiclass microstructures were not limited to simple truss structures (to bypass connectivity issues) as they can also be fully non-periodic structures, which need the consideration of compatibility between adjacent unit cells \cite{sivapuram2016simultaneous, xia2017recent}. Instead, to bypass the compatibility challenge, Zhu \etal~\cite{zhu2017two} fulfilled the non-periodic structural design by utilizing solid multi-material microstructures as the building blocks. 
Sanders \etal~\cite{sanders2021optimal} attempted to make the hierarchical structures by focusing on solving the smooth and continuous transitions between truss unit cells without breaking the desired properties. The study manually interpolates the truss unit cell geometry and composes the unit cells into a series of hybrid transitional unit cells to achieve smooth transitions. In contrast, Zheng \etal~\cite{zheng2021data} enables the multi-scale metamaterial structures relying on a data-driven topology optimization approach and the parameterized spinodoid unit cells. A conventional feedforward neural network has been used as a surrogate model to map from shape parameters to elasticity tensor.
In this paper, we use a conditional deep generative model to learn the conditional distribution of microscale unit cell geometries given the material properties obtained from the macroscale structural optimization to achieve one-to-many property to structure mapping and connect the macrostructure with microstructures in a simple yet principled way.

\subsection{Data-driven models for inverse design of microstructures}

Generative modeling, a branch of unsupervised learning techniques in machine learning, has been used to build the statistical model of data distributions. In recent years, deep generative models like VAEs or GANs have drawn much attention and are used to generate realistic data. These models are primarily applied in the computer science domain for image synthesis and nature language processing~\cite{el2019deep}. Unlike these typical applications, 
generative models have been used in the microstructural design, including crystalline porous structures, nanophotonic structures, and metamaterial structures. Kim \etal~\cite{kim2020inverse} developed a ZeoGAN model to generate realistic zeolite materials and their corresponding energy shapes via training the combined data of three grids (\ie, the energy, silicon atom, and oxygen atom grids). 
Liu \etal~\cite{liu2018generative} enabled the inverse design of nanophotonic metasurfaces (2D image data) by adopting a conditional GAN model where a pre-trained simulator was added to approximates the transmittance spectrum for a given geometric pattern at its input. Ma \etal~\cite{ma2019probabilistic} revealed the non-intuitive and non-unique relationship between metamaterial structures and their optical responses by incorporating a VAE in their generative model, which encodes the designed pattern together with the corresponding optical response into a latent space. A VAE model (combined with a regressor of the mechanical properties of interest) was also used by Wang \etal~\cite{wang2020deep} to generate the pixel-based 2D metamaterial microstructures. Kumar \etal~\cite{kumar2020inverse} designed the spinodoid non-periodic cellular structures inversely via the use of two multi-layer perceptron\textemdash an inverse network and a forward network\textemdash the former takes a queried stiffness as input and outputs design parameters defining a cellular unit cell, and the latter takes the predicted design parameters as input and reconstructs the stiffness and verifies the prediction from the inverse network. In this paper, we use a conditional GAN model to realize the inverse design of cellular structures by learning the mapping from the material properties (\ie. Young's modulus and Poisson's ratio) to the shape parameters that define the implicit surfaces of the cellular unit cells. To the best of our knowledge, the conditional GAN model has never been used in the inverse design of cellular structures.

Particularly, existing studies have used conventional feedforward neural networks \cite{kumar2020inverse} or variational autoencoders (VAEs) \cite{wang2020deep} to fulfill the inverse design of cellular unit cells. Compared to conventional feedforward neural networks, our IH-GAN approach can potentially generate multiple or diverse equally good designs given the same input properties. Compared to VAEs, one significant advantage of our IH-GAN approach is that we use a simple yet principled way of deriving a one-to-many mapping rather than using any complicated heuristics. Our IH-GAN model can also benefit the geometric compatibility between multiple types of unit cells by taking advantage of the nature of the continuous shape variation of IH-GAN's generated unit cells. Some existing studies even simplified the problems by utilizing 2D data (\eg, 2D images) while we directly take care of 3D models. In addition, another big advantage of our approach is the fast training (38 seconds) but with high prediction accuracy.

\subsection{Conditional generative adversarial networks} 

Generative adversarial networks~\cite{goodfellow2014generative} model a game between a generative model (\textit{generator}) and a discriminative model (\textit{discriminator}). The generative model maps an arbitrary noise distribution to the data distribution (\ie, the distribution of designs in our scenario), thus can generate new data; while the discriminative model tries to perform classification, \ie, to distinguish between real and generated data. The generator $G$ and the discriminator $D$ are usually built with deep neural networks. As $D$ improves its classification ability, $G$ also improves its ability to generate data that fools $D$. 
Thus, a vanilla GAN (standard GAN with no bells and whistles) has the following objective function:
\begin{equation}
\min_G\max_D V(D,G) = \mathbb{E}_{\mathbf{x}\sim P_{data}}[\log D(\mathbf{x})] + 
\mathbb{E}_{\mathbf{z}\sim P_{\mathbf{z}}}[\log(1-D(G(\mathbf{z})))],
\label{eq:gan_loss}
\end{equation}
where $\mathbf{x}$ is sampled from the data distribution $P_{data}$, $\mathbf{z}$ is sampled from the noise distribution $P_{\mathbf{z}}$, and $G(\mathbf{z})$ is the generator distribution. A trained generator thus can map from a predefined noise distribution to the distribution of designs. 

The \textit{conditional GAN} or \textit{cGAN}~\cite{mirza2014conditional} further extends GANs to allow the generator to learn a conditional distribution. This is done by simply feeding the condition, $\mathbf{y}$, to both $D$ and $G$. The loss function then becomes:
\begin{equation}
\min_G\max_D V_{\text{cGAN}}(D,G) = \mathbb{E}_{\mathbf{x}\sim P_{data}}[\log D(\mathbf{x}|\mathbf{y})] + 
\mathbb{E}_{\mathbf{z}\sim P_{\mathbf{z}}}[\log(1-D(G(\mathbf{z}|\mathbf{y})))].
\label{eq:cgan_loss}
\end{equation}
Therefore, given any conditions, cGAN can generate a set of designs that satisfy the given conditions, by feeding a set of random noise. In our case, the conditions are material properties (\ie, the Young's modulus and the Poisson's ratio).

\section{TPMS-based cellular structures} \label{sec:TPMS}

Our cellular structures are created as a composition of three different TPMS surfaces that have a cubic symmetry, namely Schwarz P (P), Diamond (D), and Schoen's F-RD \cite{blanquer2017surface}. Table~\ref{tab_tpms} lists their implicit functions and corresponding geometric models. 

\begin{table}[hbt!]
\centering
\stepcounter{equation}
\caption{TPMS surfaces with cubic symmetry}
\label{tab_tpms}
\def\arraystretch{1.5}
\resizebox{1\textwidth}{!}{%
\begin{tabular}{lllc}
\hline
Morphology  & TPMS function & & Model \\ 
\hline \\[-2ex]
Schwarz P (P) & $f_P(x, y, z) = \cos(X) + \cos(Y) + \cos(Z) + t_1=0$ &
\refstepcounter{tablesubeqn}(\thetablesubeqn) \label{eq:tpms1}& \parbox[][8em][c]{7em}{\includegraphics[width=1in]{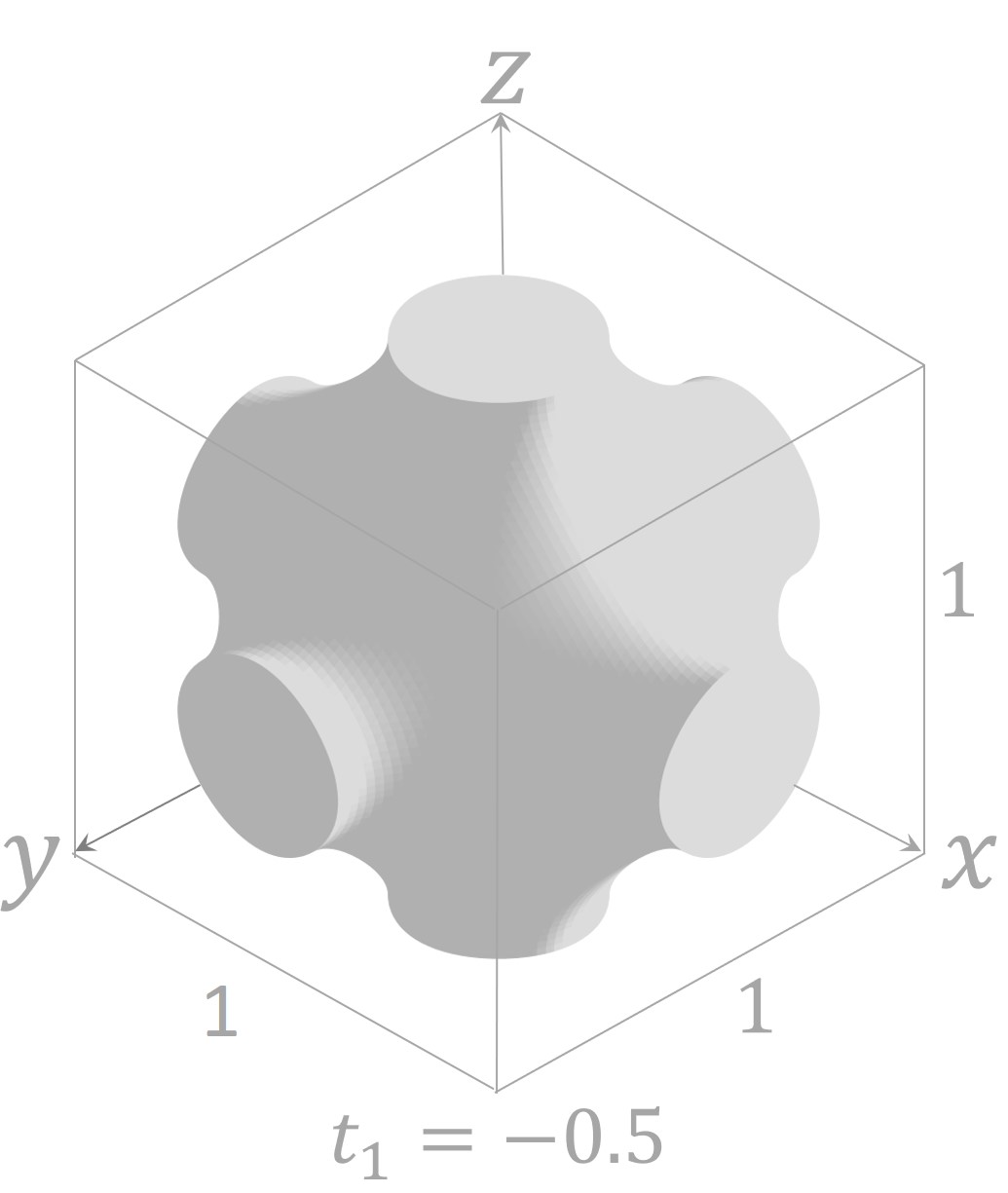}}\\[2cm]
Diamond (D) & $f_D(x, y, z) = \cos(X)\cos(Y)\cos(Z) - \sin(X)\sin(Y)\sin(Z) + t_2=0$ & 
\refstepcounter{tablesubeqn}(\thetablesubeqn) \label{eq:tpms2}& \parbox[][8em][c]{7em}{\includegraphics[width=1in]{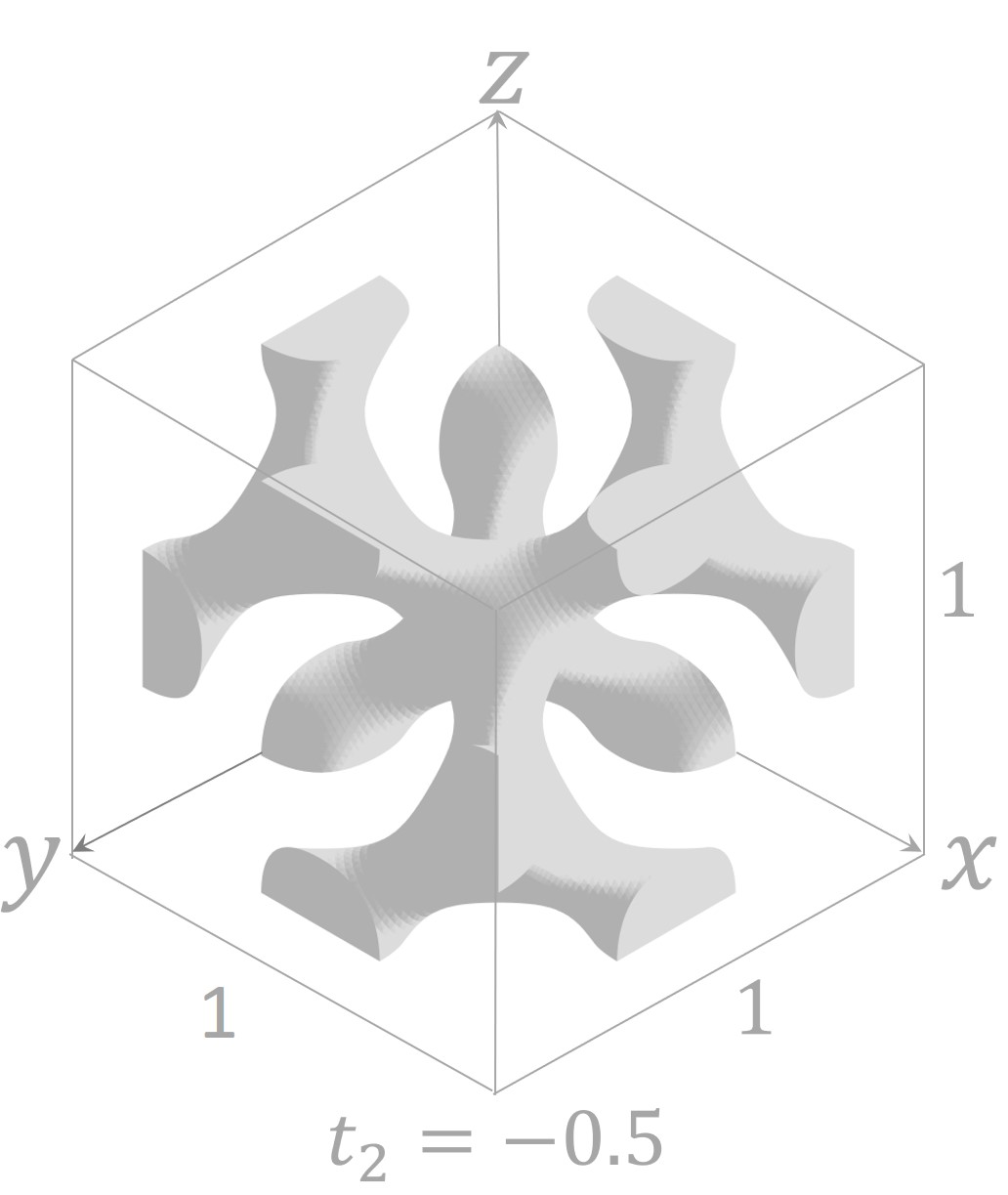}}\\[2cm]
\newline Schoen F-RD (F-RD) &\begin{tabular}[l]{@{}l@{}}$f_{FRD}(x, y, z) = 8\cos(X)\cos(Y)\cos(Z) + \cos(2X)\cos(2Y)\cos(2Z)$\\
$- \big(\cos(2X)\cos(2Y) + \cos(2Y)\cos(2Z) + \cos(2Z)\cos(2X)\big) + t_3=0$ \end{tabular} &
\refstepcounter{tablesubeqn}(\thetablesubeqn) \label{eq:tpms3}& \parbox[][8em][c]{7em}{\includegraphics[width=1in]{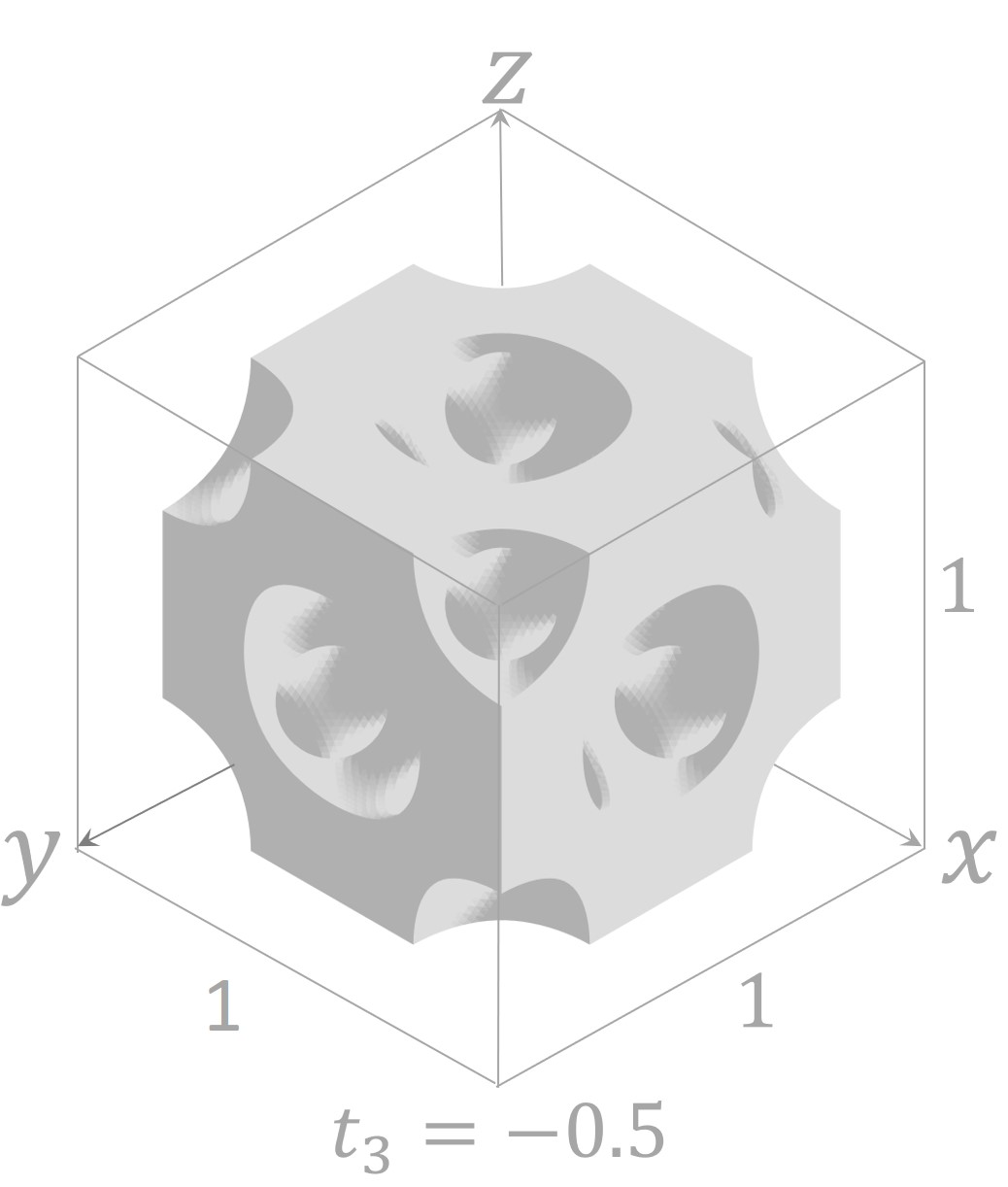}}\\[1.0cm]
{\raggedright where $X = 2\pi x$, $Y = 2\pi y$, $Z = 2\pi z$.\par} \\
\hline
\end{tabular}
}
\end{table}

\begin{figure}[hbt!]
\begin{subfigure}[b]{0.26\linewidth}
\centering
\includegraphics[width=1\linewidth]{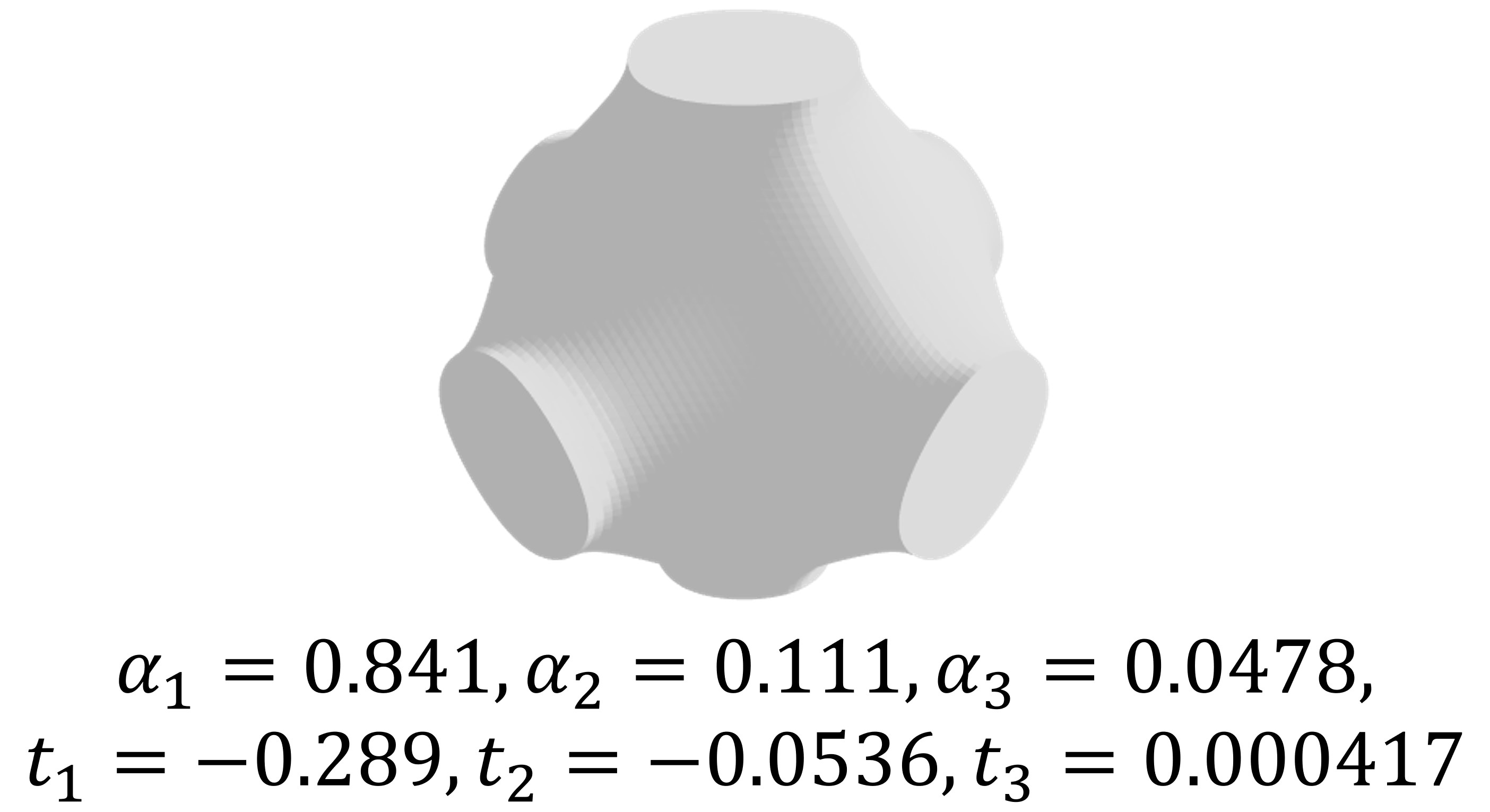}
\caption{}
\end{subfigure}%
\begin{subfigure}[b]{0.24\linewidth}
\centering
\includegraphics[width=1\linewidth]{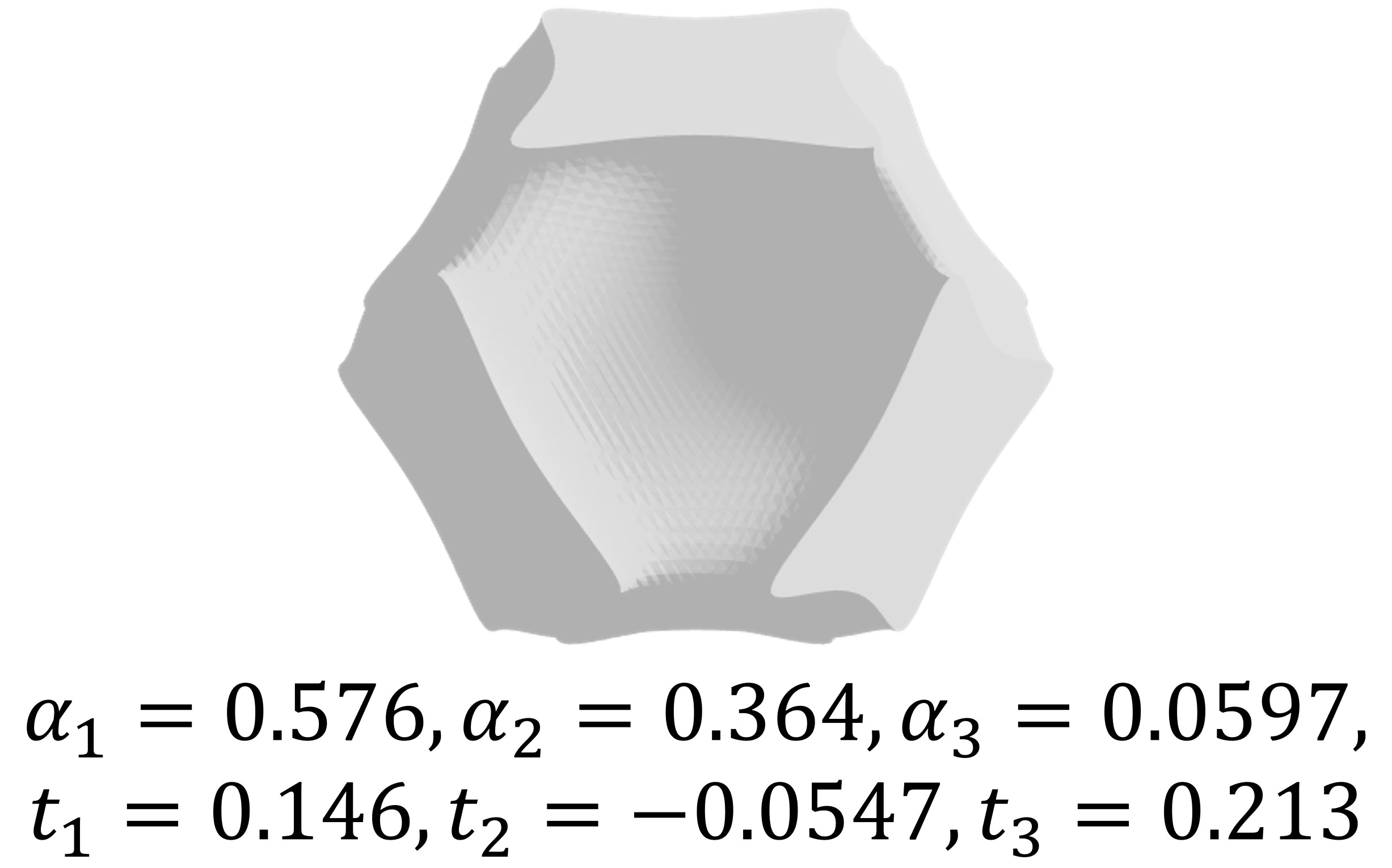}
\caption{}
\end{subfigure}%
\begin{subfigure}[b]{0.24\linewidth}
\centering
\includegraphics[width=1\linewidth]{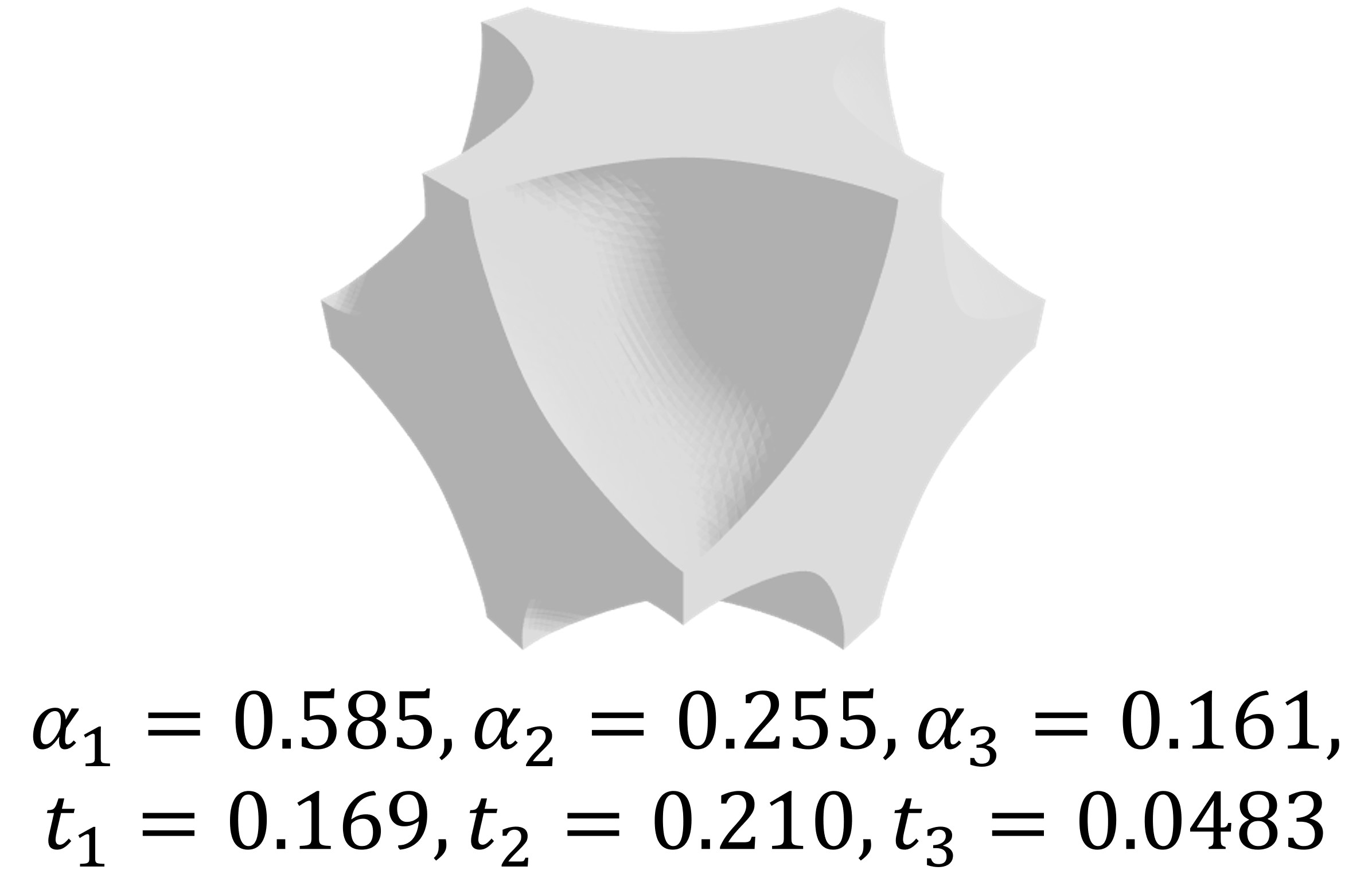}
\caption{}
\end{subfigure}%
\begin{subfigure}[b]{0.26\linewidth}
\centering
\includegraphics[width=1\linewidth]{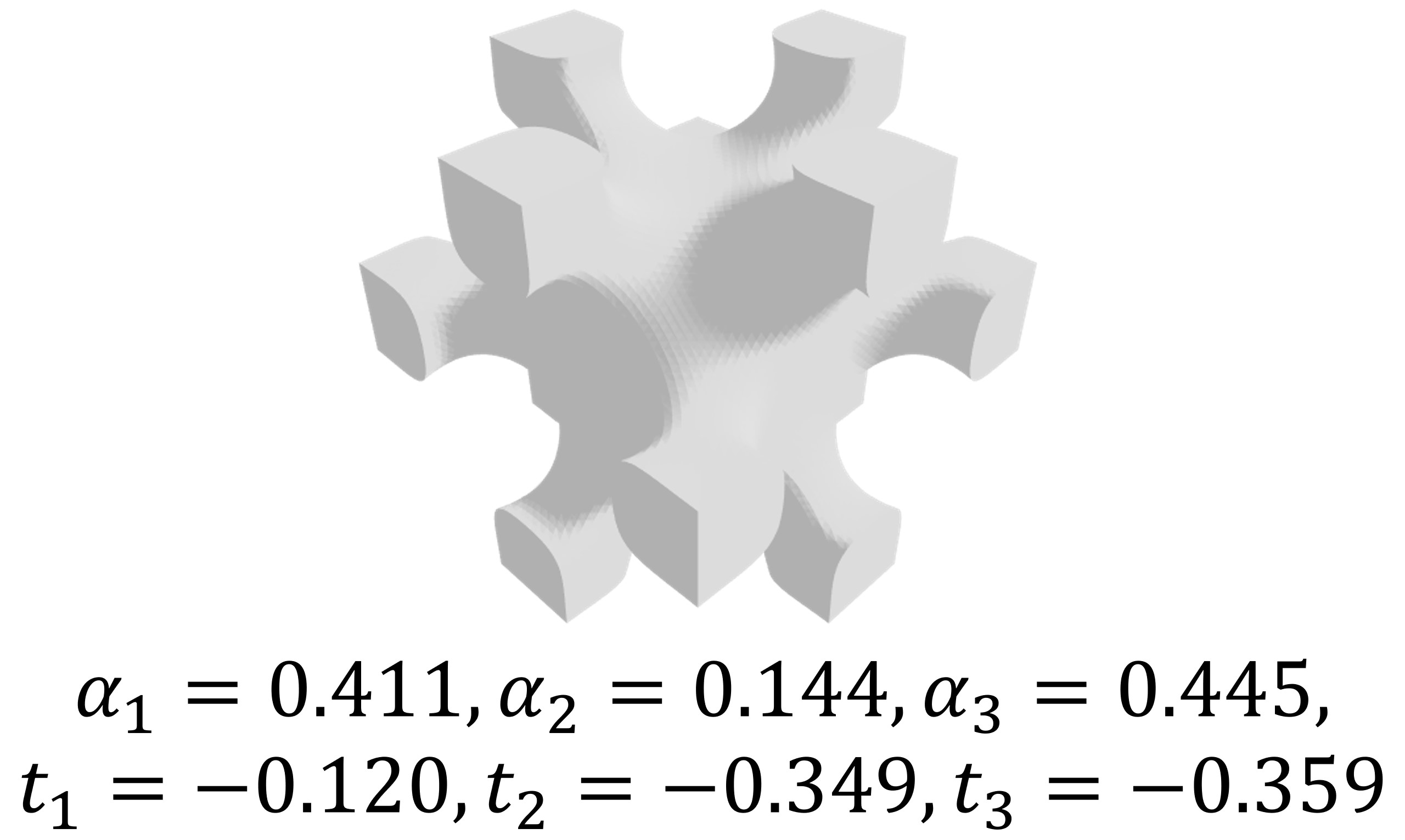}
\caption{}
\end{subfigure}
%
\begin{subfigure}[b]{0.25\linewidth}
\centering
\includegraphics[width=1\linewidth]{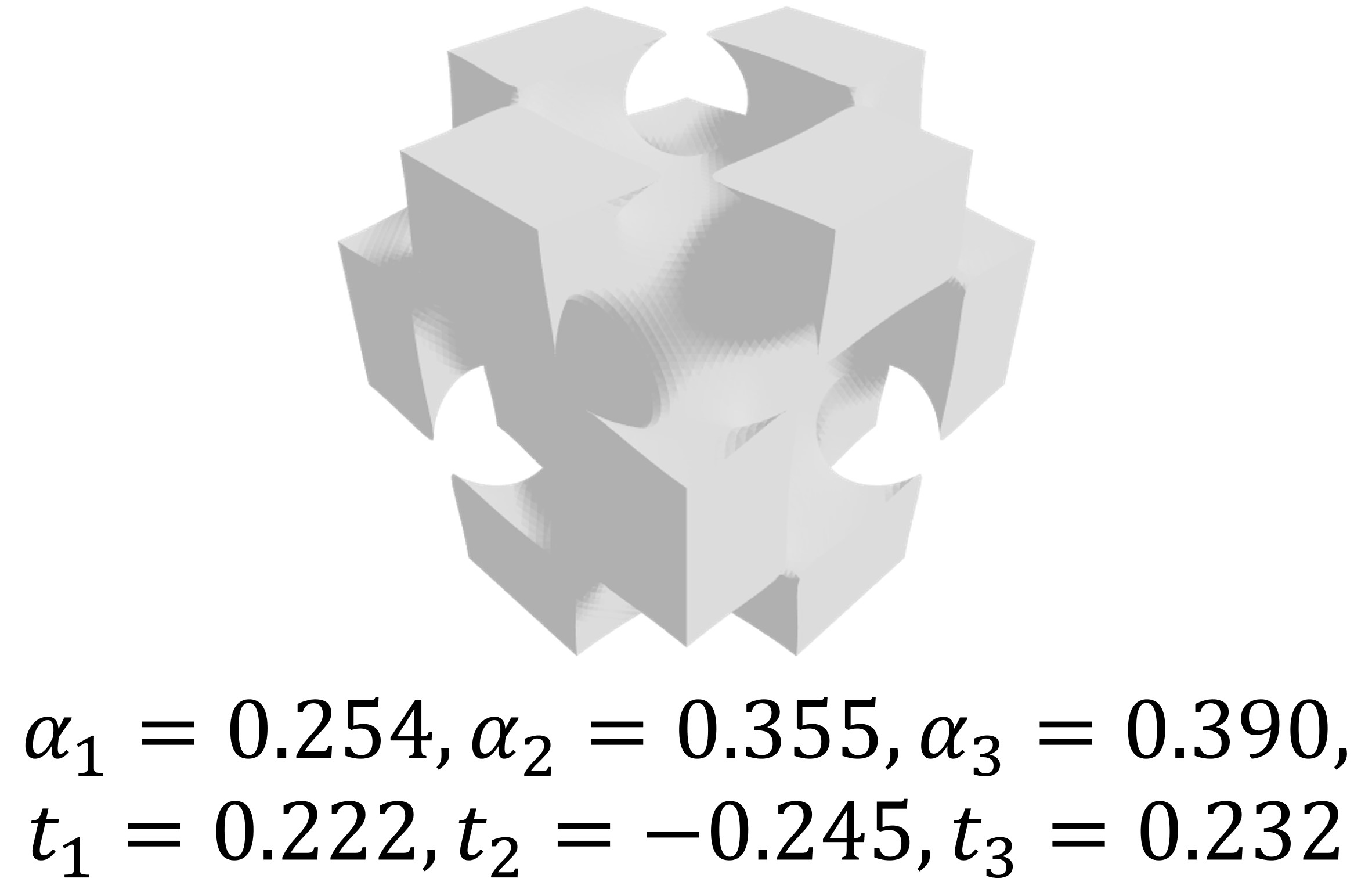}
\caption{}
\end{subfigure}%
\begin{subfigure}[b]{0.25\linewidth}
\centering
\includegraphics[width=1\linewidth]{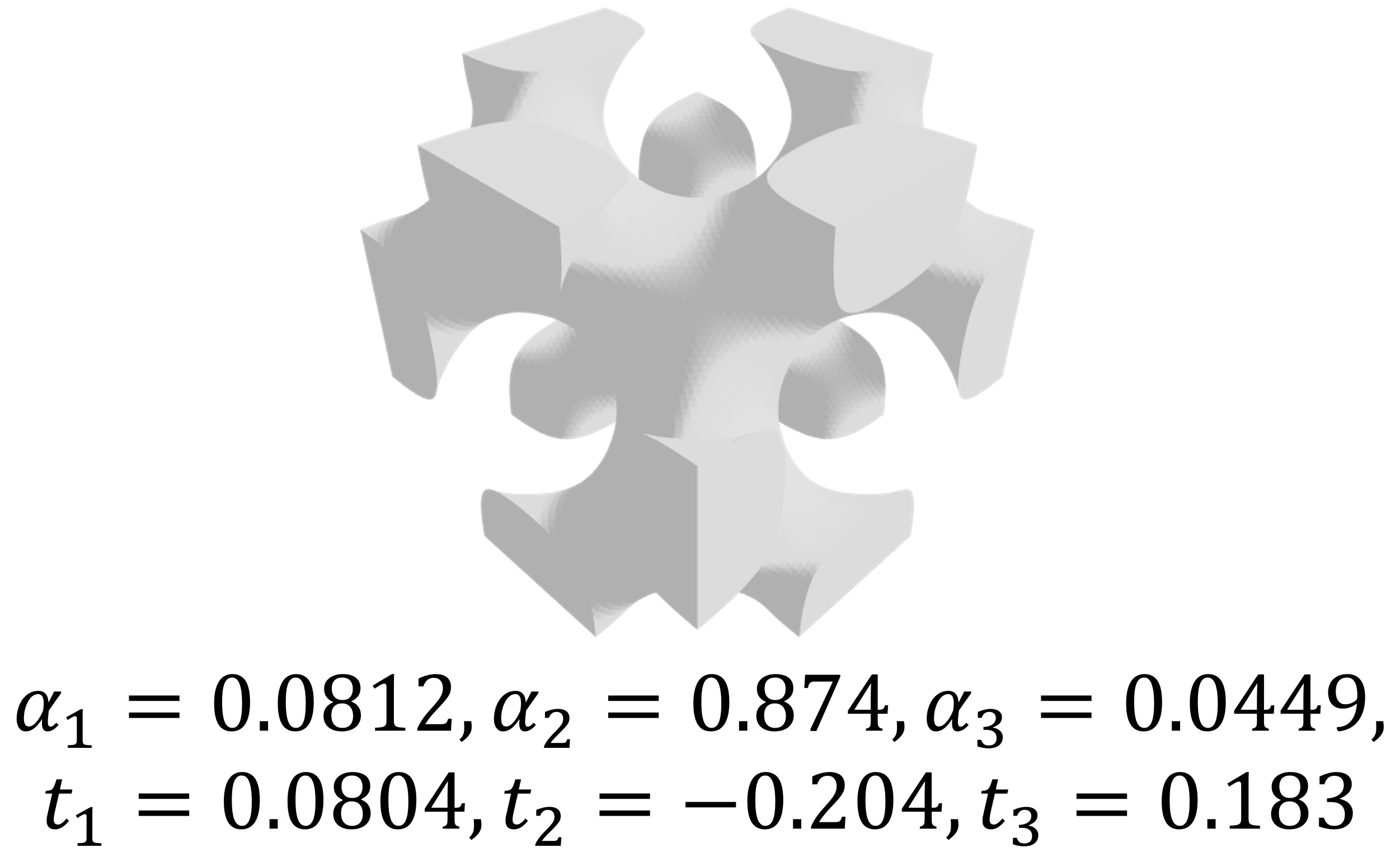}
\caption{}
\end{subfigure}%
\begin{subfigure}[b]{0.25\linewidth}
\centering
\includegraphics[width=1\linewidth]{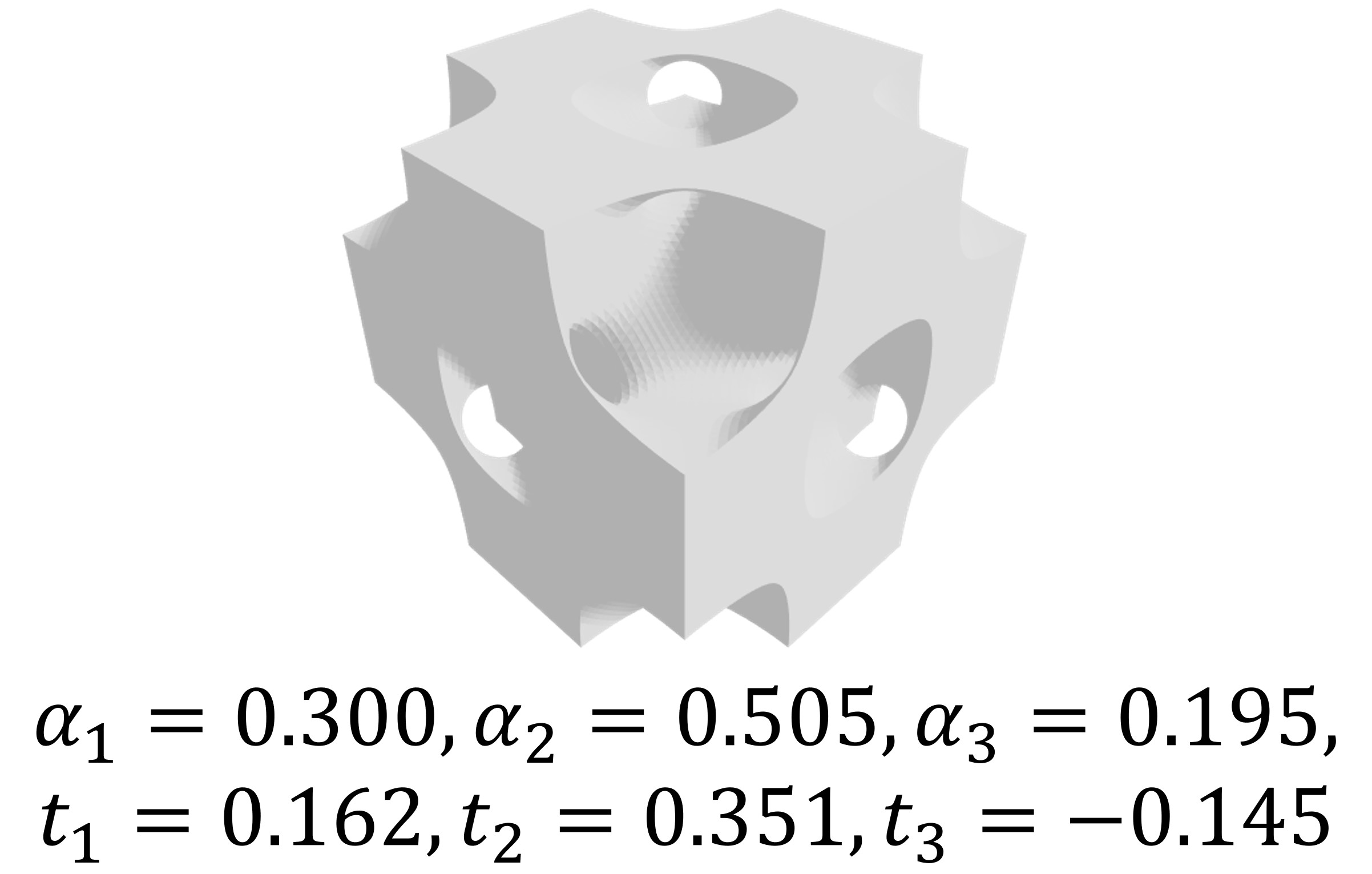}
\caption{}
\end{subfigure}%
\begin{subfigure}[b]{0.25\linewidth}
\centering
\includegraphics[width=1\linewidth]{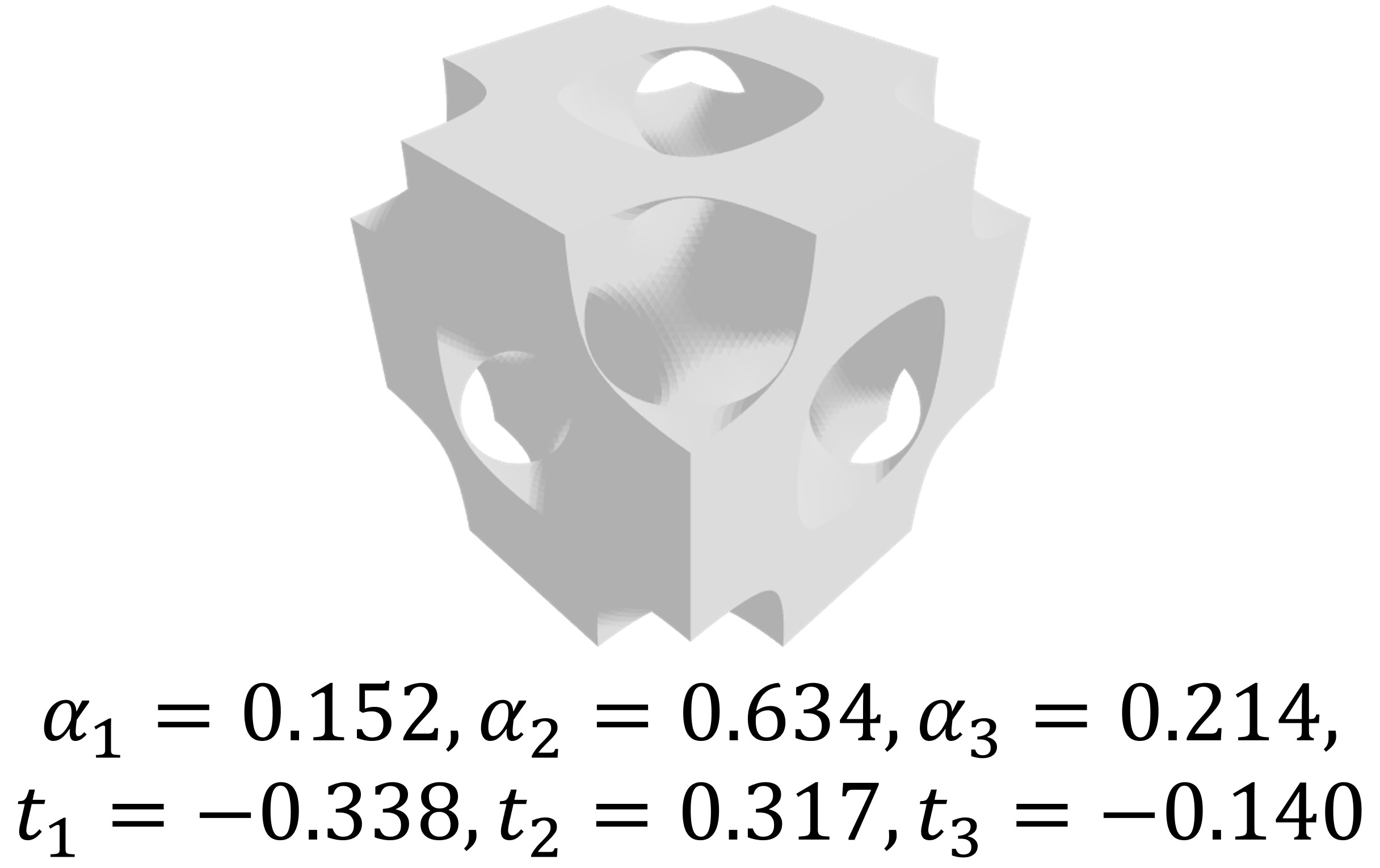}
\caption{}
\end{subfigure}

\caption{Merged unit cells using different shape parameters.}
\vspace{0pt}
\label{fig:merge}
\end{figure}

\begin{figure}[hbt!]
\begin{subfigure}[b]{0.5\linewidth}
\centering
\includegraphics[width=0.78\linewidth]{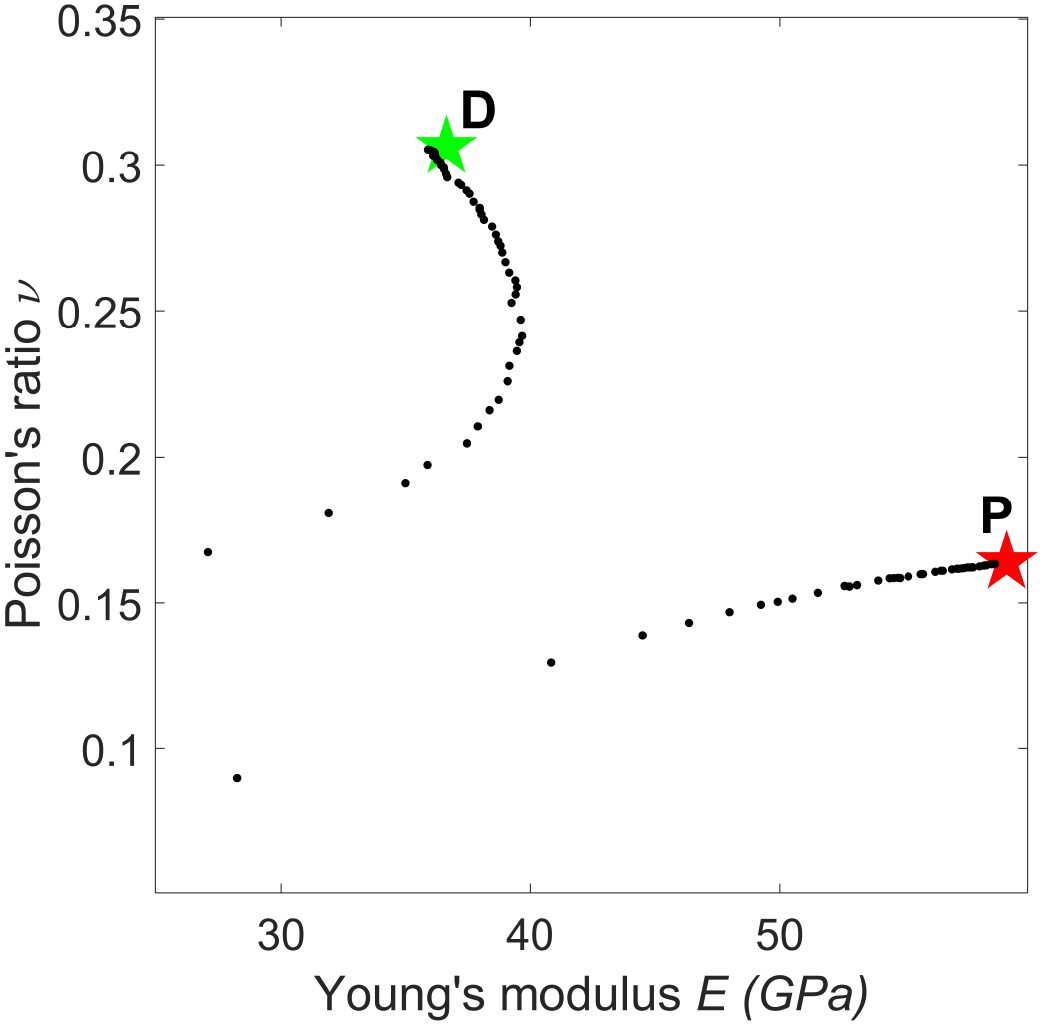}
\caption{}
\end{subfigure}%
\begin{subfigure}[b]{0.5\linewidth}
\centering
\includegraphics[width=0.78\linewidth]{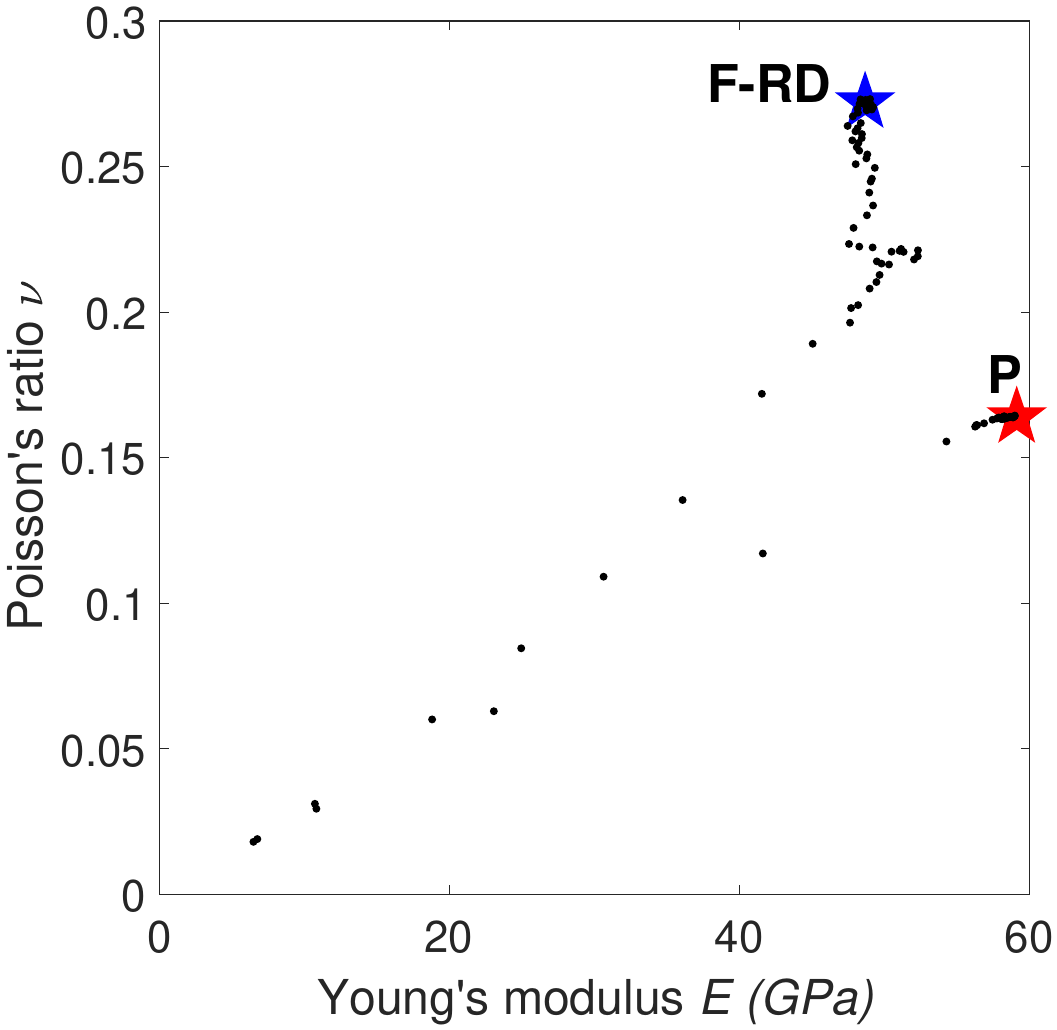}
\caption{}
\end{subfigure}
\begin{subfigure}[b]{0.5\linewidth}
\centering
\includegraphics[width=0.78\linewidth]{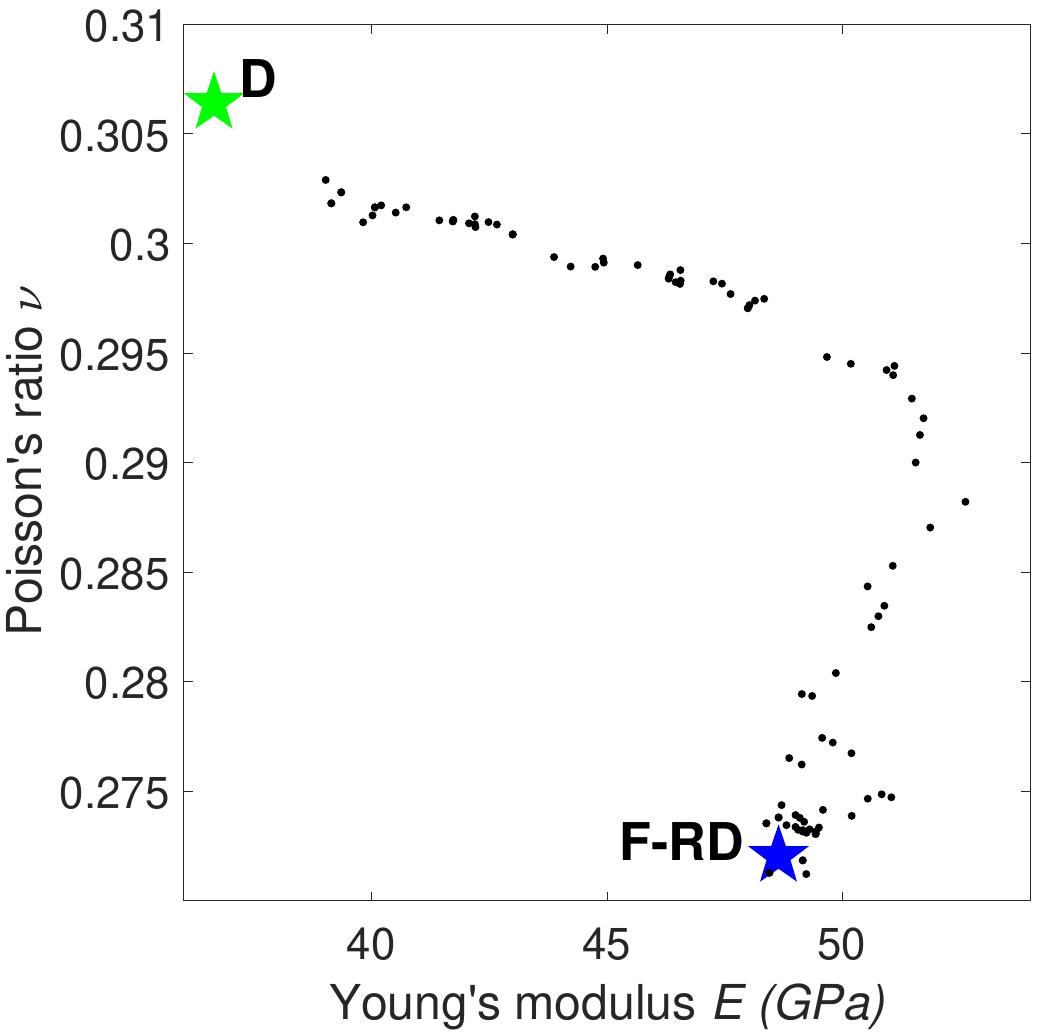}
\caption{}
\end{subfigure}%
\begin{subfigure}[b]{0.5\linewidth}
\centering
\includegraphics[width=0.78\linewidth]{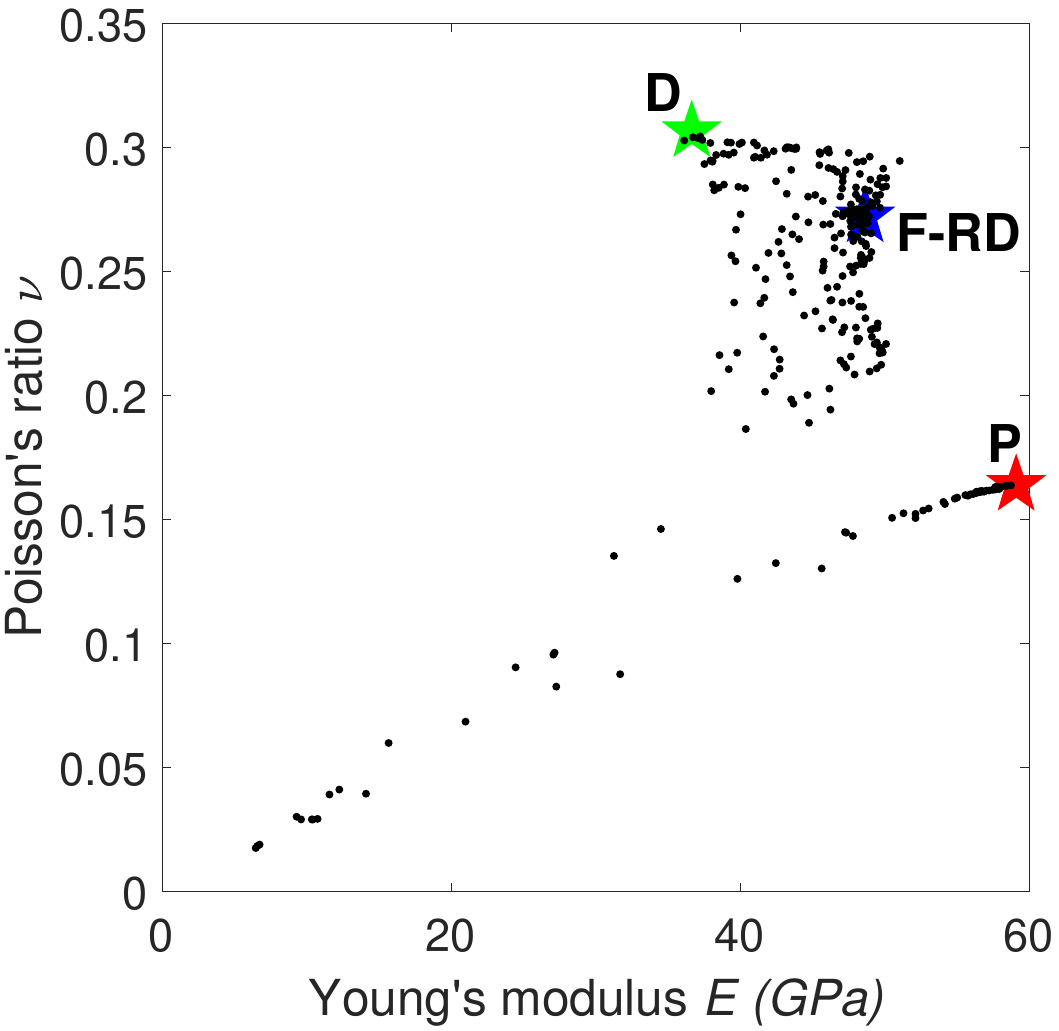}
\caption{}
\end{subfigure}

\caption{Property variation based on varying weights of the merged baseline unit cells (P, D, and F-RD with $t_1=t_2=t_3=0$). (a) Property variation of merged P and D ($\alpha f_P + (1-\alpha)f_D$) with the increasing $\alpha$ ($\Delta \alpha = 0.01$). (b) Property variation of merged P and F-RD ($\alpha f_P + (1-\alpha)f_{FRD}$) with the increasing $\alpha$ ($\Delta \alpha = 0.01$). (c) Property variation of merged D and F-RD ($\alpha f_D + (1-\alpha)f_{FRD}$) with the increasing $\alpha$ ($\Delta \alpha = 0.01$). (d) Property variation of merged P, D, and F-RD (Equation~\ref{eq:tpms}) with 300 samples. The star symbols indicate the locations of the three baseline classes.}
\vspace{0pt}
\label{fig:mergeprop}
\end{figure}

The three TPMS surfaces are merged as a weighted sum of their implicit functions using Equation~(\ref{eq:tpms}):
\begin{equation}
\begin{split}
f_{merge}(x, y, z) &= \alpha_1\big(4f_P(x, y, z)\big) + \alpha_2\big(4f_D(x, y, z)\big) + \alpha_3f_{FRD}(x, y, z), \\
\alpha_1 + \alpha_2 + \alpha_3 &= 1, \\
0 \leq \alpha_1, \alpha_2, \alpha_3 &\leq 1
\end{split}
\label{eq:tpms}
\end{equation}
where $\alpha_1$, $\alpha_2$, and $\alpha_3$ are randomized with a fixed sum of 1 to generate diverse cellular unit cells. The weights of P and D surfaces are augmented by a multiplier (= 4) to balance the proportion of P and D surfaces when merging the three basic TMPS surfaces (shown in Table~\ref{tab_tpms}). The level set value ($t_1$, $t_2$, and $t_3$) in Equation~(\ref{eq:tpms1})-(\ref{eq:tpms3}) determines the volume fraction (\ie, relative density) of a unit cell by thickening or thinning the surfaces. By picking different level set values, we can further increase the variations of these merged unit cells. Figure~\ref{fig:merge} displays some merged unit cells generated by Equation~(\ref{eq:tpms}). In addition to the shape variation results from the merging operation, we also briefly show how the merging works in the property space by combining the three baseline classes of unit cells (\ie, P, D, and F-RD with $t_1=t_2=t_3=0$) in Figure~\ref{fig:mergeprop}. It should be noted that the proposed methodology is not limited to the three classes of TPMS unit cells with cubic symmetry. The method can be generalized to more classes of unit cells (\eg, utilizing more general Fourier series-based or spinodoid unit cells \cite{maldovan2009periodic, wang2020generative, kumar2020inverse} for tunable anisotropy).

\section{Effective properties of homogenized cellular structures} \label{sec:homogenization}

In this paper, a voxel-based numerical homogenization method \cite{dong2019149} is employed to compute the effective elasticity tensor of the TPMS unit cells. Both numerical \cite{coelho2016scale,da2019topology,da2021inverse} and practical experiments \cite{jia2020multiscale} elucidate that the homogenization method could be mechanically admissible to the prediction of the equivalent moduli and validation for the inverse homogenization design even with a small number of unit cell repetitions. From the homogenized constitutive matrix, one can obtain the identical Young's modulus and Poisson's ratio along the axial directions ($x-$, $y-$, and $z-$) due to the cubic symmetry of those TPMS-based cellular structures.

\begin{figure*}[hbt!]
\begin{subfigure}[b]{0.5\linewidth}
\centering
\includegraphics[width=0.6\linewidth]{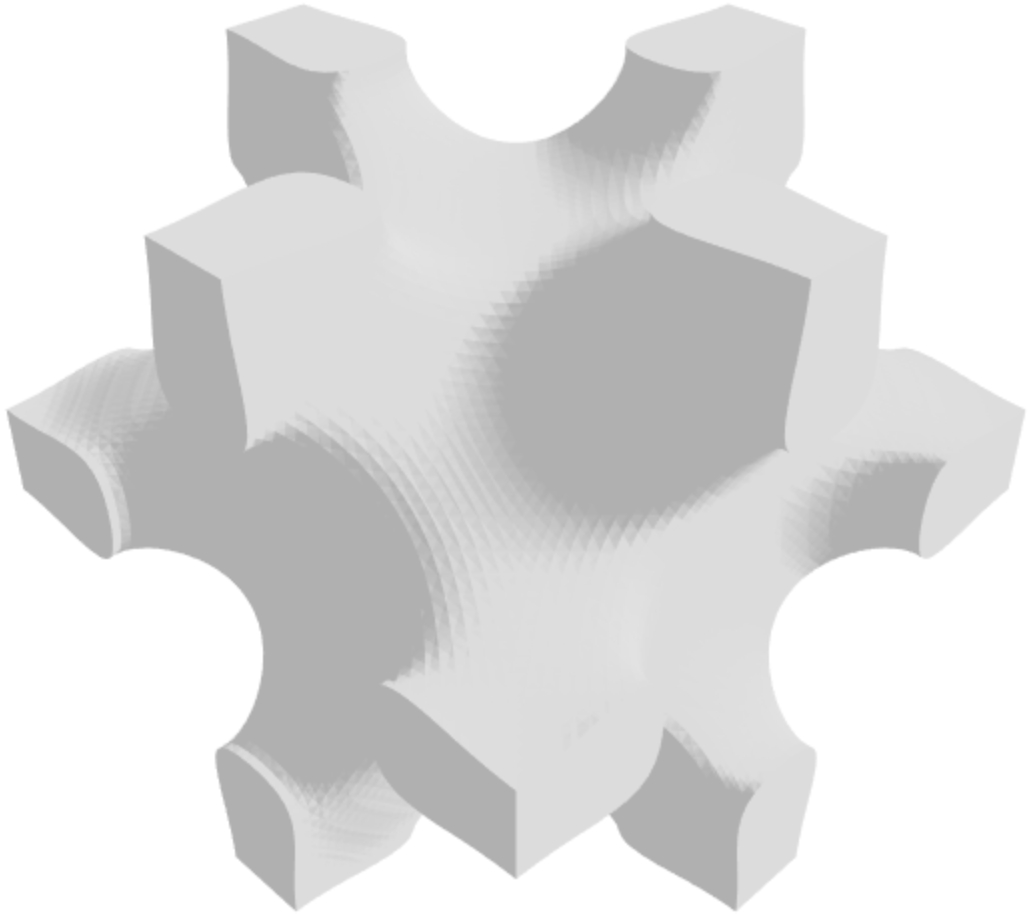}
\caption{}
\end{subfigure}%
\begin{subfigure}[b]{0.50\linewidth}
\centering
\includegraphics[width=0.6\linewidth]{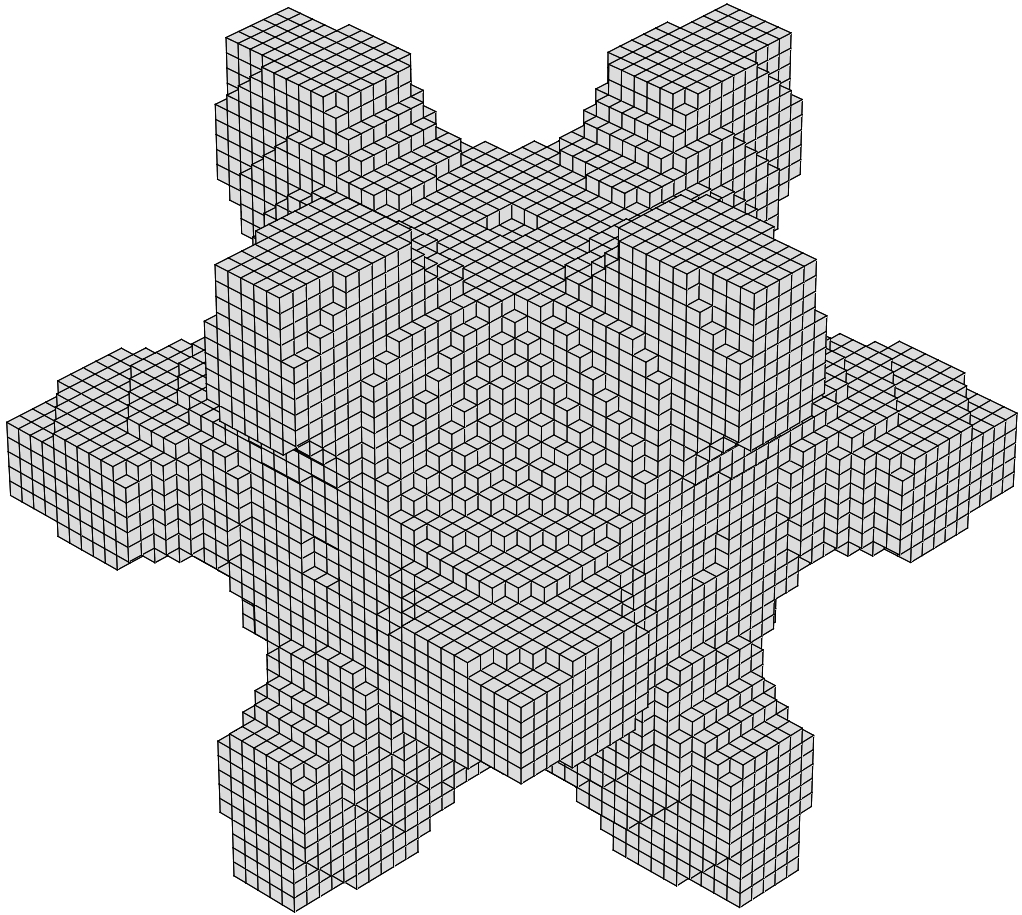}
\caption{}
\end{subfigure}

\caption{Voxelization of a TPMS-based unit cell. (a) Implicit surface. (b) Voxels.}
\label{fig:voxel}
\end{figure*}

The finite element used in this homogenization is an eight-node hexahedron. Therefore, each unit cell is voxelized into $40 \times 40 \times 40$ cubes (Figure~\ref{fig:voxel}). 

 The constitutive matrix $\boldsymbol{C}_e$ of the elementary material that is isotropic over each finite element, and thus can be expressed by Lam\'e's parameters \cite{fischer2005physics} as:
\begin{equation}
\boldsymbol{C}_e =
\begin{bmatrix}
\lambda+2\mu & \lambda     & \lambda     & 0   & 0   & 0\\
\lambda     & \lambda+2\mu & \lambda     & 0   & 0   & 0\\
\lambda     & \lambda     & \lambda+2\mu & 0   & 0   & 0\\
0           & 0           & 0           & \mu & 0   & 0\\
0           & 0           & 0           & 0   & \mu & 0\\
0           & 0           & 0           & 0   & 0   & \mu
\end{bmatrix}
\label{eq:constitutive}
\end{equation}
where the Lam\'e's first and second parameters $\lambda$ and $\mu$ are:
\begin{equation}
\begin{split}
\lambda &= \frac{\nu E}{(1+\nu)(1-2\nu)}, \\
\mu &= \frac{E}{2(1+\nu)}\\
\end{split}
\label{eq:lame}
\end{equation}
where $E$ and $\nu$ are Young's modulus and Poisson's ratio of the elementary material\footnote{In this paper, we assume the material has Young's modulus of 200 GPa and Poisson's ratio of 0.3. Using Equation~(\ref{eq:lame}), the Lam\'e's first and second parameters $\lambda$ and $\mu$ are calculated as 115.4 and 76.9, respectively.}. The elemental stiffness matrix $\boldsymbol{k_e}$ and the global stiffness matrix $\boldsymbol{K}$ assembled for the unit cell are:
\begin{equation}
\boldsymbol{k_e} = \int_{V_e} \boldsymbol{B}^T_e\boldsymbol{C}_e\boldsymbol{B}_e\,dV_e, \ \ \ \boldsymbol{K} = \displaystyle\sum_{e}^{N}\boldsymbol{k_e}
\label{eq:globalstiff}
\end{equation}
where $N$ is the number of finite element inside the unit cell. The load $\boldsymbol{f}^i$ to be used for calculating the displacement field is assembled by:
\begin{equation}
\boldsymbol{f}^i = \displaystyle\sum_{e}\int_{V_e} \boldsymbol{B}^T_e\boldsymbol{C}_e\boldsymbol{\varepsilon}^i\,dV_e
\label{eq:load}
\end{equation}
where the macroscopic strains are chosen as the unit strains $\boldsymbol{\varepsilon}^i$:
\begin{equation}
\begin{split}
\boldsymbol{\varepsilon}^1 &= (1,0,0,0,0,0)^T, \boldsymbol{\varepsilon}^2 = (0,1,0,0,0,0)^T,
\boldsymbol{\varepsilon}^3 = (0,0,1,0,0,0)^T,\\
\boldsymbol{\varepsilon}^4 &= (0,0,0,1,0,0)^T,
\boldsymbol{\varepsilon}^5 = (0,0,0,0,1,0)^T,
\boldsymbol{\varepsilon}^6 = (0,0,0,0,0,1)^T
\end{split}
\label{eq:strains}
\end{equation}
The global displacement fields $\boldsymbol{\upchi}^i$ of the unit cell are achieved by solving the following equation:
\begin{equation}
\boldsymbol{K}\boldsymbol{\upchi}^i = \boldsymbol{f}^i
\label{eq:disp}
\end{equation}
where the displacement vectors $\boldsymbol{\upchi}^i$ are assumed to be V-periodic\textemdash \ie, boundary nodes on three opposite faces of the unit cell will have the same displacement. Finally, each entry in $\boldsymbol{C}^H$ can be obtained by using the following equation:
\begin{equation}
\boldsymbol{C}^H_{ij} = \frac{1}{|V|}\displaystyle\sum_{e}\int_{V_e} (\boldsymbol{\upchi}^{0(i)}_e-\boldsymbol{\upchi}^{(i)}_e)^T\boldsymbol{k}_e(\boldsymbol{\upchi}^{0(j)}-\boldsymbol{\upchi}^{(j)}_e)\,dV_e
\label{eq:constitutiveij}
\end{equation}
where $\boldsymbol{\upchi}^{0(i)}_e$ is the element displacement field that corresponds to the $i$-th unit strain in Equation~(\ref{eq:strains}), and $\boldsymbol{\upchi}^{(i)}_e$ is the corresponding displacement field obtained from globally enforcing the unit strains in Equation~(\ref{eq:disp}). After iterating the six unit strains, the $6 \times 6$ homogenized constitutive matrix $\boldsymbol{C}^H$ is attained. To achieve the unit cell's effective Young's modulus and Poisson's ratio, one still needs to find the inverse of $\boldsymbol{C}^H$\textemdash \ie, the homogenized compliance matrix $\boldsymbol{S}^H$. From the compliance matrix, the cubic symmetric unit cells have:
\begin{equation}
\begin{split}
E^H &= E_x = E_y = E_z = \frac{1}{S^H_{11}} = \frac{1}{S^H_{22}} = \frac{1}{S^H_{33}}, \\
\nu^H &= -S^H_{12}E = -S^H_{13}E = -S^H_{21}E = -S^H_{23}E = -S^H_{31}E = -S^H_{32}E
\end{split}
\label{eq:compliance}
\end{equation}

The relative density ($\rho$) of a unit cell can be computed as a ratio between the number of voxels containing materials (voxel = 1) and the total number of voxels ($40 \times 40 \times 40$).

\begin{figure}[hbt!]
\begin{center}
\includegraphics[width=0.5\linewidth]{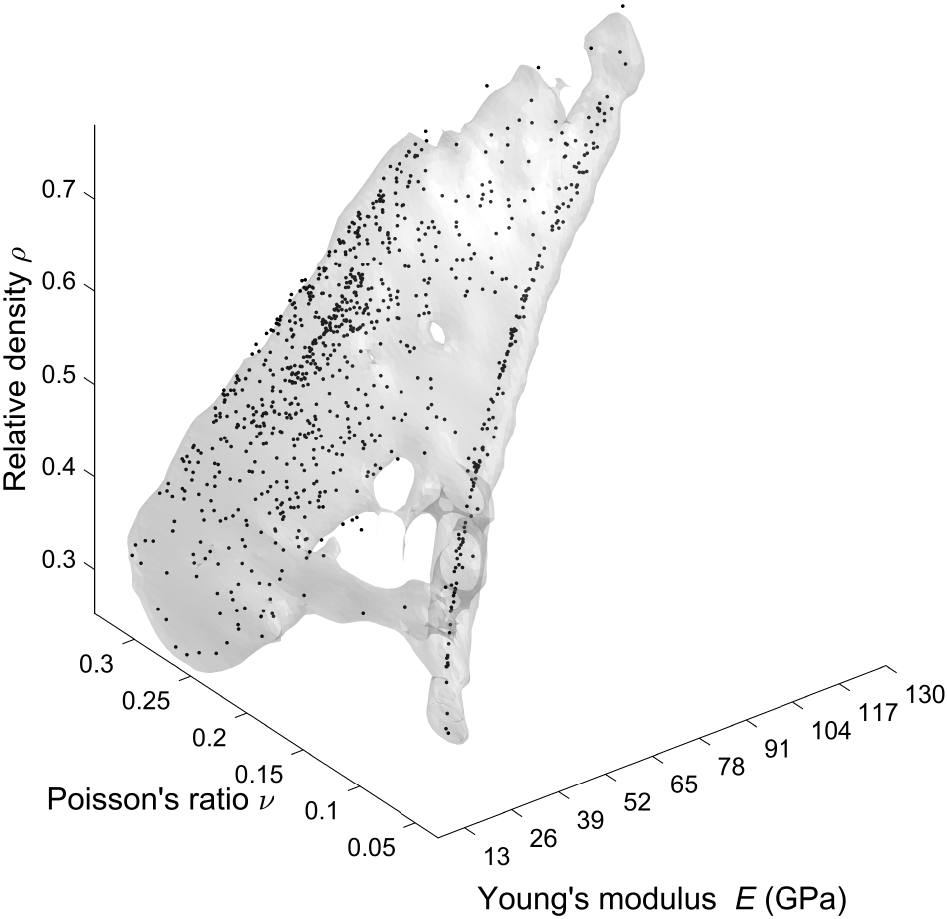}
\caption{Effective elastic properties and relative density of training data.}
\label{fig:training_data}
\end{center}
\end{figure}

Figure~\ref{fig:training_data} shows the coverage of effective elastic properties and relative density of our training data. Note that sparse or no training data is present in some regions (\eg, the lower right region) of the property space. This means that IH-GAN cannot faithfully generate the unit cells with the properties from those regions since it has not seen any data there during training. Therefore, in downstream demos like structural optimizations, we need to constrain the solution space of the properties based on the distribution of our training data. To capture the distribution, we represent the property space using a signed distance field, which is visualized as the gray envelope of 0-level set surface in Figure~\ref{fig:training_data}. We will describe later how we constrain the solution space to the envelope while solving structural optimization problems. 

\section{Inverse homogenization GAN}

\begin{figure}[hbt!]
\begin{center}
\includegraphics[width=0.9\linewidth]{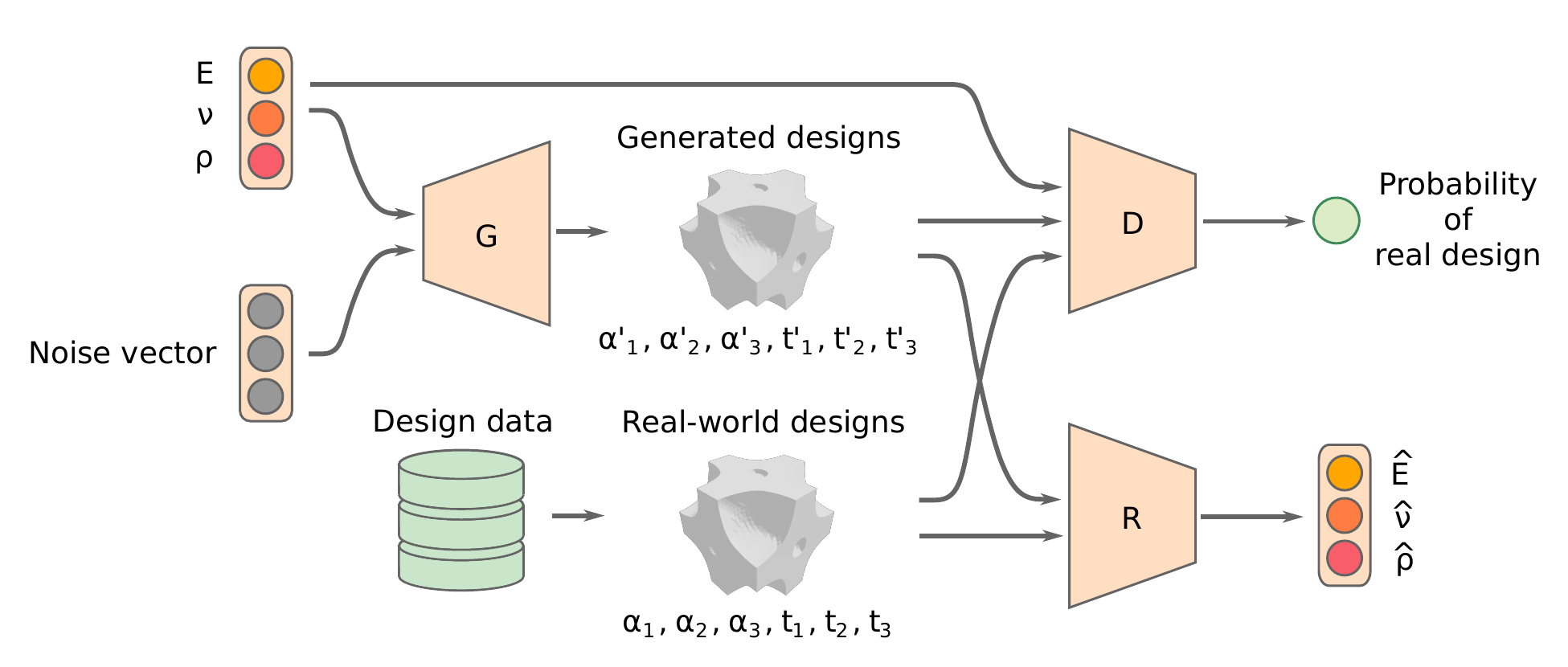}
\caption{Architecture of the IH-GAN. Notations $G$, $D$, and $R$ represent the generator, the discriminator, and the auxiliary property regressor, respectively.}
\label{fig:architecture}
\end{center}
\end{figure}

We propose a conditional GAN, Inverse Homogenization GAN (IH-GAN), to learn the IH mapping~\textemdash~a one-to-many mapping from the property space $\mathcal{M}$ in $\mathbb{R}^3$ to the shape parameter space $\mathcal{S}$ in $\mathbb{R}^6$ (Figure~\ref{fig:architecture}). Specifically, $\mathcal{M}$ is defined by the Young's modulus $E$, Poisson's ratio $\nu$, and relative density $\rho$ of a unit cell; while $\mathcal{S}$ is defined by the six shape parameters $\alpha_1$, $\alpha_2$, $\alpha_3$, $t_1$, $t_2$, and $t_3$ from Equation~(\ref{eq:tpms1})-(\ref{eq:tpms3}) and (\ref{eq:tpms}). The generator $G$ has both the properties and the noise vector $\mathbf{z}$ as inputs. The noise is drawn from a predefined prior distribution $P_{\mathbf{z}}$ (\eg, multivariate normal distribution). Given $E$, $\nu$, and $\rho$, we can draw multiple unit cell shapes as the output of $G$ by sampling noise vectors from $P_{\mathbf{z}}$ as the input to $G$.

We further improve the IH mapping by ensuring the generated shape parameters can be accurately mapped back to their corresponding properties. Specifically, we add an \textit{Auxiliary Property Regressor} ($R$) to predict each design's properties (Figure~\ref{fig:architecture}). This leads to two additional loss terms:
\begin{equation}
L_r(R) = \mathbb{E}_{\mathbf{x}\sim P_{data}}\left[|\mathbf{y}-R(\mathbf{x})|\right],
\label{eq:l_r}
\end{equation}
and
\begin{equation}
L_g(G) = \mathbb{E}_{\mathbf{z}\sim P_{\mathbf{z}}}\left[|\mathbf{y}-R(G(\mathbf{z}|\mathbf{y}))|\right],
\label{eq:l_g}
\end{equation}
where $\mathbf{x}=(\alpha_1, \alpha_2, \alpha_3, t_1, t_2, t_3)$, and $\mathbf{y}=(E, \nu, \rho)$. Note that we cannot update $R$ in Eq.~(\ref{eq:l_g}) because, as a regression model, $R$ requires labeled data during training, but the geometries generated by $G$ do not have pre-existing labels (\ie, the actual properties\footnote{Ideally, we could apply homogenization for each generated geometry during training to obtain the actual material properties, but this would tremendously increase the training time.}).

Similar architecture is also used in two existing GAN variants: (1)~To maximize the mutual information between the generated sample $\mathbf{x}$ and its latent code $\mathbf{c}$, InfoGAN~\cite{chen2016infogan} uses an auxiliary network to approximate the conditional latent code distribution $P(\mathbf{c}|\mathbf{x})$; and (2)~To improve sample quality, the auxiliary classifier GAN (AC-GAN)~\cite{odena2017conditional} uses an auxiliary classifier to predict the probability distribution $P(\mathbf{l}|\mathbf{x})$ over the class labels $\mathbf{l}$.

The overall loss function of IH-GAN thus combines the conditional GAN's loss, Equation~(\ref{eq:l_g}), Equation~(\ref{eq:l_r}) with a hyperparameter $\gamma$:
\begin{equation}
\min_{G,R}\max_D ~~ V_{\text{IH-GAN}}(D,G,R) = V_{\text{cGAN}}(D,G) + \gamma \left(L_g(G) + L_r(R)\right).
\label{eq:ihgan_loss}
\end{equation}

There are two ways of training IH-GAN:
\begin{enumerate}
    \item Fix $G$ when updating $D$ and $R$, and fix $D$ and $R$ when updating $G$.
    \item Use a pre-trained $R$. Fix $G$ when updating $D$, and fix $D$ when updating $G$ (standard GAN training).
\end{enumerate}

We used the first approach in our experiments. As mentioned earlier, we cannot train $R$ together with $G$ because the geometries generated by $G$ do not have pre-existing labels (\ie, the actual properties), but $R$ as a regression model requires labeled data during training.

\section{IH-GAN for functionally graded cellular structural design} \label{sec:TO}

As the proposed IH-GAN model can accurately generate cellular structures possessing the desired properties, it can be useful in designing functionally graded cellular structures composed of multiple types of cellular unit cells. Rather than typical TO algorithms (\eg, SIMP) that optimize a density map by assuming an approximate relation between Young's modulus ($E$) and the density ($\rho$), our IH-GAN model uses a modified version of the SIMP algorithm that can output all the three maps of $E$, $\nu$, and $\rho$. Based on the $E$, $\nu$, and $\rho$ maps, our IH-GAN generates corresponding cellular unit cells to replace the three maps and end up with a functionally graded cellular structure. We use two different elasticity objective functions\textemdash \ie, minimum compliance and target deformation\textemdash in our optimization problems.

\subsection{Minimum compliance}

We first use the same objective ($\boldsymbol{C_c}$) as the one used in the standard TO algorithm. With $E$, $\nu$, and $\rho$ as the design variables, our optimization problem is defined as:
\begin{mini}|l|
{E, \nu, \rho}{\boldsymbol{C_c} = \boldsymbol{u}^T\boldsymbol{K}(\boldsymbol{E},\boldsymbol{\nu})\boldsymbol{u}}{}{}{}
\addConstraint{&\boldsymbol{K}(\boldsymbol{E},\boldsymbol{\nu})\boldsymbol{u}} {\hspace{-50pt} - \boldsymbol{f}_{ext} = 0} {}
\addConstraint{&\Phi(E_i,\nu_i,\rho_i) \leq 0,} {\hspace{15pt} i = 1,..., N_e} {}
\addConstraint{&\displaystyle\sum_{i=1}^{N_e}\rho_iv_i} {\hspace{-35pt} \leq \hat{V},} {}
\label{eq:TO_cp}
\end{mini}
where $\boldsymbol{E}$ and $\boldsymbol{\nu}$ are the vectors of the element Young's modulus and Poisson's ratio, $\boldsymbol{K}$ is the global stiffness matrix, $\boldsymbol{u}$ is the displacement vector, and $\boldsymbol{f}_{ext}$ represents the external loads applied to the object. The equality constraint is the static elasticity (Hooke's law) equilibrium equation. 

The first inequality constraint $\Phi \leq 0$ guarantees each cellular unit cell's properties have high probability density in the training data distribution (Figure~\ref{fig:training_data}) so that IH-GAN can faithfully generate shape parameters given $E$, $\nu$, and $\rho$. It can have the following form:
\begin{equation}
\Phi(E_i,\nu_i, \rho_i) = \tau - \text{Pr}(E_i,\nu_i, \rho_i;\theta),
\end{equation}
where the probability density function {$\text{Pr}(E_i,\nu_i,\rho_i;\theta)$} can be estimated by methods like kernel density estimation (KDE), $\theta$ denotes the parameters of the estimated distribution, and $\tau$ is a threshold on the probability density. {In this paper, we use the implicit distance function $\Phi$ \cite{zhu2005animating} of the training data's property space $\mathcal{M}$ instead to form nonlinear constraints:}
\begin{equation}
\Phi(p) = \|p - \Bar{p}\| - r, \hspace{15pt} p = (E_i,\nu_i, \rho_i) \in \mathcal{M},
\end{equation}
{where $\|\cdot\|$ calculates the Euclidean distance between two points in $\mathcal{M}$, and $\Bar{p}$ is the average position of the neighboring points of $p$ with a range of $2r$, where $r$ is typically the average spacing of the points in $\mathcal{M}$.}

{The second inequality constraint controls the overall mass $\hat{V}$, where $v_i$ denotes the $i$-th element volume.} 

\subsection{Target deformation}

We then take a vector of nodal target displacements and boundary conditions as input. The target deformation problem optimizes the material distributions (\ie, $E$ and $\nu$ maps) over the design domain to attain the desired linear deformation assuming a linear elastic behavior. The deformation objective function is defined as:
\begin{mini}|l|
{E, \nu}{\boldsymbol{C_d} = (\boldsymbol{u-\hat{u}})^T\boldsymbol{D}(\boldsymbol{u-\hat{u}}),}{}{}{}
\addConstraint{&\Phi(E_i,\nu_i) \leq 0,} {\hspace{15pt} i = 1,..., N_e} {}
\label{eq:TO_df}
\end{mini}
where $\boldsymbol{\hat{u}}$ is the vector of the target displacements, and $\boldsymbol{D}$ is a diagonal matrix that is used to define the query nodes of interest. For example, we can focus on a certain portion of the design domain by uniformly defining 31 query points at the bottom of the beam (Table~\ref{tableDeformation}). 

To achieve the desired displacement, we relax the constraints by removing the overall mass constraint, and thus $\Phi(E_i,\nu_i)$ becomes a 2D distance function. The solutions of the target deformation problem contain two maps (\ie, $\boldsymbol{E}$ and $\boldsymbol{\nu}$), which are combined and fed into the IH-GAN model to generate the 6 shape parameters ($\alpha_1, \alpha_2, \alpha_3, t_1, t_2, t_3$). The corresponding unit cell shapes are then created using Equation~(\ref{eq:tpms}).

\section{Numerical experiments} \label{result}

In this section, we introduce a new unit cell shape dataset and our experimental settings. We release both our dataset and code at \url{https://github.com/IDEALLab/IH-GAN_CMAME_2022}.

\subsection{Datasets}

To train the IH-GAN, we use a unit cell shape database containing the shape parameters and the properties (\ie, effective elastic properties and relative density) of 924 unit cells.

\paragraph{Shape parameters} Using the design method in Section~\ref{sec:TPMS}, each cellular unit cell can be represented by six parameters $(\alpha_1, \alpha_2, \alpha_3, t_1, t_2, t_3)$. To produce a diverse dataset, we first randomly generate $N$ groups of $\alpha_1$, $\alpha_2$, and $\alpha_3$ with a fixed sum of 1. Next, we generated $N$ groups of $t_1$, $t_2$, and $t_3$ through a Latin hypercube sampling \cite{menvcik2016latin} strategy to evenly cover the level set space. In this paper, we have $t_1$, $t_2$, $t_3$ $\in$ $[-0.4, 0.4]$ to avoid modeling failures\textemdash \ie, breaks due to a low density and fully solid due to a high density. We excluded shapes having small cross section areas or having zero contact areas with neighboring cells. The final dataset contains 924 different unit cells.

\paragraph{Properties} The dataset of effective elastic properties is collected by homogenizing each of the 924 unit cells using the approach described in Section~\ref{sec:homogenization}. In our property dataset, each unit cell's properties are represented as a set of three parameters (\ie, $E^H$, $\nu^H$, and $\rho$).

The database is then split with a ratio of 8:2 for training (739) and evaluation (185), respectively. 

\subsection{IH-GAN model configuration and training}

\begin{figure}[hbt!]
\begin{center}
\includegraphics[width=1\linewidth]{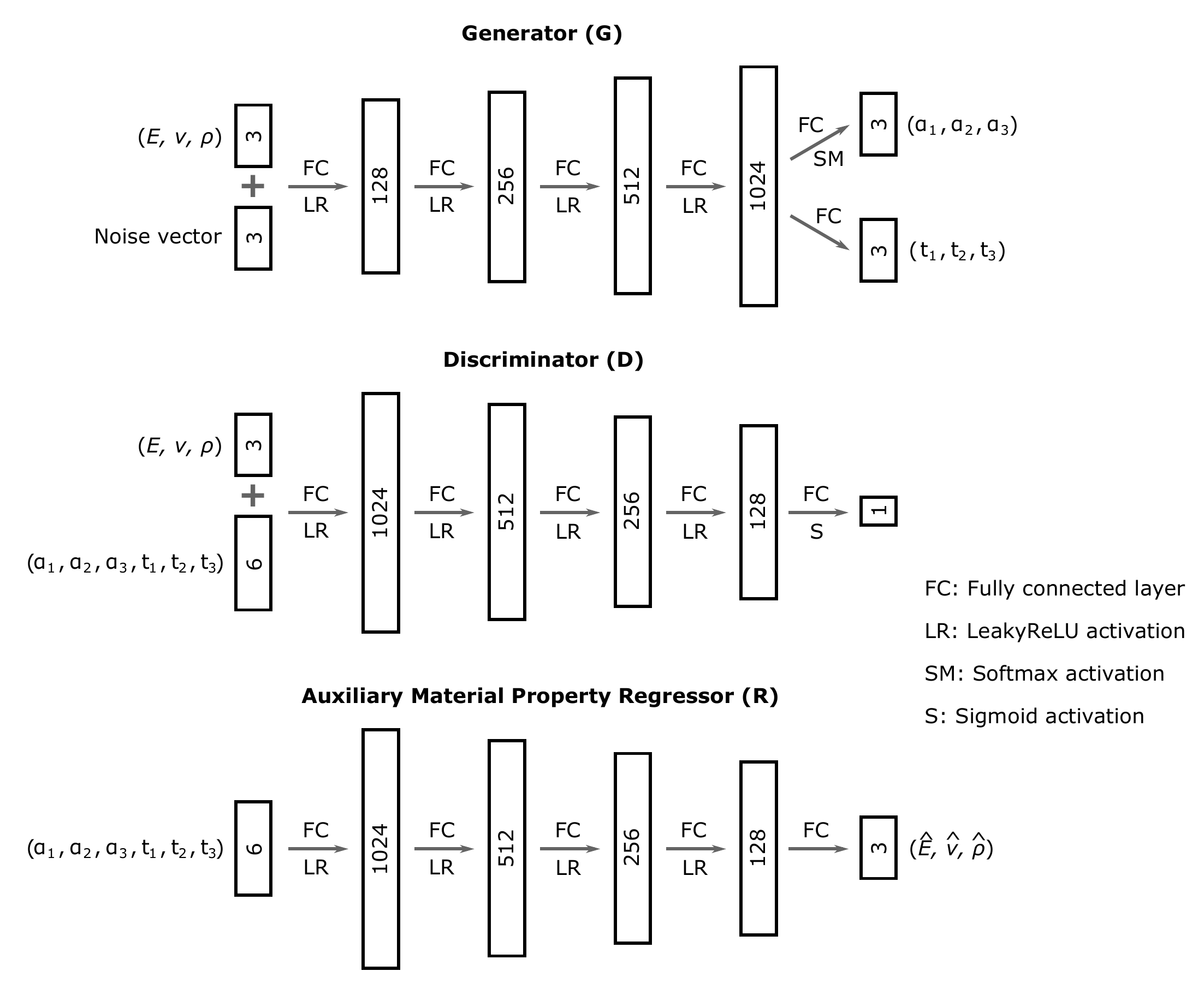}
\caption{Neural network configuration used for experiments.}
\label{fig:configuration}
\end{center}
\end{figure}

The generator, the discriminator, and the auxiliary regressor are fully-connected neural networks with the architectures shown in Figure~\ref{fig:configuration}. The unit cell shapes are represented as 6-dimensional vectors. The condition vectors (\ie, properties $E$, $\nu$, and $\rho$) are 3-dimensional vectors. We set the noise input $\mathbf{z}$ as a 3-dimensional vector drawn from a standard multivariate normal distribution. We set $\lambda=20$ in Equation~(\ref{eq:ihgan_loss}). We used 5000 training iterations during training, with each iteration randomly sampling 32 examples as a mini-batch. The same learning rate of 0.0002 is used for optimizing the generator, the discriminator, and the auxiliary regressor. This simple neural network configuration allows a wall-clock training time of 38 seconds on a GeForce GTX TITAN X. The inference time is less than one second.



\section{Results and discussion} \label{results_discussion}

\subsection{Performance of IH-GAN}
\label{sec:ihgan_perf}


\begin{figure}[hbt!]
\begin{center}
\includegraphics[width=1\linewidth]{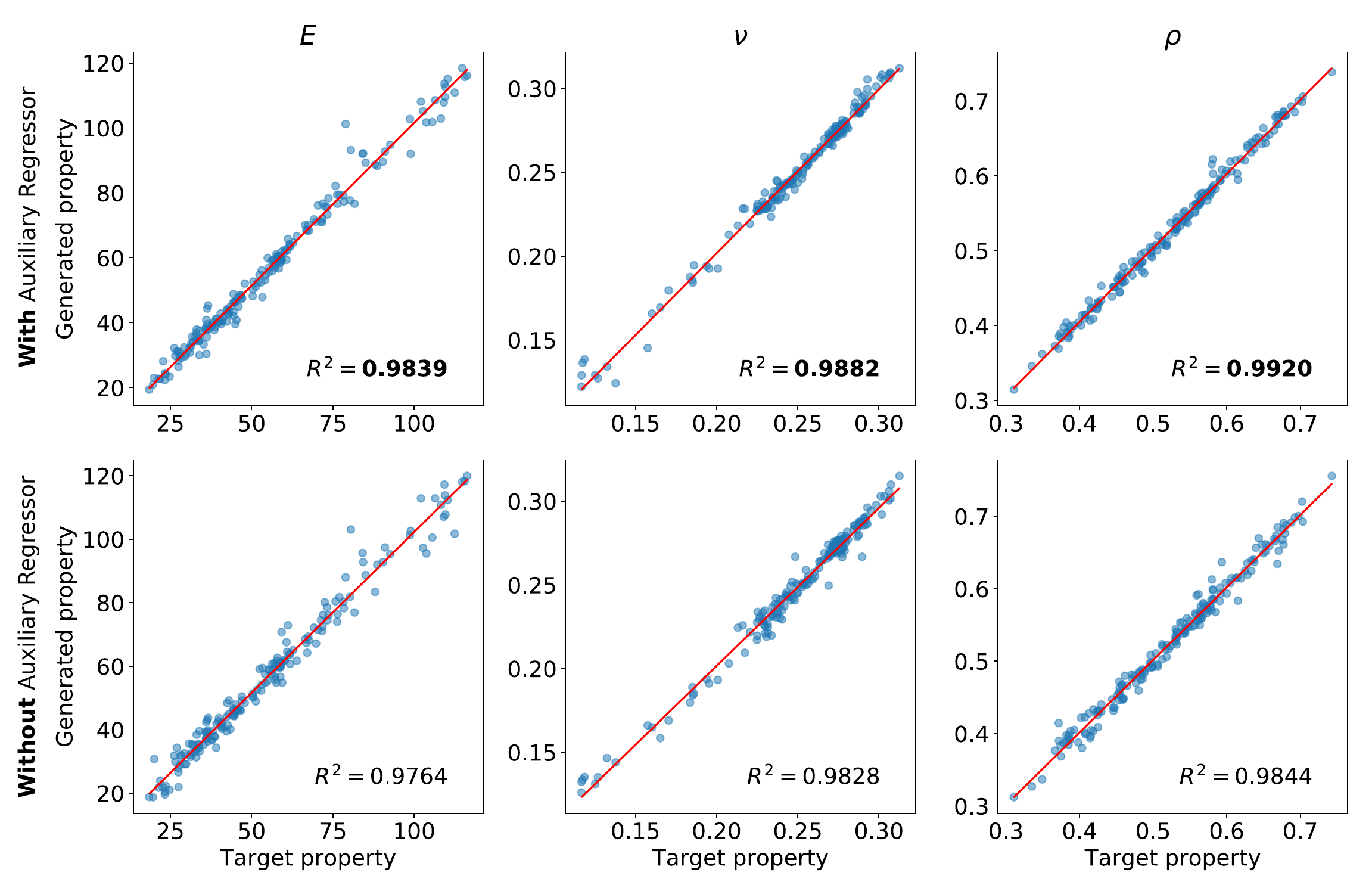}
\caption{Coefficients of determination ($R^2$-scores) showing how well the actual properties ($E'$, $\nu'$, and $\rho'$) of generated unit cells match the target ($E$, $\nu$, and $\rho$). Unit cells generated by the GAN with an auxiliary regressor show higher $R^2$-scores.}
\label{fig:test_r2}
\end{center}
\end{figure}



To evaluate whether the generated unit cells have the exact properties on which the geometries are conditioned, we compute the \textit{property error} on the test dataset.

Specifically, we use the properties $(E, \nu, \rho)$ from the test set as conditions for IH-GAN to generate corresponding unit cell geometries. Then we evaluate those geometries for their actual effective material properties $(E', \nu')$ via homogenization and compute their actual volume fractions $\rho'$. We quantify how well the actual properties $(E', \nu', \rho')$ match the target ($E$, $\nu$, and $\rho$) using the coefficient of determination ($R^2$-score):
\begin{equation}
R^2 = 1 - \frac{\sum^N_{i=1}\|y_i-y'_i\|^2}{\sum^N_{i=1}\|y_i-\bar{y}_i\|^2},
\label{eq:r_squared}
\end{equation}
where $y_i$ denotes the target property ($E$, $\nu$, or $\rho$), $y'_i$ denotes the actual property ($E'$, $\nu'$, or $\rho'$) of generated unit cells, and $\bar{y}_i$ denotes the mean of the target property. Results are shown in Figures~\ref{fig:test_r2}. To demonstrate the effects of the auxiliary regressor, we also compute the errors when it is removed. Figure~\ref{fig:test_r2} shows that all the $R^2$-scores are higher than 0.97, and using an auxiliary regressor improves the $R^2$-scores in general.
\color{black}






\begin{figure}[hbt!]
\begin{center}
\includegraphics[width=1\linewidth]{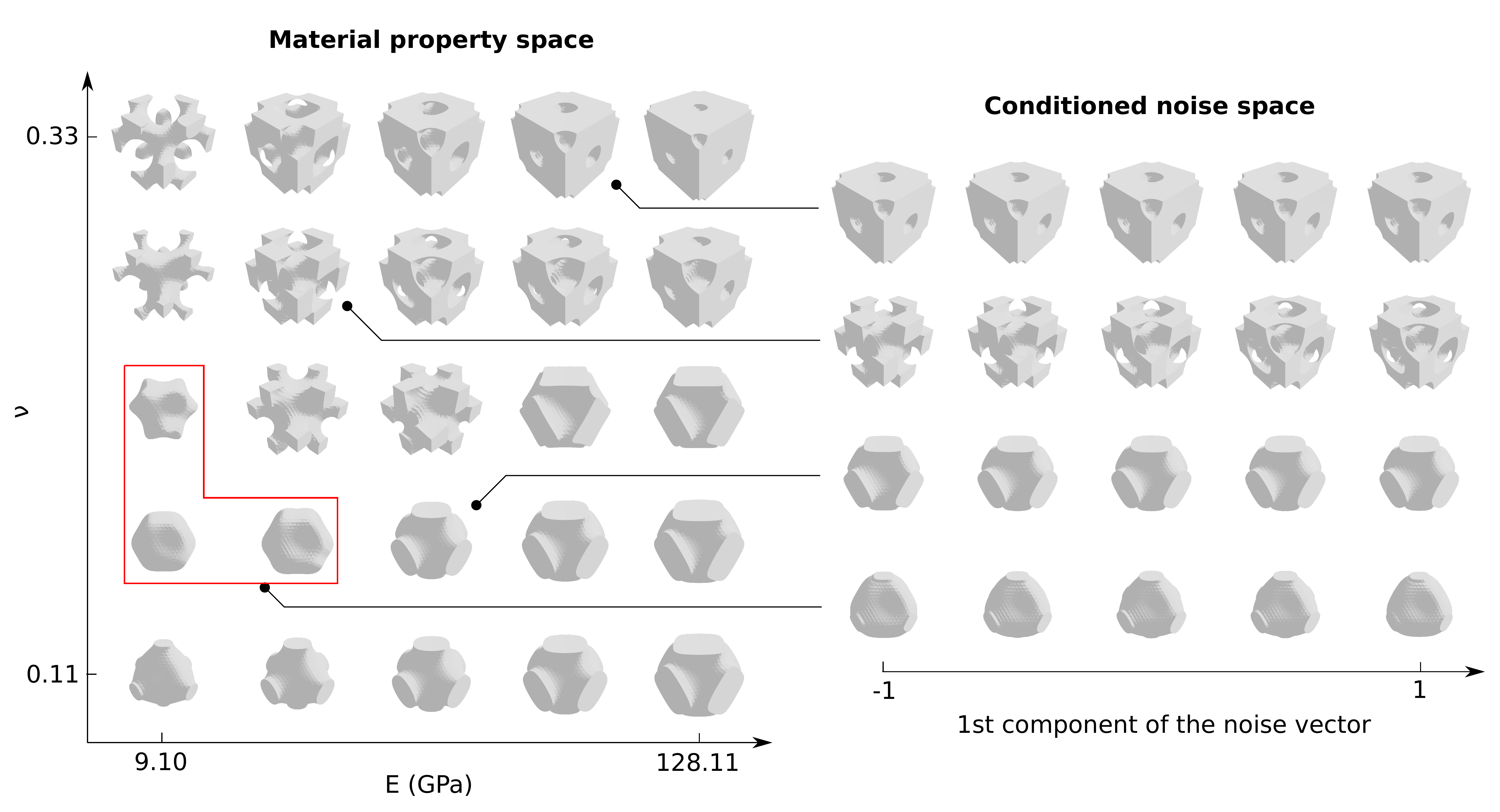}
\caption{The effects of material properties ($E$ and $\nu$ as conditions) and noise on synthesized unit cells. \textit{Left}: synthesized unit cells conditioned on different material properties and fixed noise. Unit cells in the red box are generated in the region of the property space where there are no real-world property data. \textit{Right}: each row shows synthesized unit cells by conditioning on certain properties while varying the first noise variable.}
\label{fig:condition_space}
\end{center}
\end{figure}

\begin{figure}[hbt!]
\begin{center}
\includegraphics[width=1\linewidth]{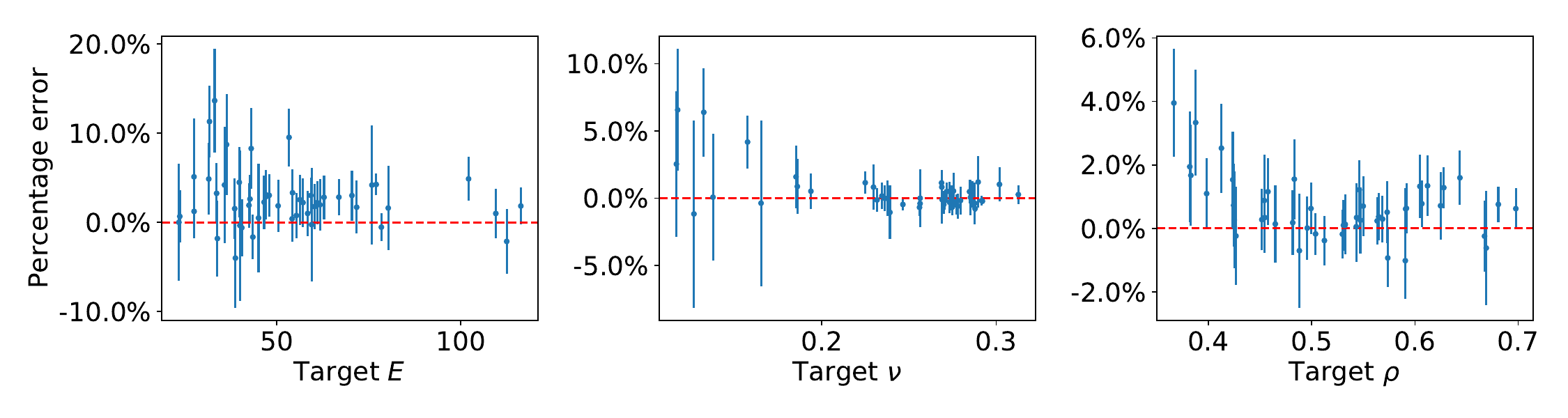}
\caption{Distribution of property errors when randomly perturbing the noise input (points and vertical lines denote mean and standard deviation, respectively).}
\label{fig:test_dev}
\end{center}
\end{figure}

On the left of Figure~\ref{fig:condition_space}, we visualize the generated shape variation in a continuous 2D material property space, \ie, how the generated unit cell shape changes when varying Young's modulus $E$ and Poisson's ratio $\nu$. It shows that there is a strong correlation between $E$ and the unit cell's volume fraction. Meanwhile, as $\nu$ increases, mass transports from the center of the unit cell to the periphery. These observations are consistent with physical intuition. Note that the true property space may contain infeasible regions where there are no real-world data (\eg, the lower left and lower right regions). In those regions, IH-GAN generates either invalid unit cells (\ie, round-shaped unit cells at the lower left, which have zero contact areas with neighboring cells, as indicated by the red box) or invariant shapes (\ie, shapes at the lower right corner). When IH-GAN is used for downstream tasks, we need to exclude those infeasible regions by bounding $E$ and $\nu$. On the right of Figure~\ref{fig:condition_space}, we show how shape varies with the noise vector given fixed $E$ and $\nu$ as the condition. We obtain different unit cells by perturbing the noise input to the generator. This results in a one-to-many mapping from the property space to the shape parameter space. The magnitude of shape variation, however, differs when conditioned on different properties. 

Ideally, for an inverse homogenization mapping, given fixed properties $(E, \nu, \rho)$ as conditions, the noise input would only change the geometry of generated unit cells; whereas the properties they actually possess would be fixed as $(E, \nu, \rho)$. In reality, given fixed $(E, \nu, \rho)$ as conditions (targets) of generated geometries, their actual properties $(E', \nu', \rho')$ may deviate from $(E, \nu, \rho)$ as we perturb the noise input. To test how far $(E', \nu', \rho')$ deviates from $(E, \nu, \rho)$, we take 50 sets of $(E, \nu, \rho)$ as conditions, under each of which we randomly perturb the noise inputs to generate 50 unit cells and compute the percentage errors between $(E, \nu, \rho)$ and $(E', \nu', \rho')$:
\begin{equation}
\epsilon_i = (y'_i - y_i)/y_i,
\label{eq:error}
\end{equation}
where $y_i$ and $y'_i$ are the target and the actual properties, respectively. The results are shown in Figure~\ref{fig:test_dev}. The errors for Young's modulus have a larger variation caused by noise perturbation than Poisson's ratio. Despite some outliers, most percentage errors for $E$, $\nu$, and $\rho$ have mean values within [-3\%, 5\%], [-2\%, 2\%], and [-2\%, 3\%], respectively; and standard deviations within 7\%, 4\%, and 2\%, respectively.

\subsection{Functionally graded cellular structural design}

\begin{figure}[hbt!]
\begin{center}
\includegraphics[width=1\linewidth]{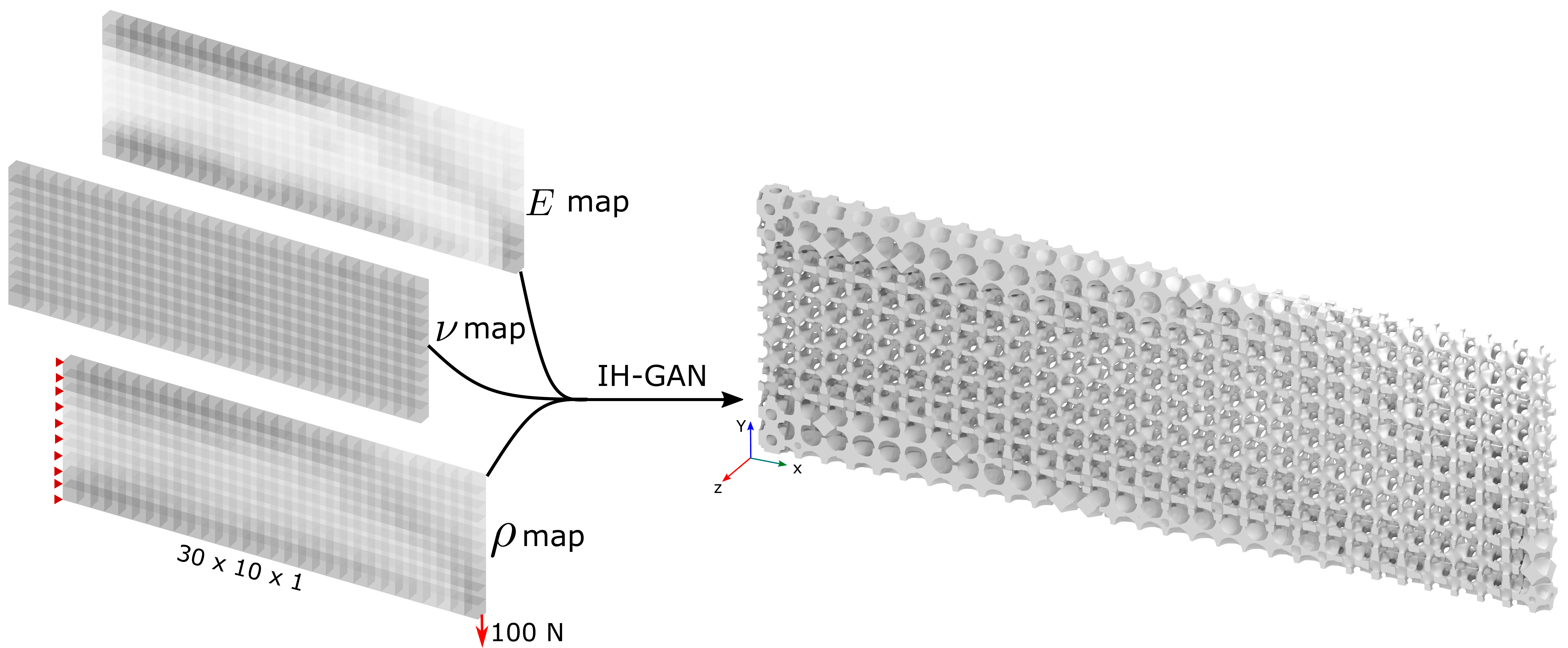}
\caption{An illustrative example of a cantilever beam built by functionally graded cellular unit cells using IH-GAN. The beam has a dimension of $30\ mm \times 10\ mm \times 1\ mm$ ($x \times y \times z$). {\it Left:} optimized Young's modulus map, Poisson's ratio map, and density map with boundary conditions indicated. The red triangles (\mytriangle{red}) symbolizes a fixed boundary condition along all three coordinates. The red arrows (\textcolor{red}{$\rightarrow$}) represent a point force loading condition and the direction it is applied. {\it Right:} the synthesized structure with multiple types of cellular unit cells.}
\label{fig:TOMulti}
\end{center}
\end{figure}

\begin{figure}[hbt!]
\begin{center}
\includegraphics[width=0.5\linewidth]{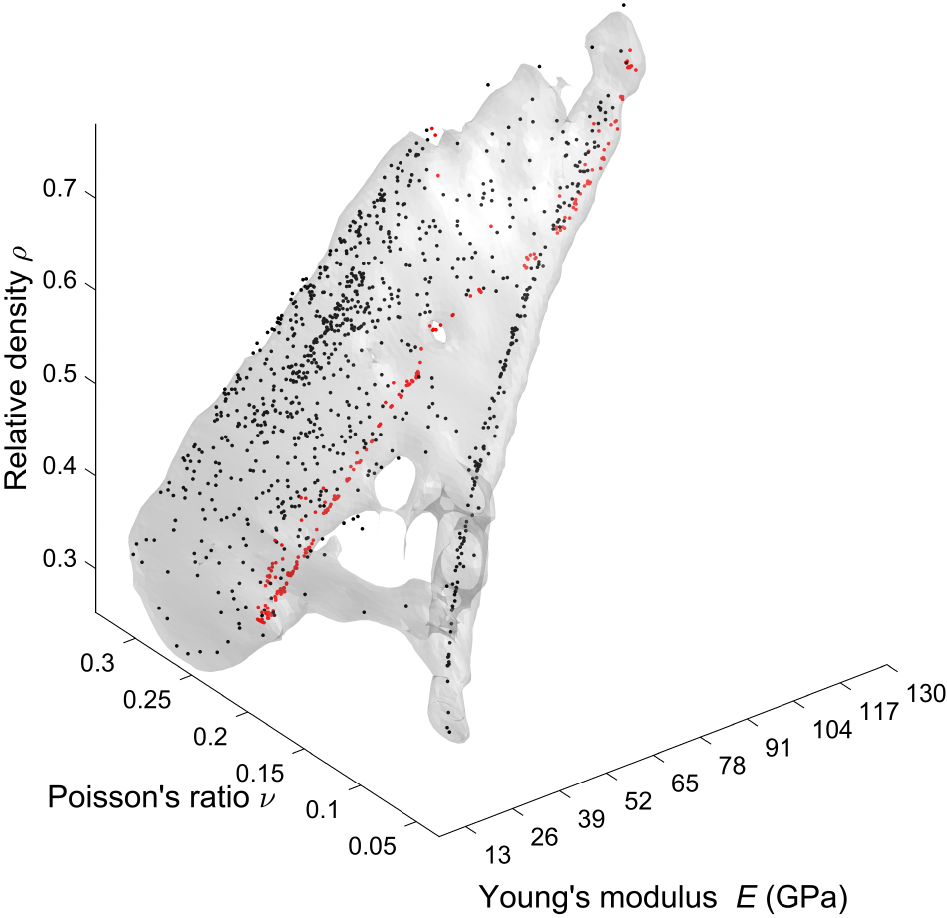}
\caption{Optimized properties (highlighted in red dots) in the property space.}
\label{fig:opt_dots}
\end{center}
\end{figure}

This paper optimizes a cantilever beam as an illustrative example to demonstrate the functionally graded cellular structural design using IH-GAN as shown in Figure~\ref{fig:TOMulti}. We obtain the optimized $E$, $\nu$, and $\rho$ maps by solving the optimization problem described in Section~\ref{sec:TO} (with the overall mass $\hat{V}=45\%$). The resultant three maps of properties are highlighted as red dots in the property space of the training data in Figure~\ref{fig:opt_dots}. We then combine the three maps and feed them into IH-GAN to generate corresponding cellular unit cells. We finally assemble these cellular unit cells to make a beam.

\subsection{Connectivity}

\begin{figure}[hbt!]
\begin{center}
\includegraphics[width=1\linewidth]{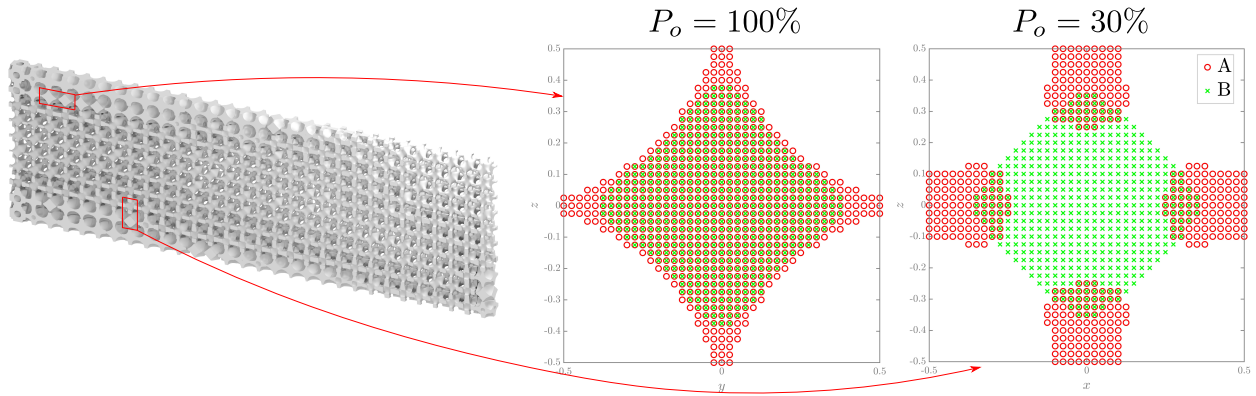}
\caption{Interface boundaries for adjacent unit cells (A and B) at two locations.}
\label{fig:connectivity}
\end{center}
\end{figure}

The major risk that arises when merging different types of cellular unit cells is the lack of sufficient interface connection area \cite{wang2021hierarchical}. If the geometries at the intersection between two adjacent unit cells are significantly different, the common face's potential overlap can be low, leading to poor connectivity. Such poor connectivity can result in a weak link that makes the entire structure vulnerable to failure (under functional needs) and eliminates any advantage obtained using multi-type cellular structures. In this paper, we validate the quality of interface connectivity by computing the percentage of overlap area:
\begin{equation}
P_o = \frac{|A_A \cap A_B|}{\min\{|A_A|, |A_B|\}}\times 100 \%,
\label{eqOL}
\end{equation}
where $A_A$ and $A_B$ are sets of pixels of two connecting unit cell faces after discretization. Figure~\ref{fig:connectivity} illustrates the boundary faces of two different unit cells, A and B, with the faces in contact vertically or horizontally when the structure is assembled (\ie, the IH-GAN structure for compliance minimization problem in Section~\ref{sec:comp_mini}). The face of unit cell A is represented by red circles, the face of unit cell B is depicted by green crosses, and the overlap areas are shown in red circles with green crosses infilled. In this paper, we utilize TPMS-based unit cells that have cubic symmetry. Therefore, each unit cell has six identical boundary faces. The average percentage of overlap areas within the whole structure is $96.43\%$.

\begin{figure*}[hbt!]
\begin{subfigure}[b]{0.5\linewidth}
\centering
\includegraphics[width=0.7\linewidth]{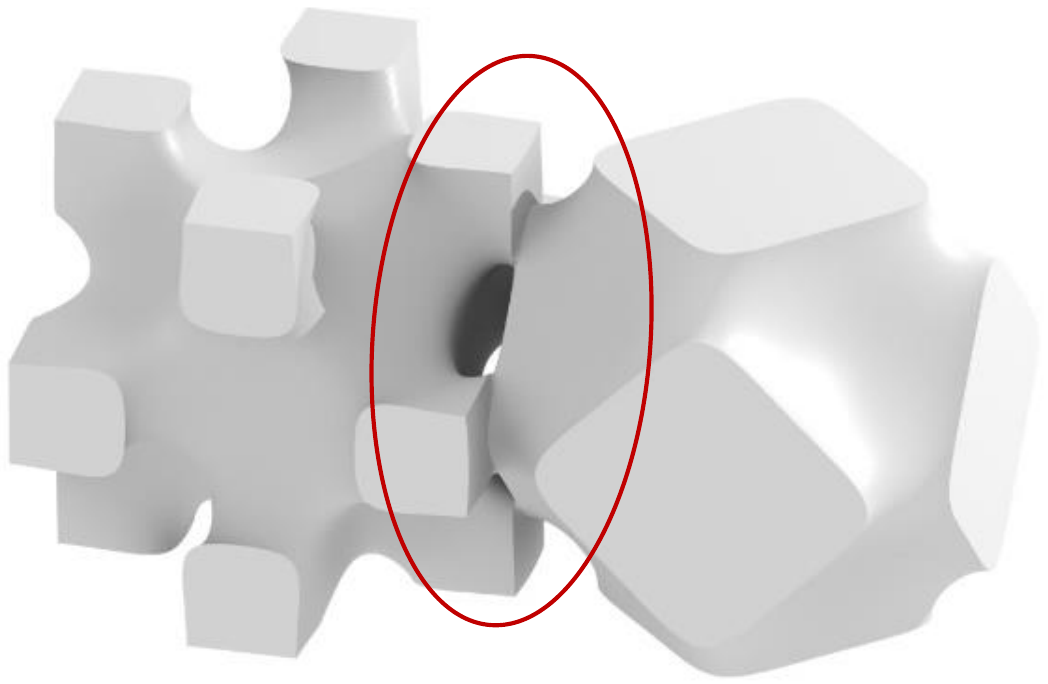}
\caption{}
\end{subfigure}%
\begin{subfigure}[b]{0.5\linewidth}
\centering
\includegraphics[width=0.7\linewidth]{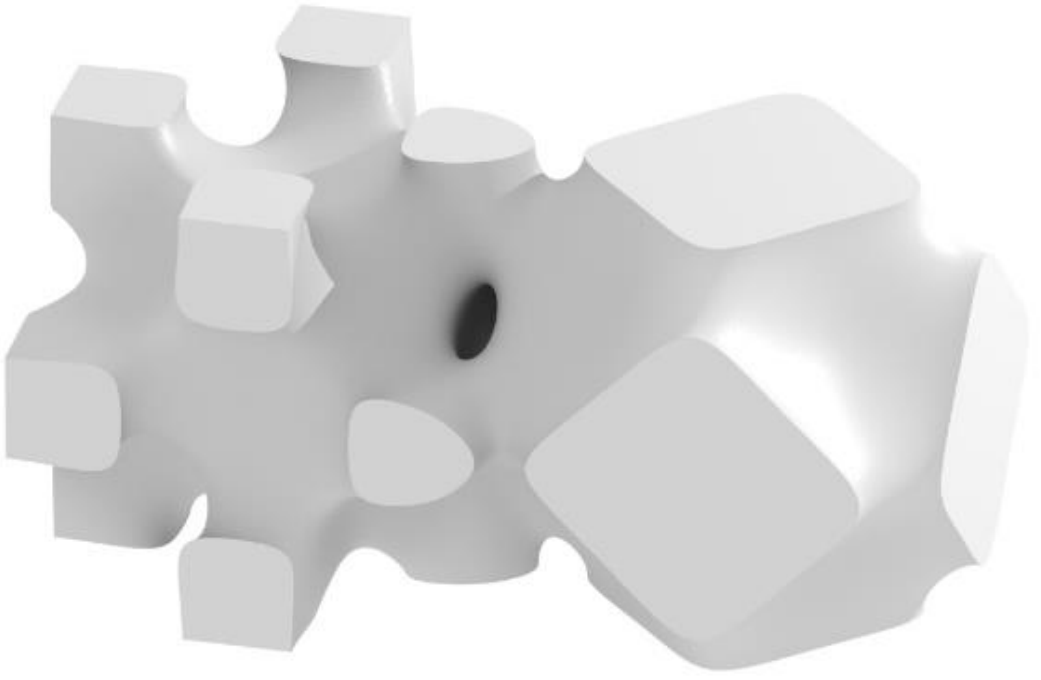}
\caption{}
\end{subfigure}

\caption{Geometric compatibility and connectivity in cellular structures. (a) The connection between two different types of unit cells without smoothing the transition layer (\ie, the connection interface). (b) A smooth connection ($100\%$ overlap) between the two different types of unit cells using interpolation.}
\label{fig:connect}
\end{figure*}

By taking advantage of the material property filter kernel used in TO and the continuous shape variation of IH-GAN's generated unit cells in the 3D material property space, IH-GAN's generated unit cells can naturally guarantee quality connectivity between different unit cells without additional compatibility optimization efforts \cite{schumacher2015microstructures, garner2019compatibility}. Most of the boundary faces have overlap areas close to $100\%$. There are still some adjacent unit cells that can have relatively low overlap areas. For example, the rightmost subfigure in Figure~\ref{fig:connectivity} shows the lowest overlap between two unit cells with a percentage of $30.0\%$. To avoid possible failure due to insufficient interface connection, we can further mitigate the connectivity issue by blending each pair of two adjacent unit cells via an interpolation operation similar to the linear interpolation used in the variable-density structure \cite{li2019design}. Figure~\ref{fig:connect} illustrates an example of the smooth transition between two different types of unit cells (\ie, the two unit cells with $30\%$ overlap in Figure~\ref{fig:connectivity}).

\subsection{Validation of structural performance}

To demonstrate the efficacy of IH-GAN in designing functional structures, we validate the cantilever beam's structural performance (Figure~\ref{fig:TOMulti}) by performing an FE simulation\footnote{In this paper, the finally assembled structures are formed by implicit surfaces (\eg, all-triangular surface mesh). In order to do FE simulations, we convert the all-triangular surface mesh into an all-tetrahedral volumetric mesh using the \emph{iso2mesh} 3D volumetric mesh generator \cite{fang2009tetrahedral}. The generated volumetric mesh consists of 650,000+ tetrahedra and takes a total generation time of 20+ seconds on average. The generated mesh was then rewritten and saved as an ABAQUS input file for the FE simulation. The FE simulation is performed using ABAQUS 2017 and takes an average of 80+ seconds to complete a job. The material properties used in the simulation are the same ones used in the homogenization process with $E=200$ GPa and $\nu=0.3$. A PC with a 2.9 GHz Intel Core i9-8950HK CPU and 32GB RAM was used for the mesh generation and FE simulation.} and comparing it with the single-type cellular structural designs. Taking Table~\ref{tableFEA} as an example, the solid beam is also redesigned into a variable-density single-type cellular structure and a uniform single-type cellular structure. For the variable-density structure, the density of D surface unit cells is optimized by the state-of-the-art proposed in \cite{li2019design}. In comparison, the uniform structure is formed by D surface unit cells with identical densities.

\subsubsection{Minimum compliance}
\label{sec:comp_mini}

\begin{figure}[hbt!]
\begin{subfigure}{0.33\linewidth}
\centering
\includegraphics[width=1\linewidth]{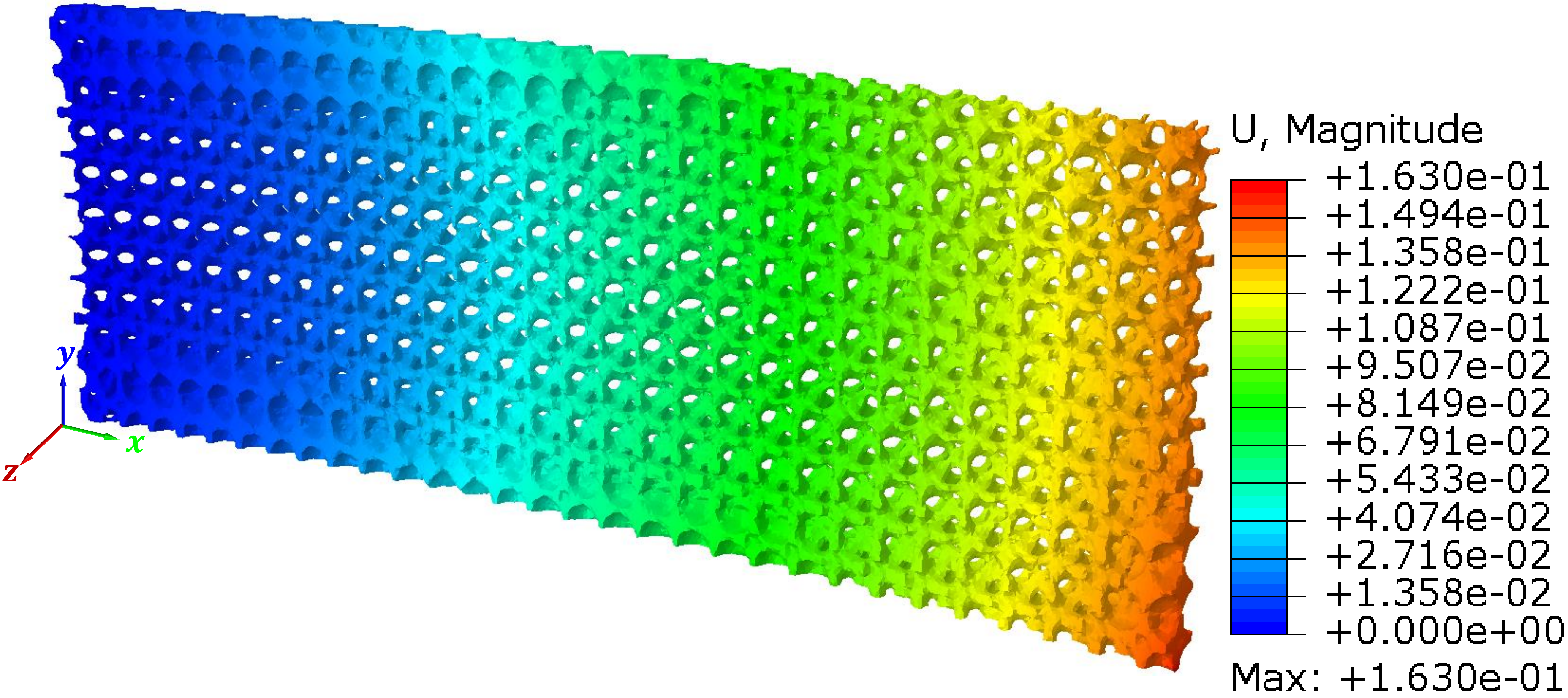}
\caption{}
\end{subfigure}%
\begin{subfigure}{0.33\linewidth}
\centering
\includegraphics[width=1\linewidth]{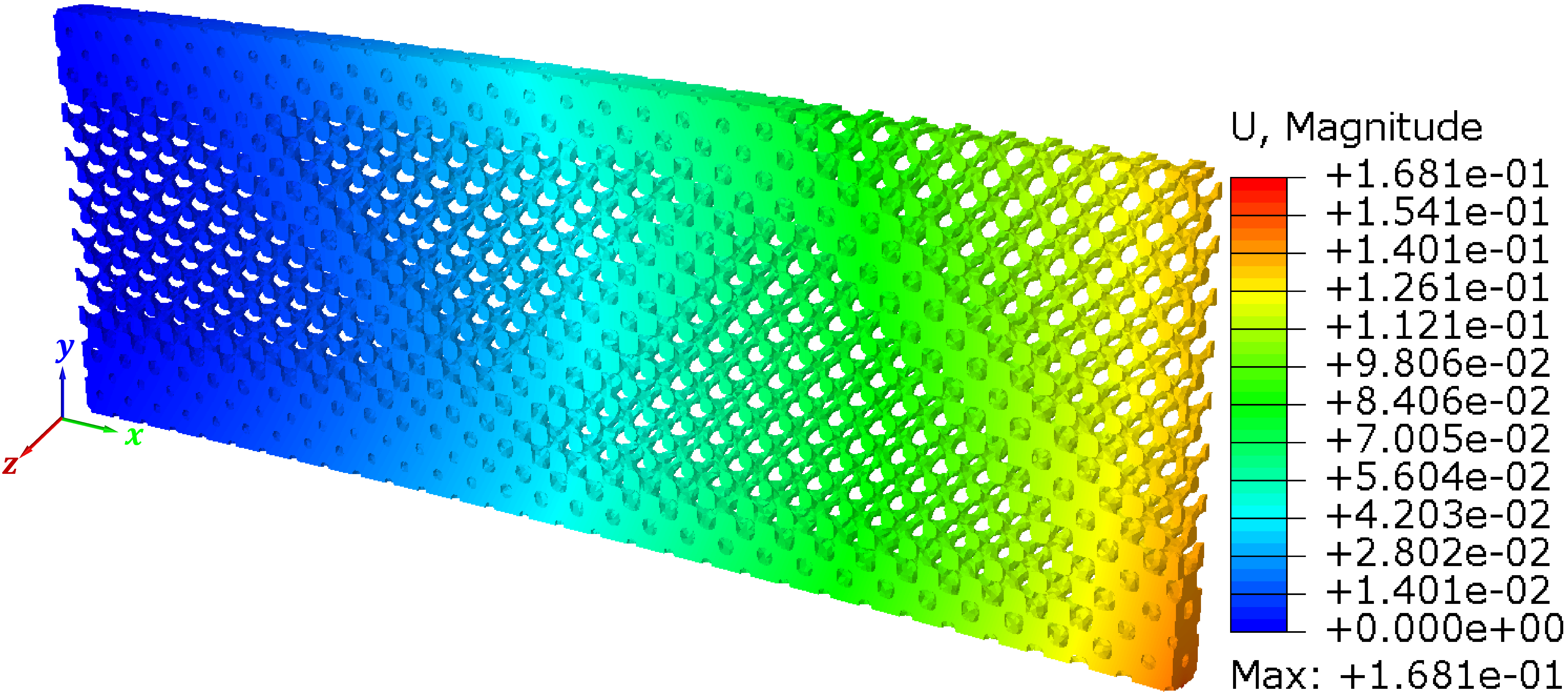}
\caption{}
\end{subfigure}%
\begin{subfigure}{0.33\linewidth}
\centering
\includegraphics[width=1\linewidth]{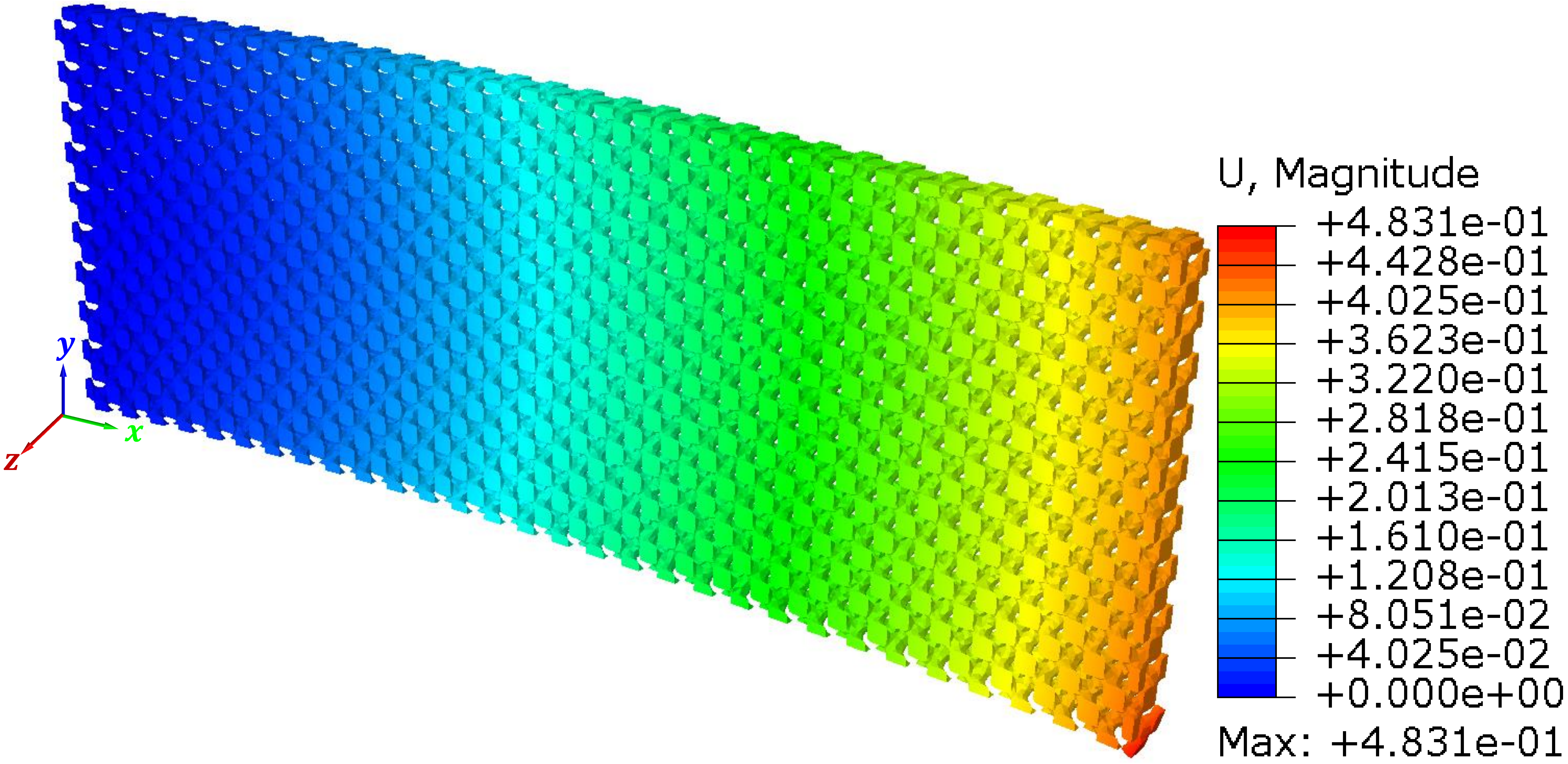}
\caption{}
\end{subfigure}
\begin{subfigure}{0.33\linewidth}
\centering
\includegraphics[width=1\linewidth]{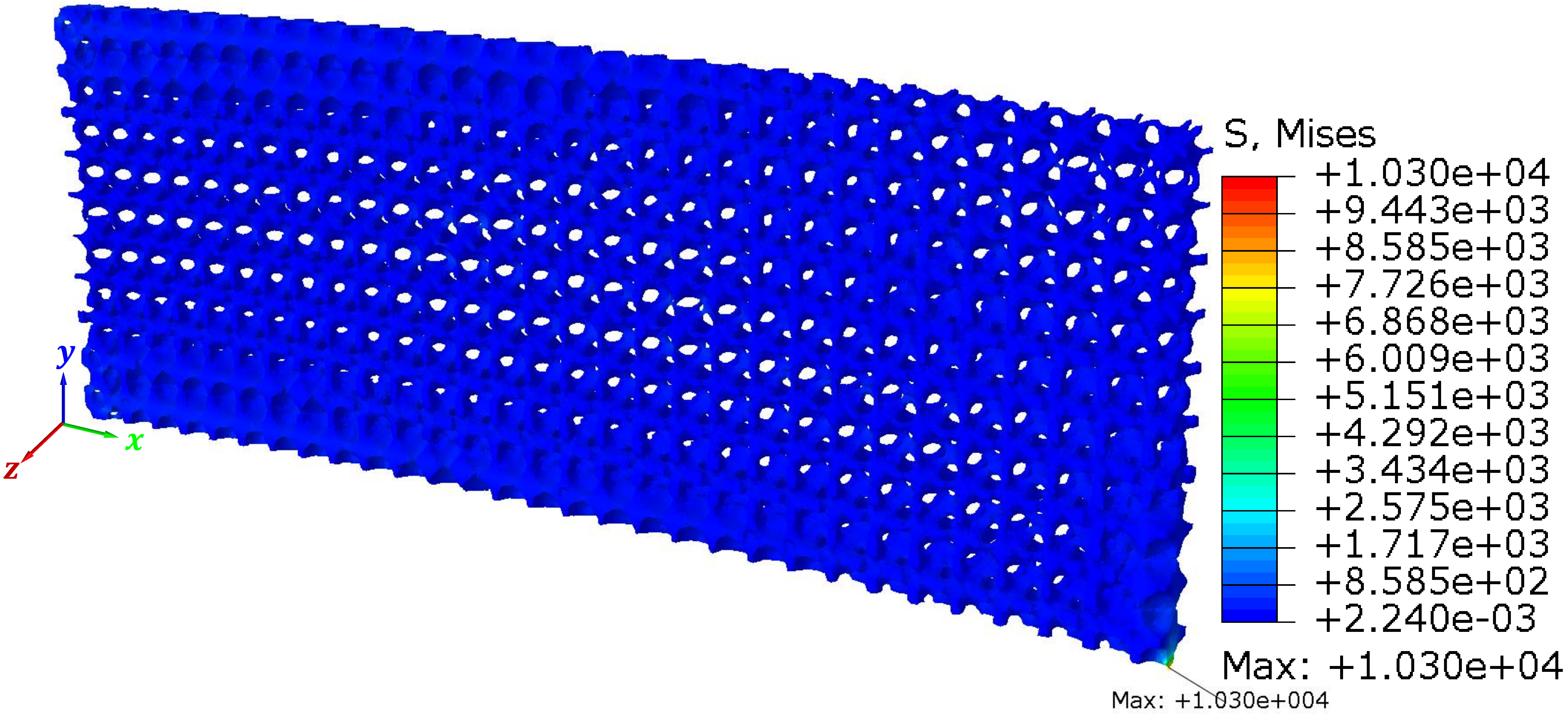}
\caption{}
\end{subfigure}%
\begin{subfigure}{0.33\linewidth}
\centering
\includegraphics[width=1\linewidth]{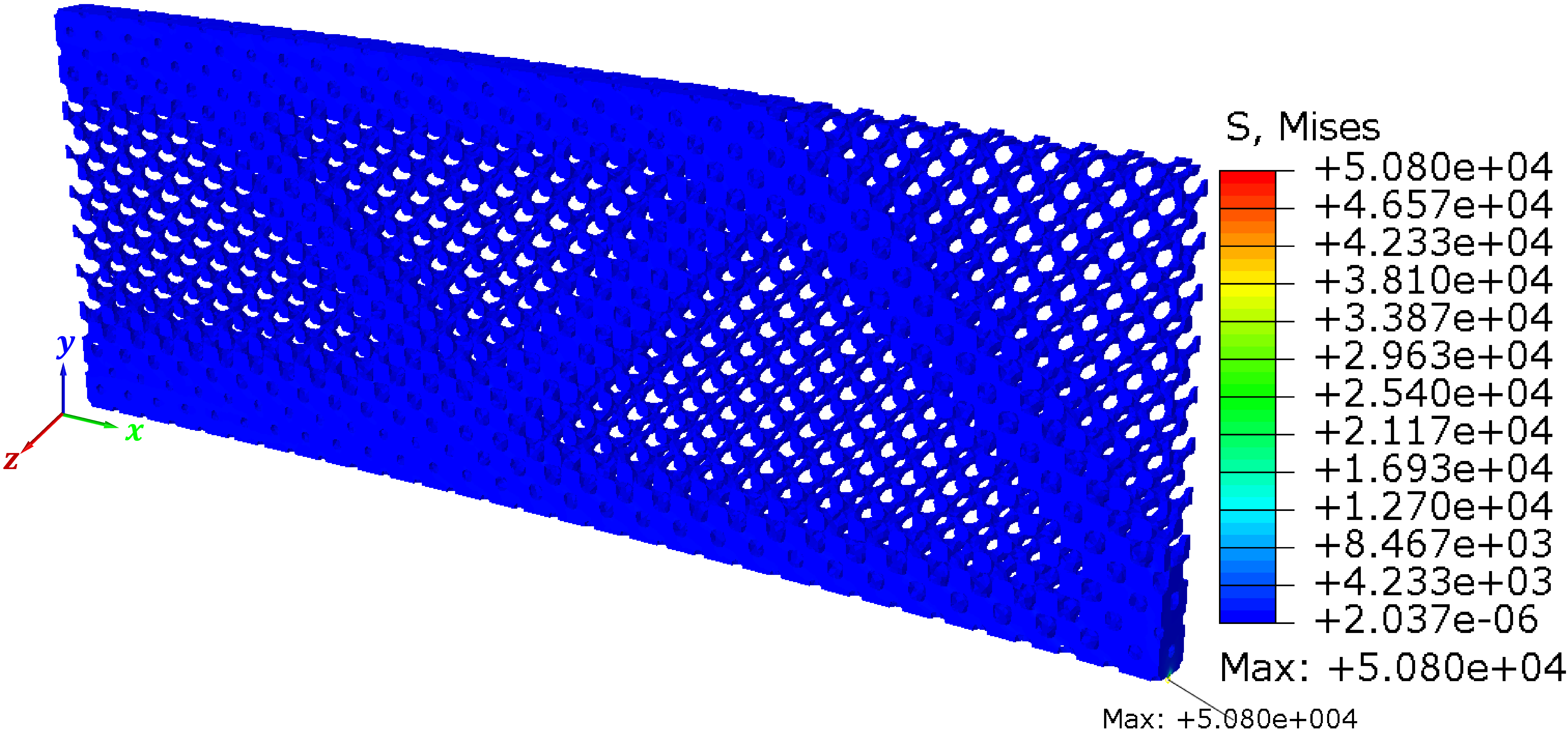}
\caption{}
\end{subfigure}%
\begin{subfigure}{0.33\linewidth}
\centering
\includegraphics[width=1\linewidth]{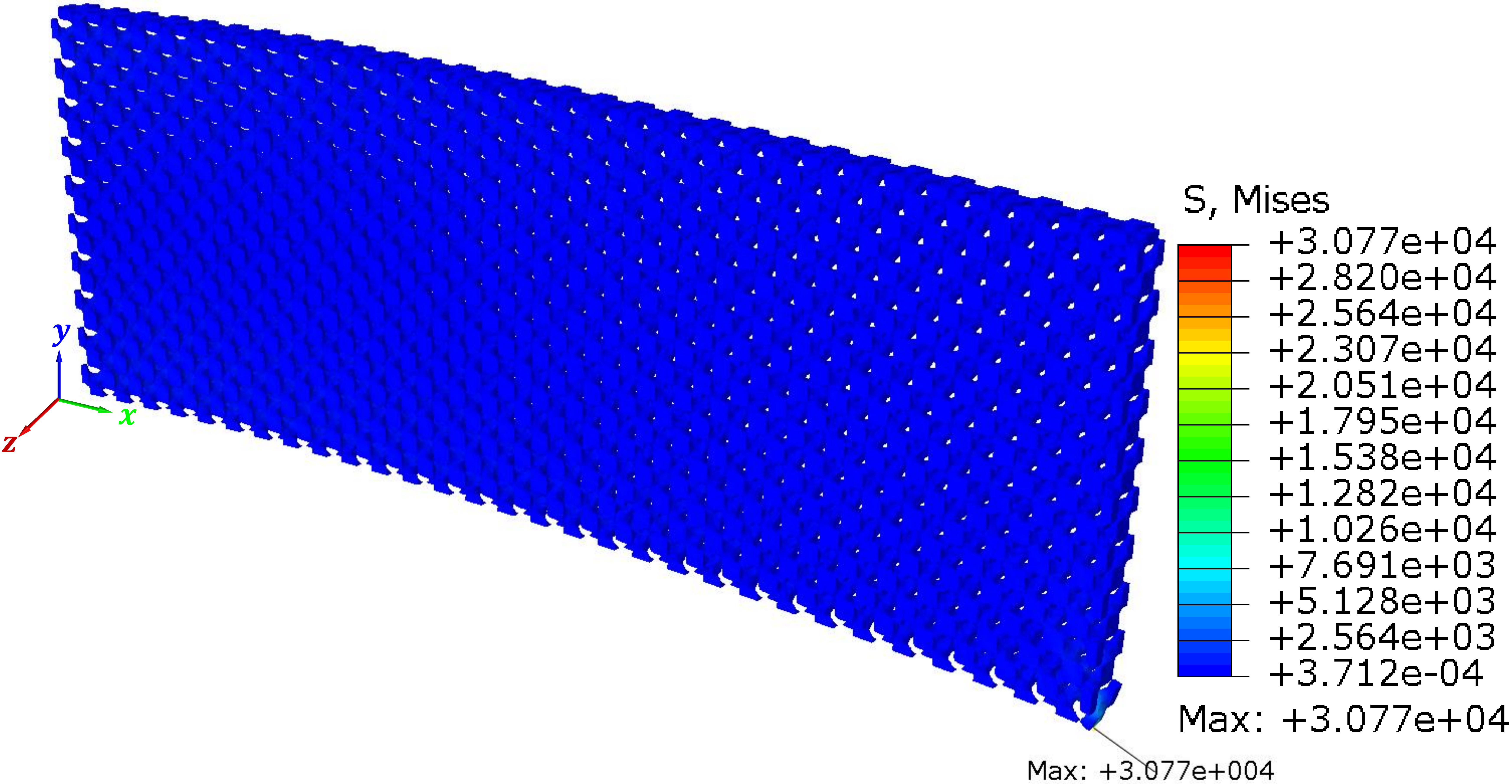}
\caption{}
\end{subfigure}
\caption{FE simulated displacement (\textit{first row:} (a), (b), and (c)) and stress (\textit{second row:} (d), (e), and (f)) contours of the beam redesigned for minimum compliance. \textit{Left:} IH-GAN-based multi-type cellular structure, \textit{Middle:} variable-density single-type cellular structure, and \textit{Right:} uniform single-type cellular structure.}
\vspace{0pt}
\label{figureFEA}
\end{figure}

To enable a fair comparison between different design methods, we apply the same load (a point force of 100 N) and keep the density consistent ($45\%$) for the three structures (Table~\ref{tableFEA}). 
Our proposed IH-GAN-based method slightly reduces the beam's maximum displacement by $3.03\%$ (from 0.1681 $mm$ to 0.1630 $mm$) compared to the variable-density structure. In contrast, the uniform structure (without optimization) results in a much larger displacement (0.4831 $mm$) and compliance (21.1058 $N\cdot mm$). While maintaining the displacement and compliance performance, the IH-GAN-based method can lower the maximum von-Mises stress ($\sigma_{max}$) significantly compared to the other two methods. The IH-GAN structure's maximum stress is 10,302.0 MPa, which is $79.7\%$ lower than the maximum stress of the variable-density structure (50,800.8 MPa) and $66.5\%$ lower than the maximum stress of the uniform structure (30,765.7 MPa). The FE simulated displacement (magnitude) and stress contours of the three structures are compared in Figure~\ref{figureFEA}. The FE simulation results show that our IH-GAN model can successfully generate graded cellular structures with functional performance exceeding current implicit function-based single-type methods.

\begin{table*}[hbt!]
\caption{Structural performance comparison for compliance minimization with different design methods}
\begin{center}
\label{tableFEA}
\begin{adjustbox}{width=1\textwidth}
\begin{tabular}{p{4.5cm}  p{4cm}  p{4cm}  p{4cm}}
& \\ [-6ex] 
\hline
Design domain & \hspace{15pt} IH-GAN & \hspace{15pt} Variable-density & \hspace{15pt} Uniform\\
\hline \\[-3ex]
\parbox[][9em][c]{9em}{\includegraphics[width=1.6in]{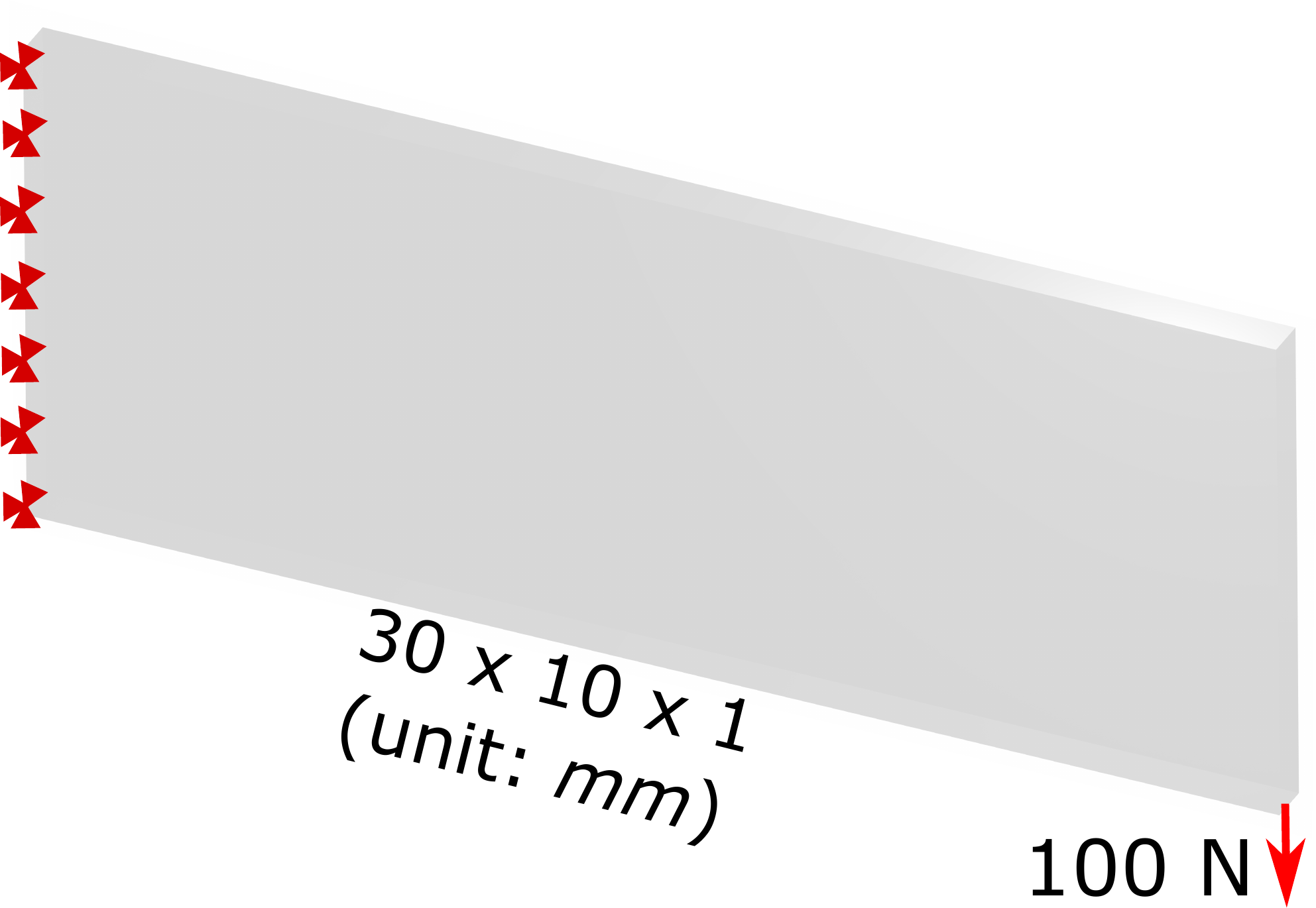}} & 
\parbox[][9em][c]{9em}{\includegraphics[width=1.7in]{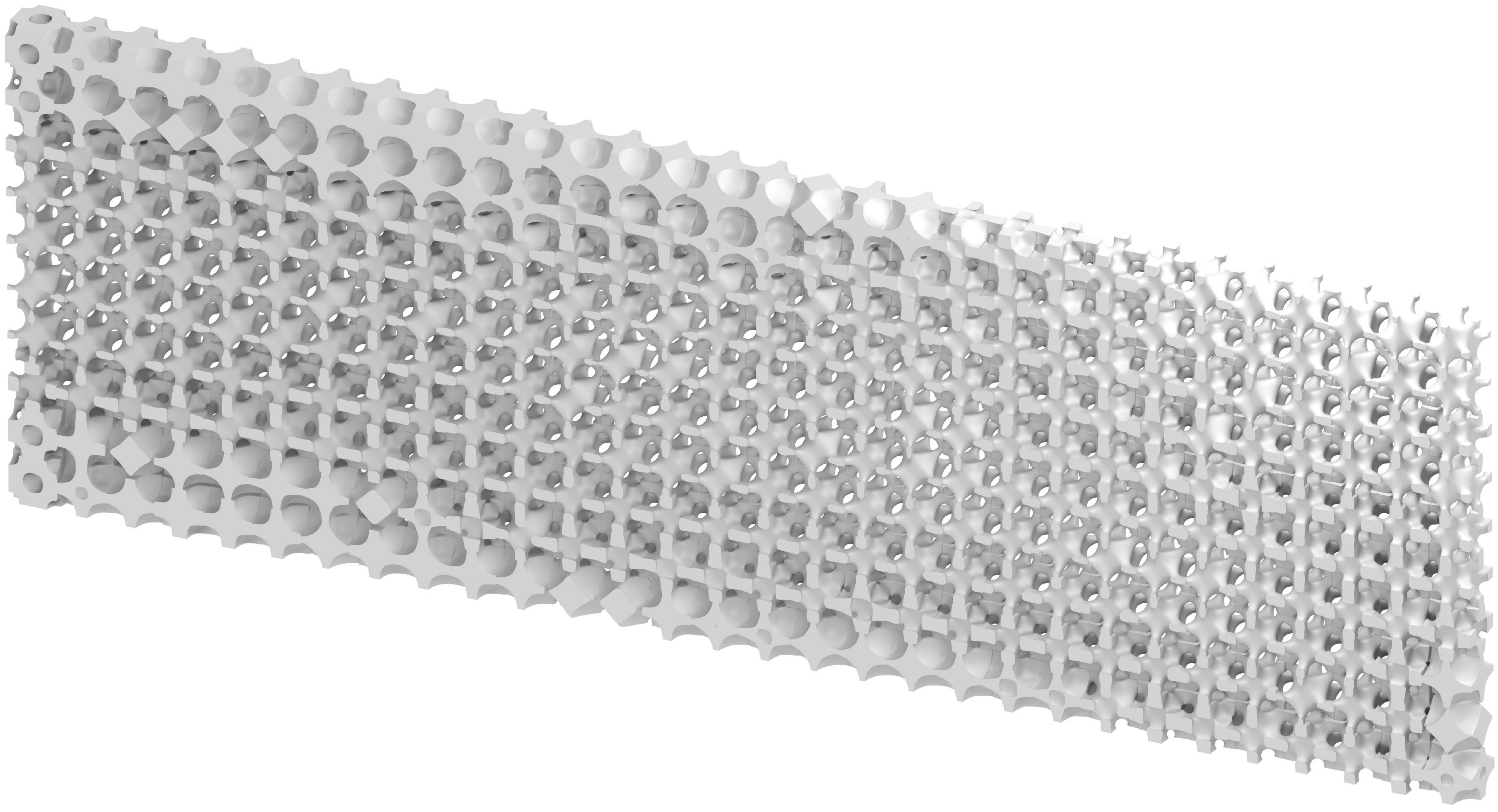}} & \parbox[][9em][c]{9em}{\includegraphics[width=1.7in]{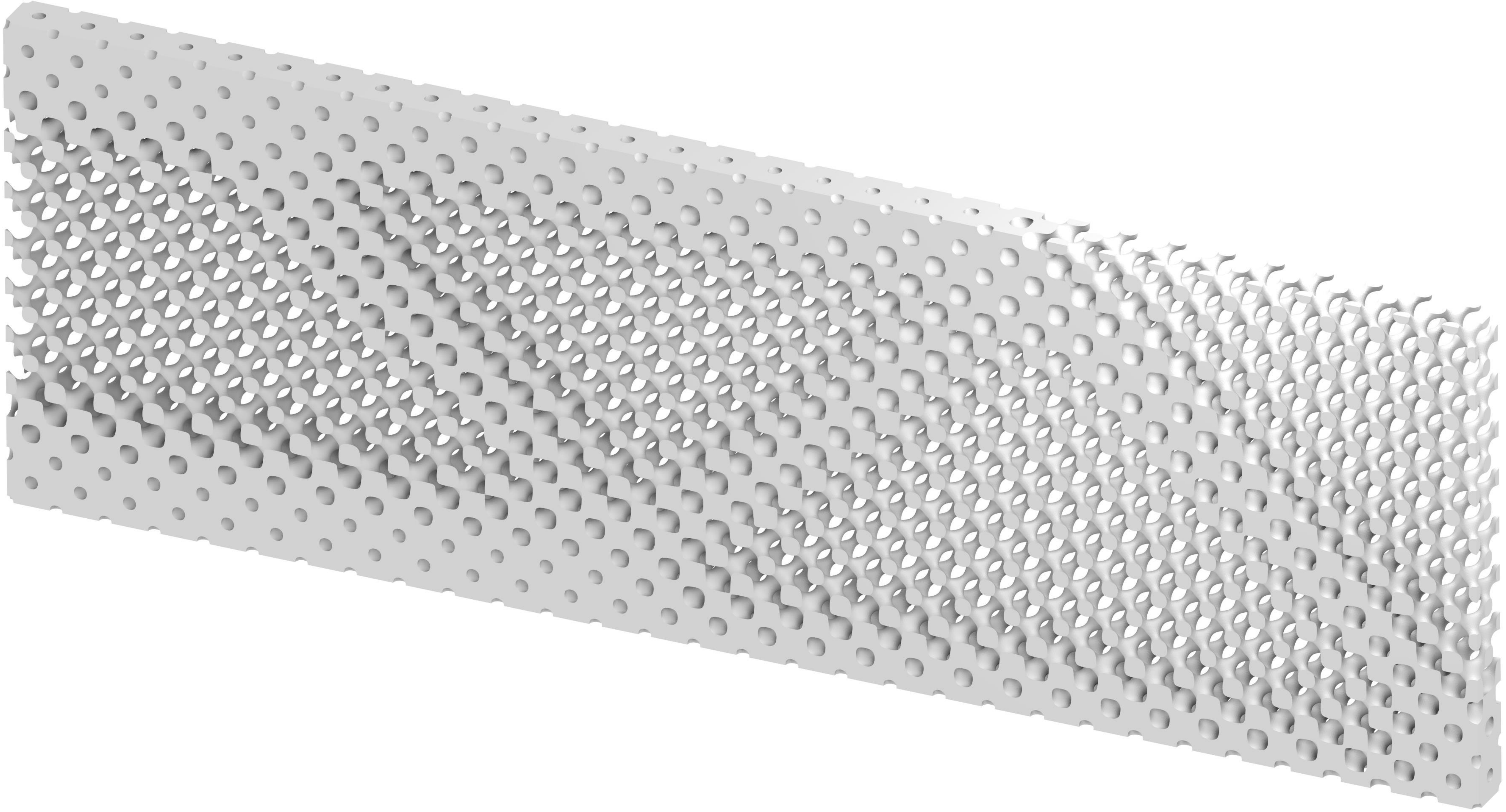}}  &
\parbox[][9em][c]{9em}{\includegraphics[width=1.7in]{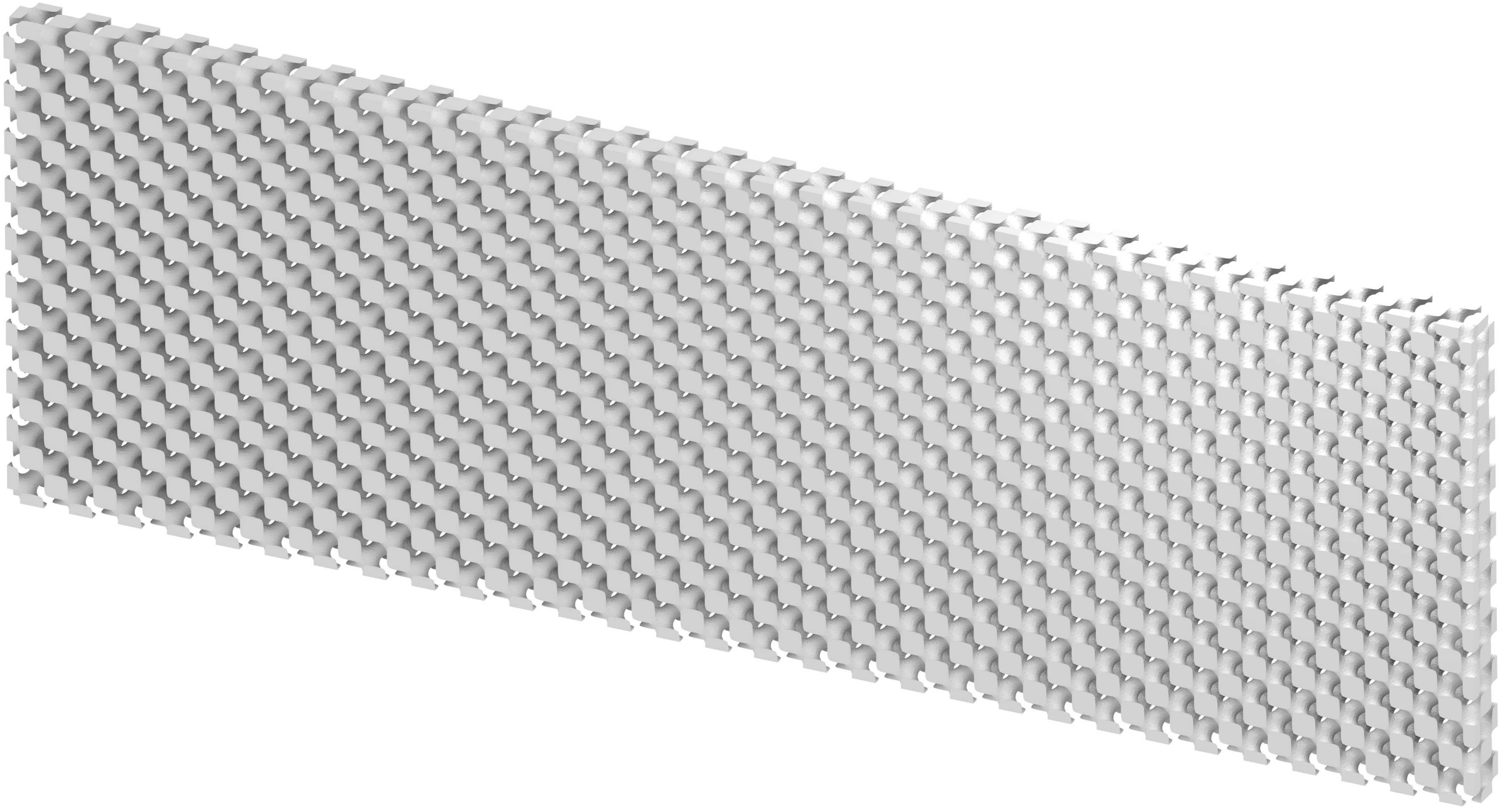}} \\
Density ($\rho$) &\hspace{1cm} $45\%$ &\hspace{1cm} $45\%$ &\hspace{1cm} $45\%$ \\[2ex]
Max. displacement ({\it mm}) &\hspace{1cm} 0.1630 &\hspace{1cm} 0.1681 &\hspace{1cm} 0.4831 \\[2ex]
$\sigma_{max}$ (MPa) &\hspace{1cm} 10,302.0 &\hspace{1cm} 50,800.8 &\hspace{1cm} 30,765.7 \\[2ex] 
Compliance ($N\cdot mm$) &\hspace{1cm} 7.3876 &\hspace{1cm} 7.4155 &\hspace{1cm} 21.1058 \\[1ex] 
\hline
\end{tabular}
\end{adjustbox}
\end{center}
\end{table*}

\subsubsection{Target deformation}
\begin{figure}[hbt!]
\begin{subfigure}{0.5\linewidth}
\centering
\includegraphics[width=0.72\linewidth]{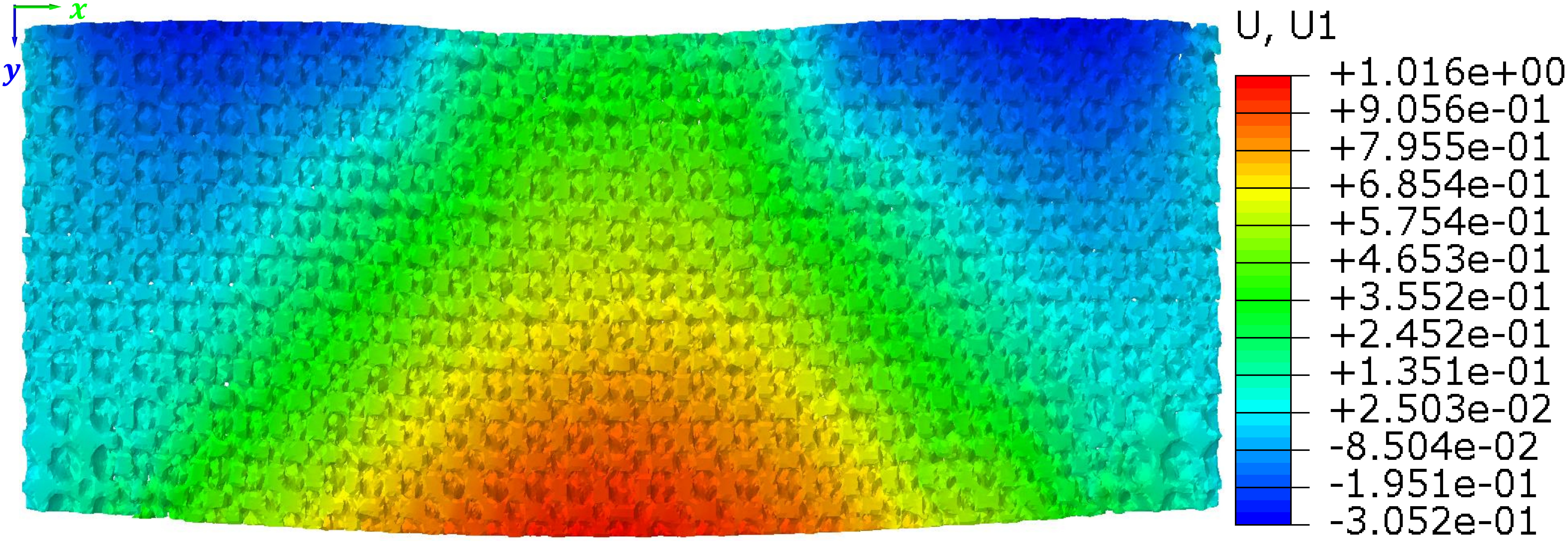}
\caption{}
\end{subfigure}%
\begin{subfigure}{0.5\linewidth}
\centering
\includegraphics[width=0.72\linewidth]{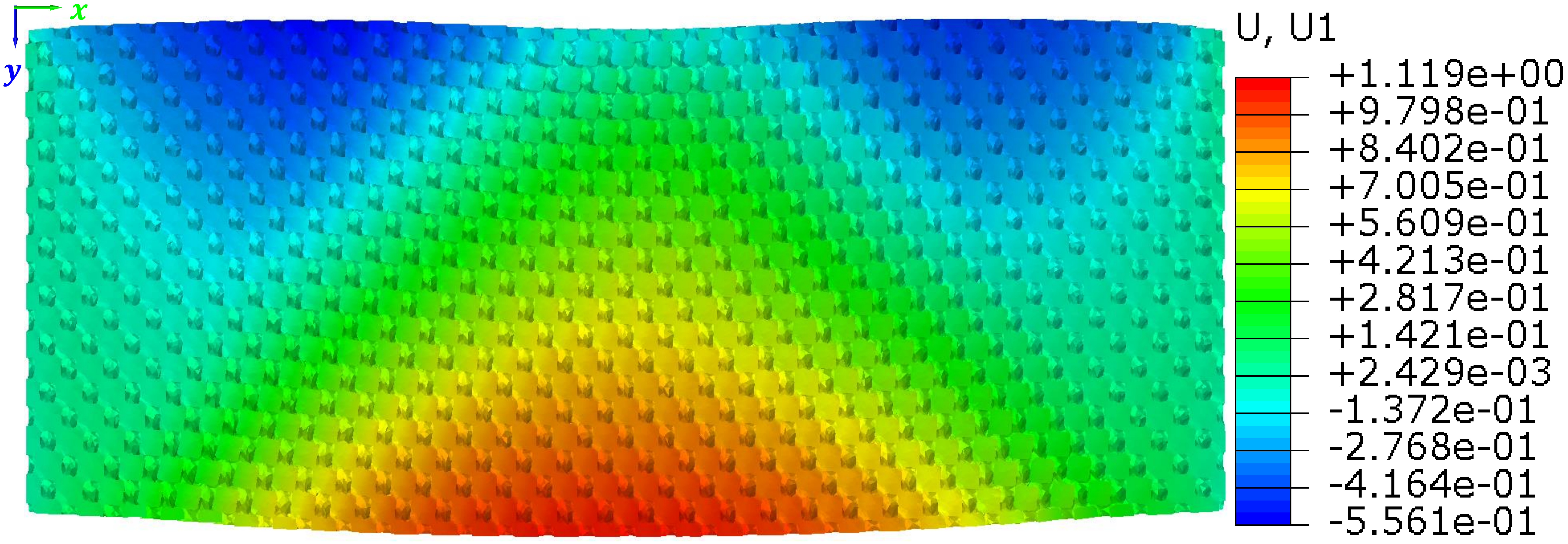}
\caption{}
\end{subfigure}
\begin{subfigure}{0.5\linewidth}
\centering
\includegraphics[width=0.72\linewidth]{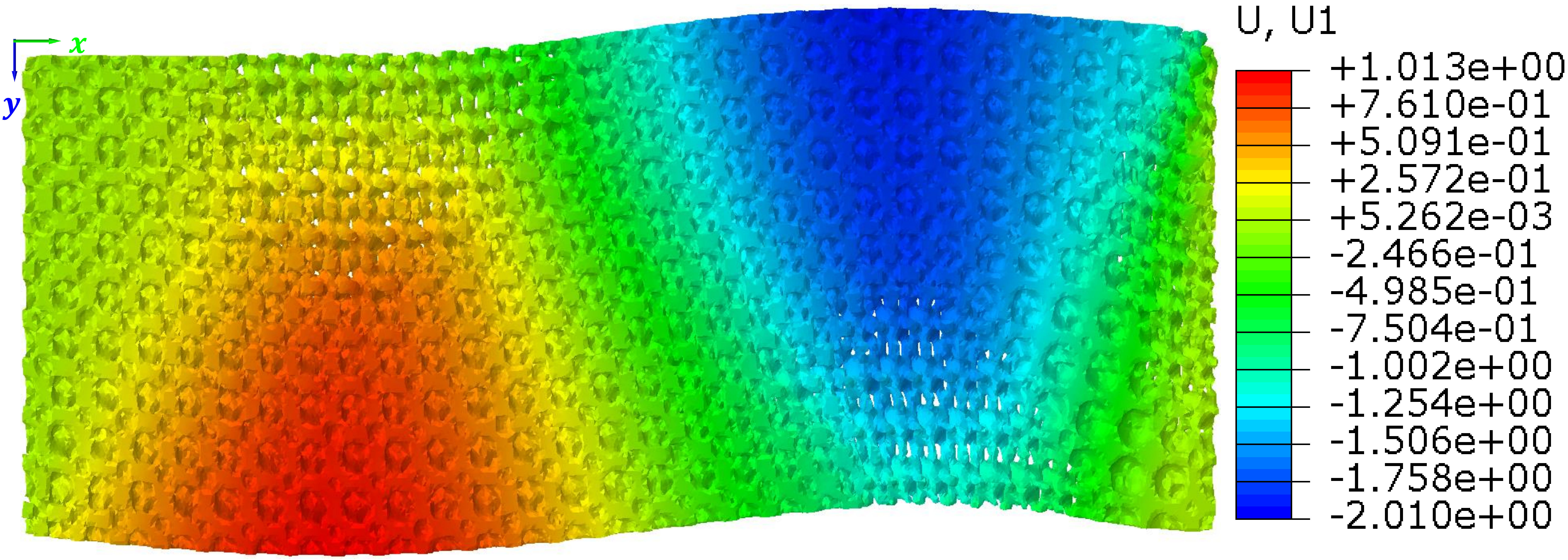}
\caption{}
\end{subfigure}%
\begin{subfigure}{0.5\linewidth}
\centering
\includegraphics[width=0.72\linewidth]{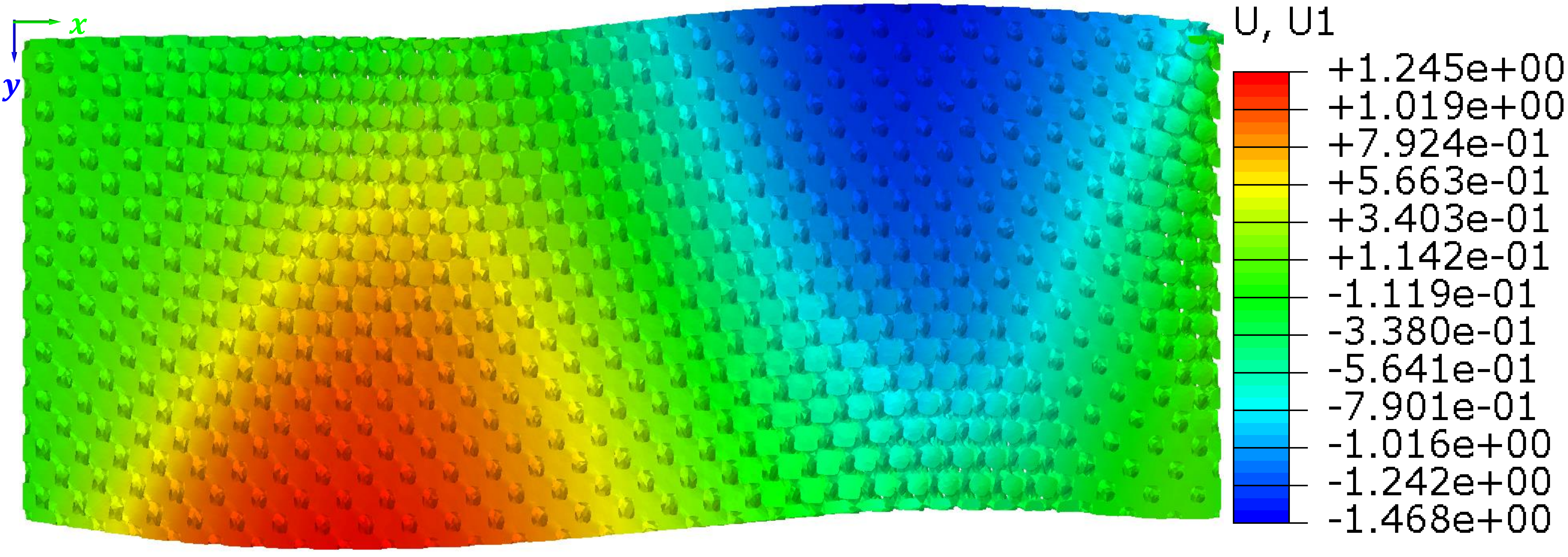}
\caption{}
\end{subfigure}
\caption{FE simulated $y$-axis (signed) displacement contours of the beam redesigned for target deformations. \textit{Left:} IH-GAN-based multi-type cellular structures for (a) target half $sin$ curve and (c) full $sin$ curve and \textit{Right:} variable-density single-type cellular structures for (b) target half $sin$ curve and (d) full $sin$ curve.}
\vspace{0pt}
\label{figureFEA_df}
\end{figure}

For the target deformation objective, we fix the left side of the beam, apply 10 $mm$ displacement along the negative $x$-axis, and fix the displacement along $y$-axis (vertical displacement) on the right side of the beam. The design domain in Table~\ref{tableDeformation} illustrates the boundary conditions where the red dotted line indicates the target deformation locations of interest (bottom edge of the beam). We experiment on two target deformations\textemdash \ie, displacements of a period of full $sin$ curve and a half $sin$ curve. The 31 nodes uniformly located on the bottom edge are used as the query points for evaluating the deformation performance.
In Table~\ref{tableDeformation}, the 3D models exhibit the final designs for the target deformations using our IH-GAN method and the variable-density single-type cellular structure.
The graphs compare the target displacements with the optimized displacements directly from structural optimization (\ie, scale separation) and the simulated displacements from ABAQUS (\ie, no scale separation).
To quantify the shape similarity between curves, we employ \emph{Mean Squared Error} (MSE) as the measuring metric\footnote{$MSE = \frac{1}{n}\sum\limits_{i=1}^n (u_i-\hat{u}_i)^2.$}. A lower MSE value corresponds to higher similarity. As seen in Table~\ref{tableDeformation}, the displacements optimized by structural optimization (scale separation) are close to the target full $sin$ and half $sin$ curves (with low MSE values), while the numerical experimental curves show some shifts and discrepancies in the FE simulations of the finally assembled structures. For both target curves (full $sin$ and half $sin$), the multi-type functionally graded cellular structures created by our IH-GAN show higher similarities with lower MSE values (0.00084 and 0.0409 for half $sin$ and full $sin$, respectively) compared to the curves generated by the variable-density single-type cellular structure (0.0062 and 0.2001 for half $sin$ and full $sin$, respectively).
The FE simulated (signed) displacement contours along $y$-axis are compared in Figure~\ref{figureFEA_df}.

\begin{table*}[hbt!]
\caption{Structural performance comparison for target deformation with different design methods}
\begin{center}
\label{tableDeformation}
\begin{adjustbox}{width=1\textwidth}
\begin{tabular}{p{4.5cm}  p{6cm}  p{6cm}}
& \\ [-6ex] 
\hline
Design domain & \hspace{15pt} IH-GAN & \hspace{15pt} Variable-density \\
\hline \\[-3ex]
\parbox[][9em][c]{9em}{\includegraphics[width=1.9in]{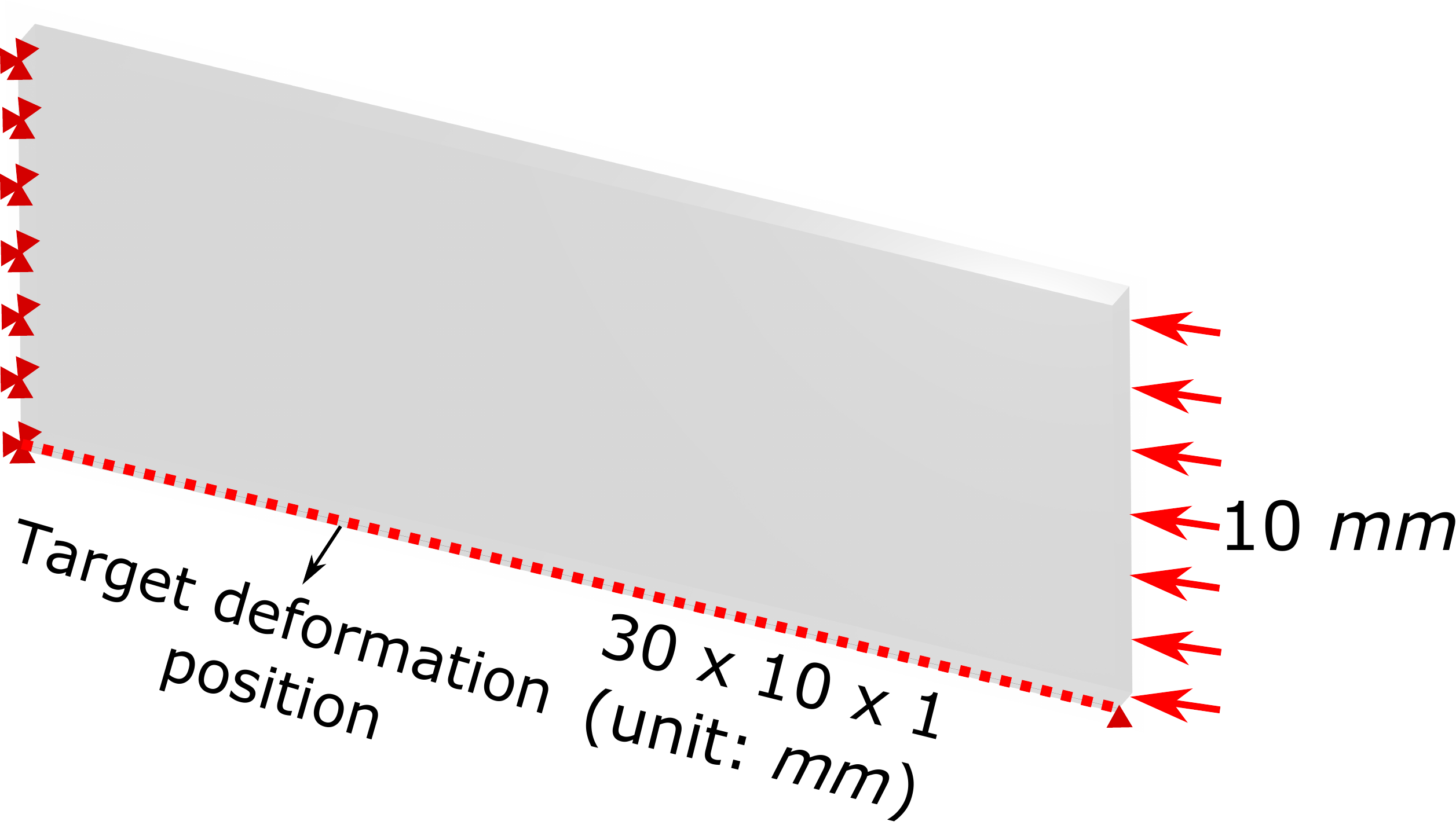}} & 
\parbox[][9em][c]{9em}{\includegraphics[width=2.0in]{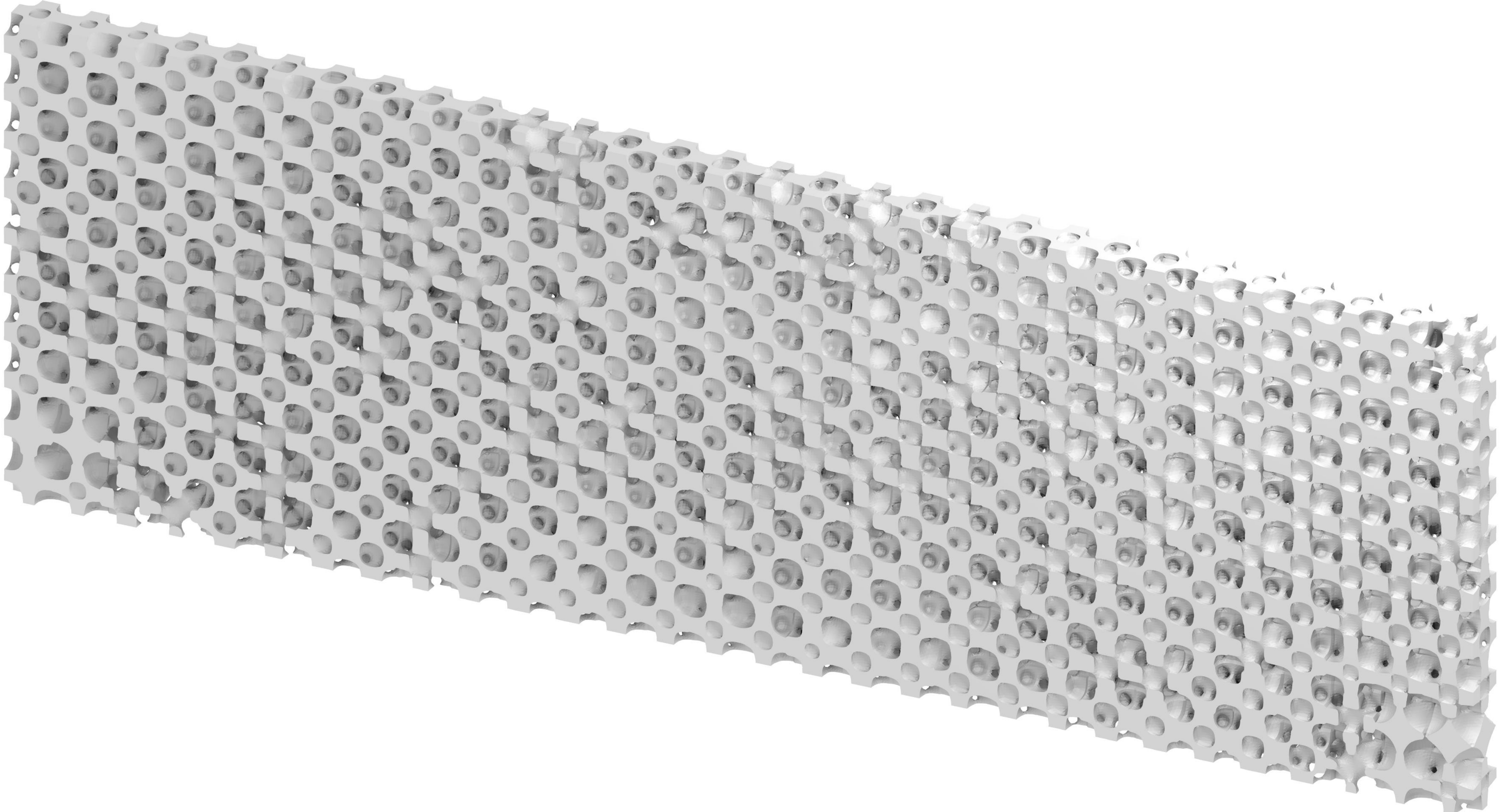}} & \parbox[][9em][c]{9em}{\includegraphics[width=2.0in]{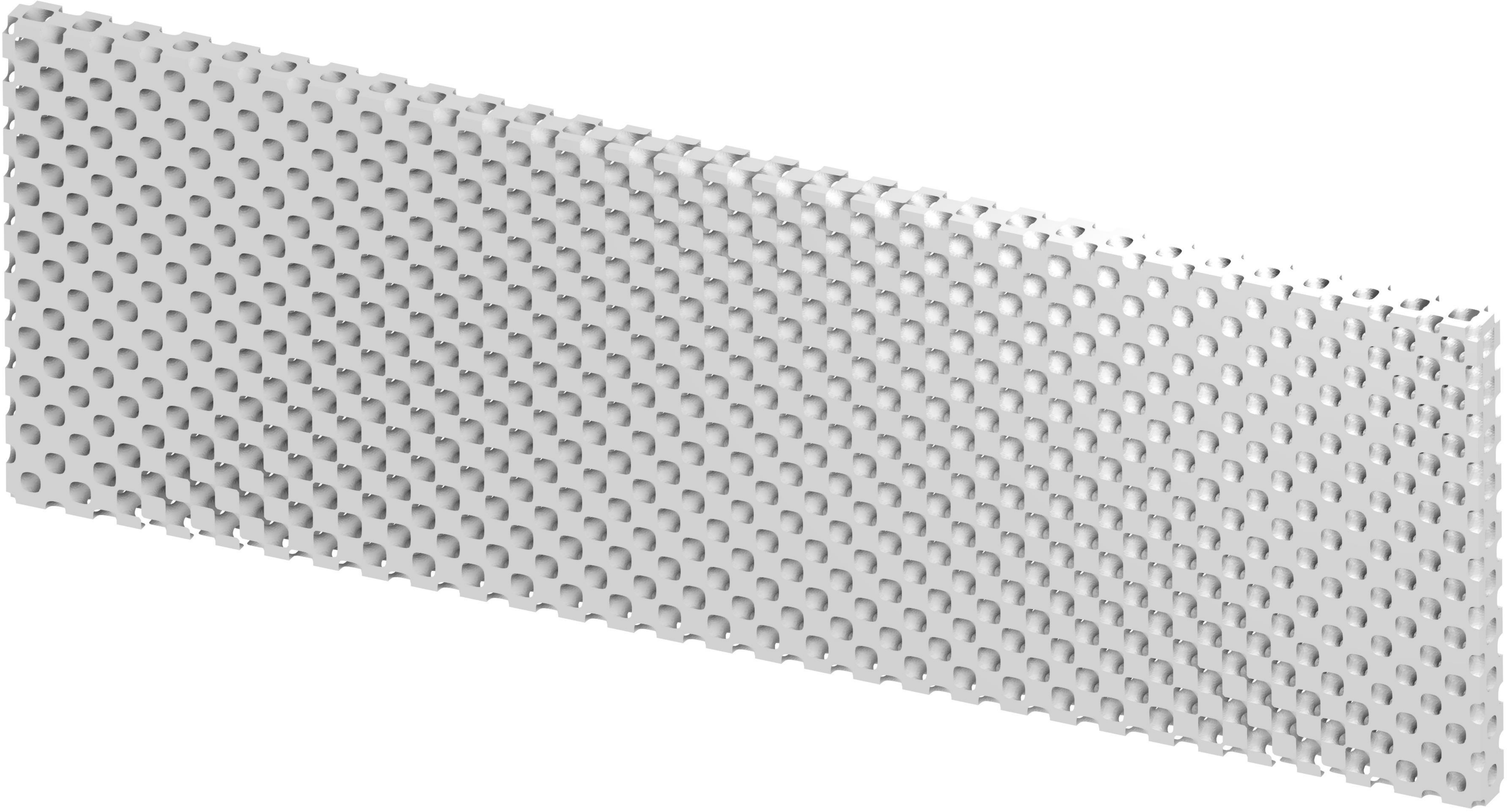}} \\[-1ex]
Half $sin$ curve &
\parbox[][9em][c]{9em}{\includegraphics[width=2.4in]{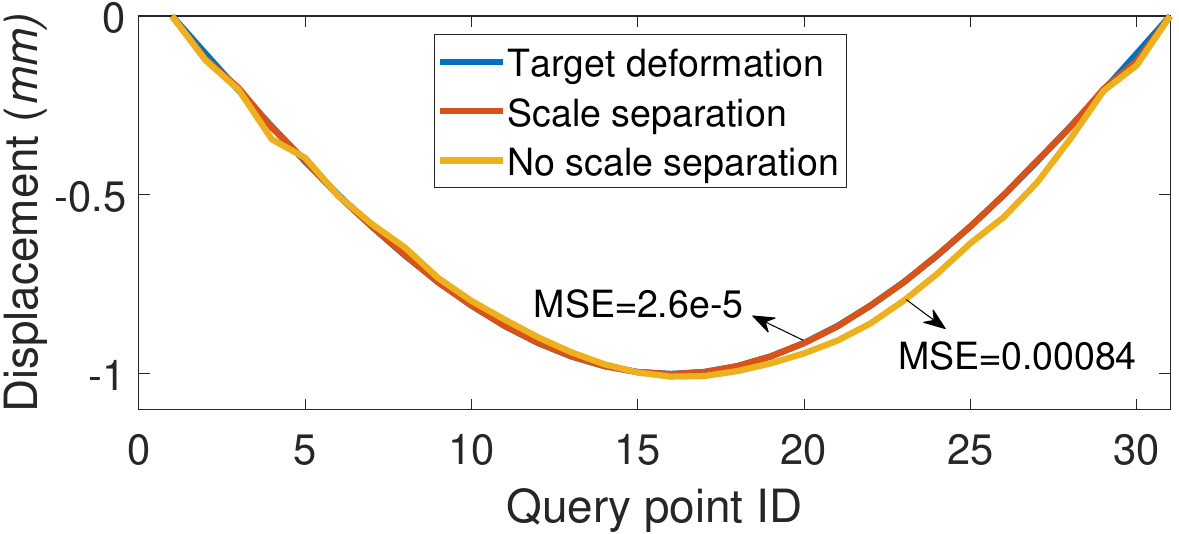}} & \parbox[][9em][c]{9em}{\includegraphics[width=2.4in]{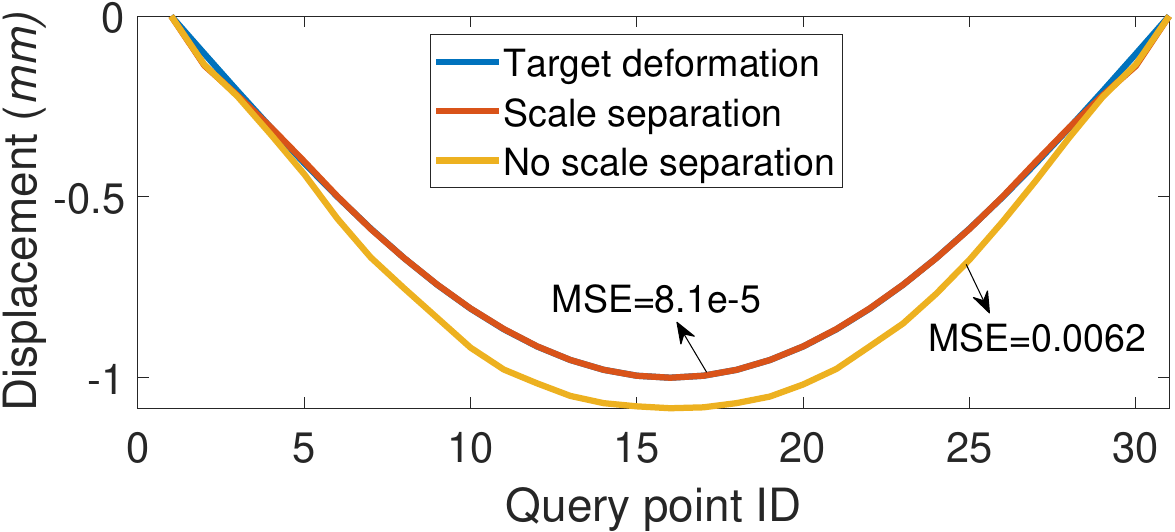}} \\[-0ex]
%
\hline \\[-3ex]
                                                                                   & 
\parbox[][9em][c]{9em}{\includegraphics[width=2.0in]{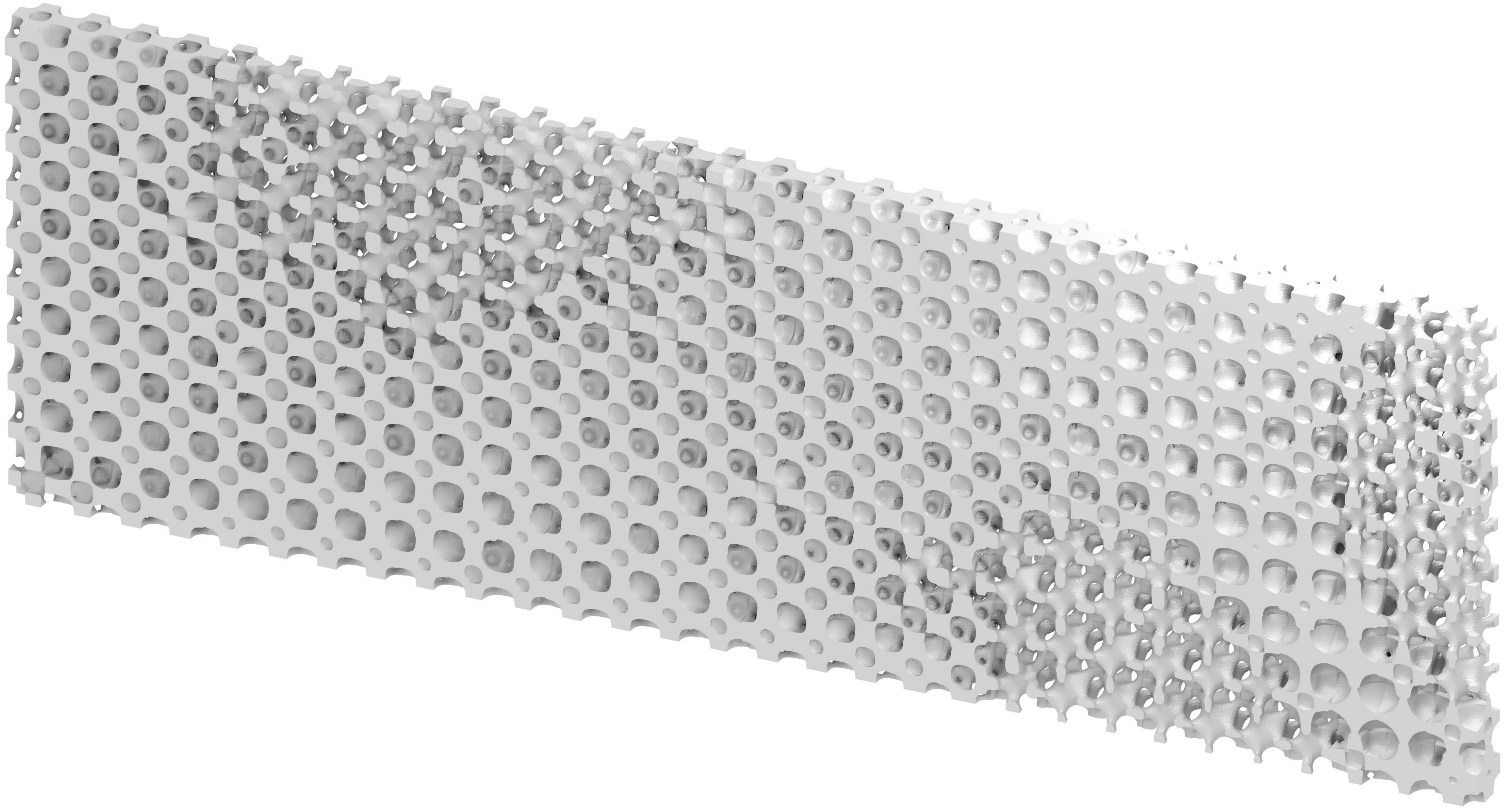}} & \parbox[][9em][c]{9em}{\includegraphics[width=2.0in]{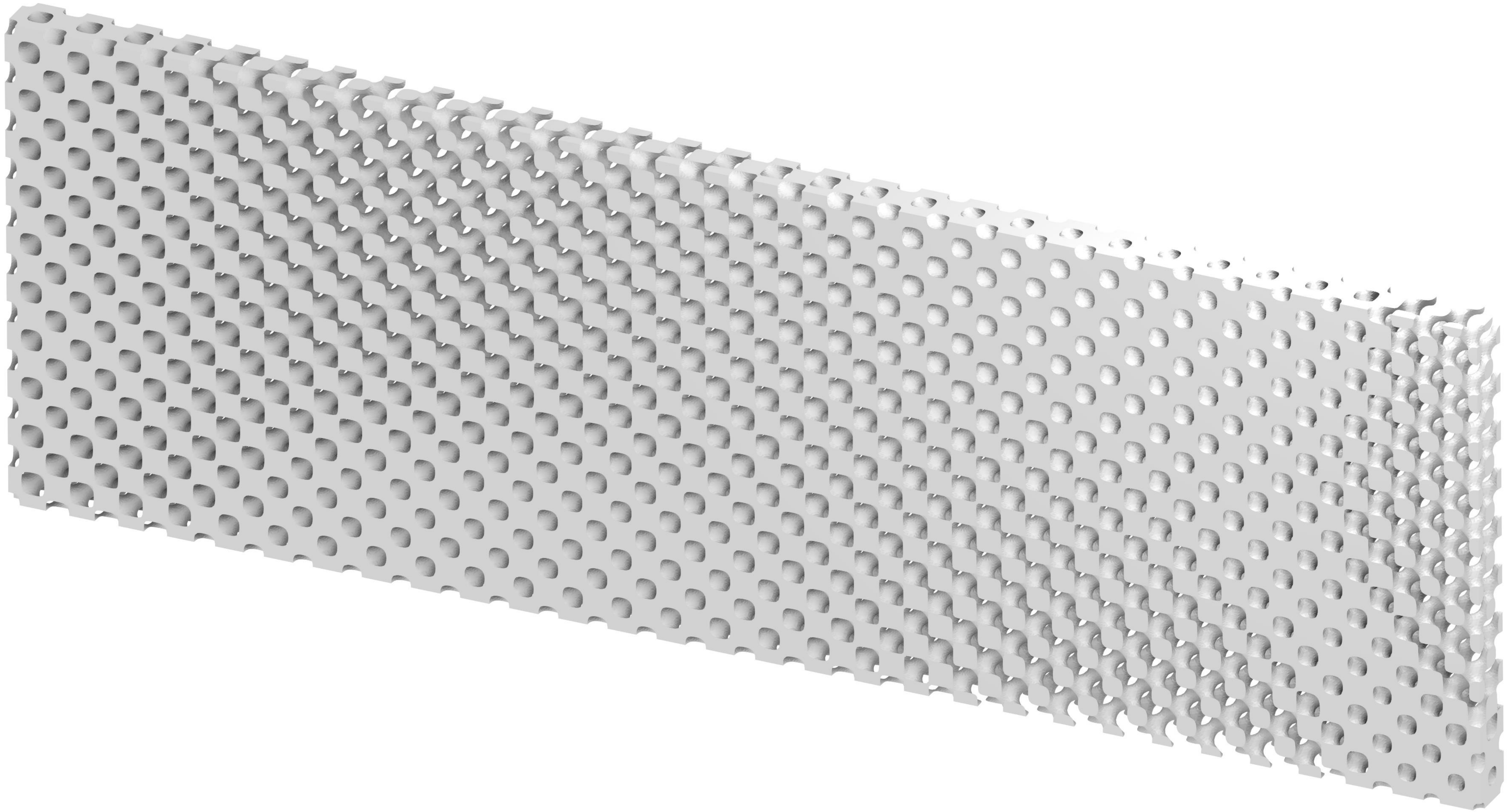}} \\[-1ex] 
Full $sin$ curve &
\parbox[][9em][c]{9em}{\includegraphics[width=2.4in]{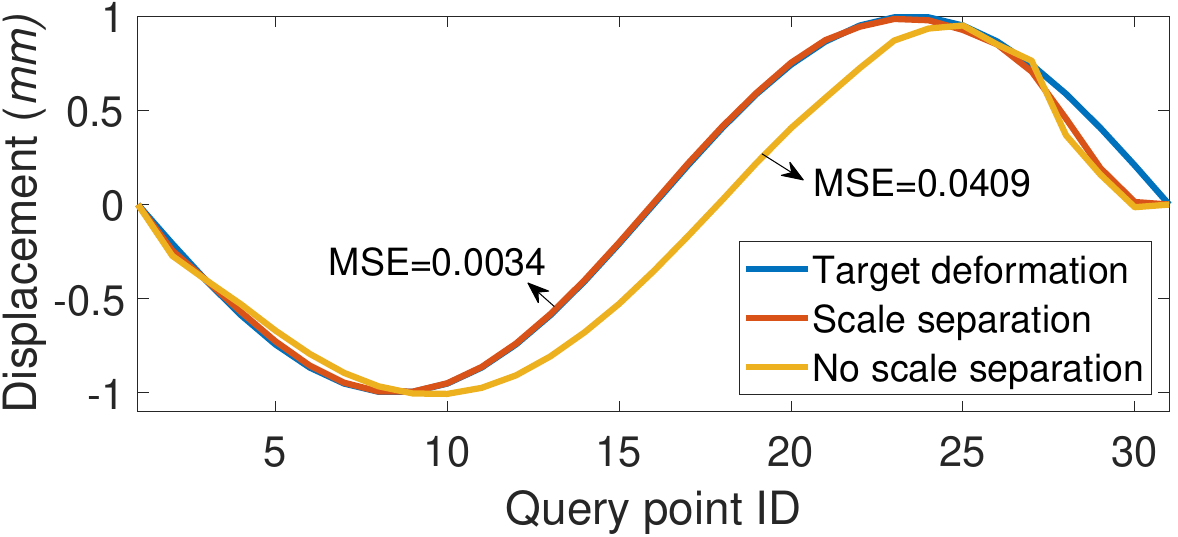}} & \parbox[][9em][c]{9em}{\includegraphics[width=2.4in]{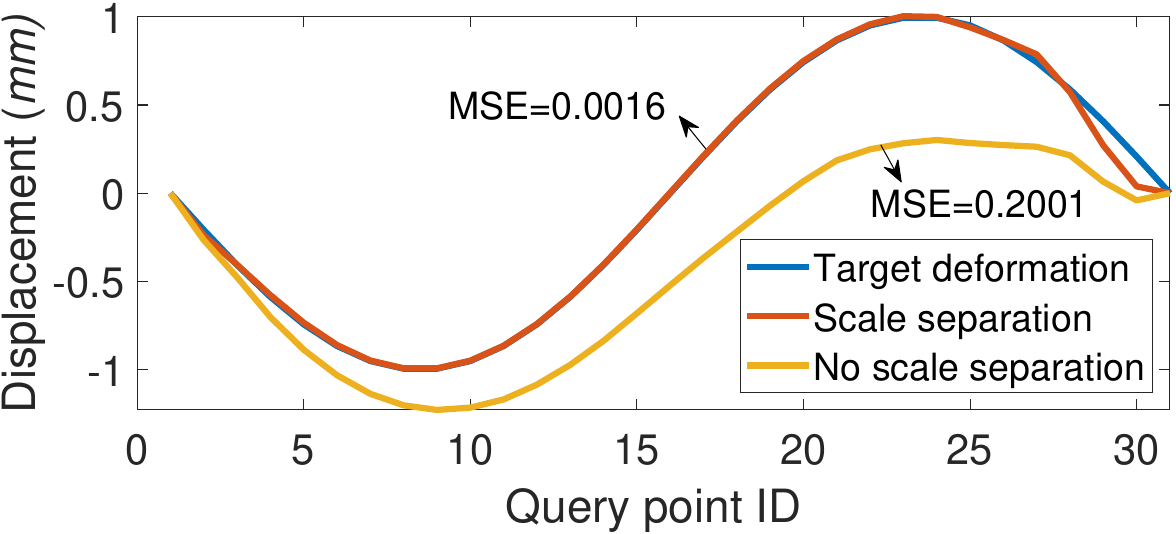}} \\[-0ex]
%
\hline
\end{tabular}
\end{adjustbox}
\end{center}
\end{table*}

\section{ Discussion} \label{discussion}

This section discusses the pros and cons of our proposed IH-GAN approach to designing multi-scale cellular structures in the present settings and scenarios. We also provided an in-depth vision of the future research directions and how we plan to explore and investigate. 

\subsection{Porous unit cells in structural performance}

\begin{figure}[hbt!]
\begin{subfigure}{0.5\linewidth}
\centering
\includegraphics[width=0.72\linewidth]{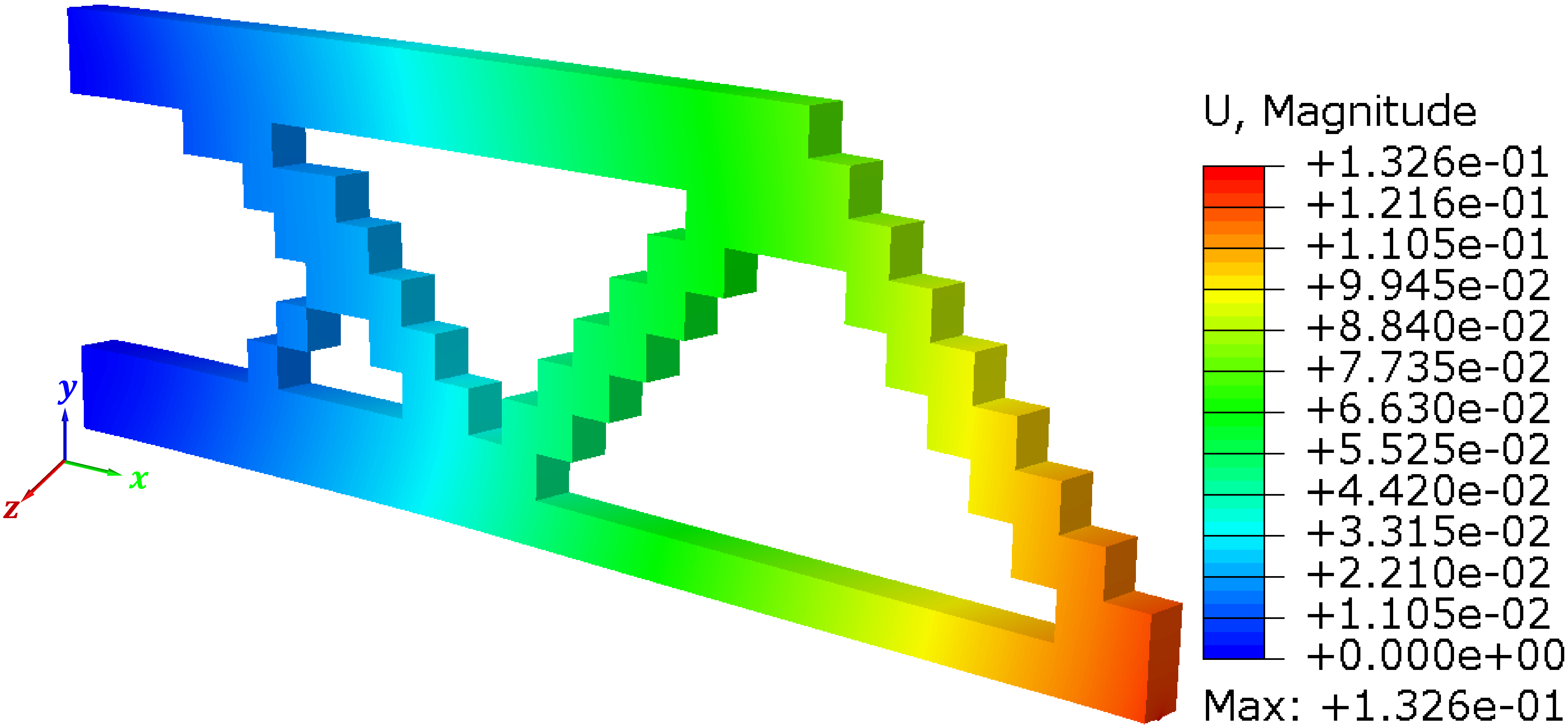}
\caption{}
\end{subfigure}%
\begin{subfigure}{0.5\linewidth}
\centering
\includegraphics[width=0.73\linewidth]{rebuttal3_figure/IH-GAN_disp_v45_r3.pdf}
\caption{}
\end{subfigure}
\begin{subfigure}{0.5\linewidth}
\centering
\includegraphics[width=0.72\linewidth]{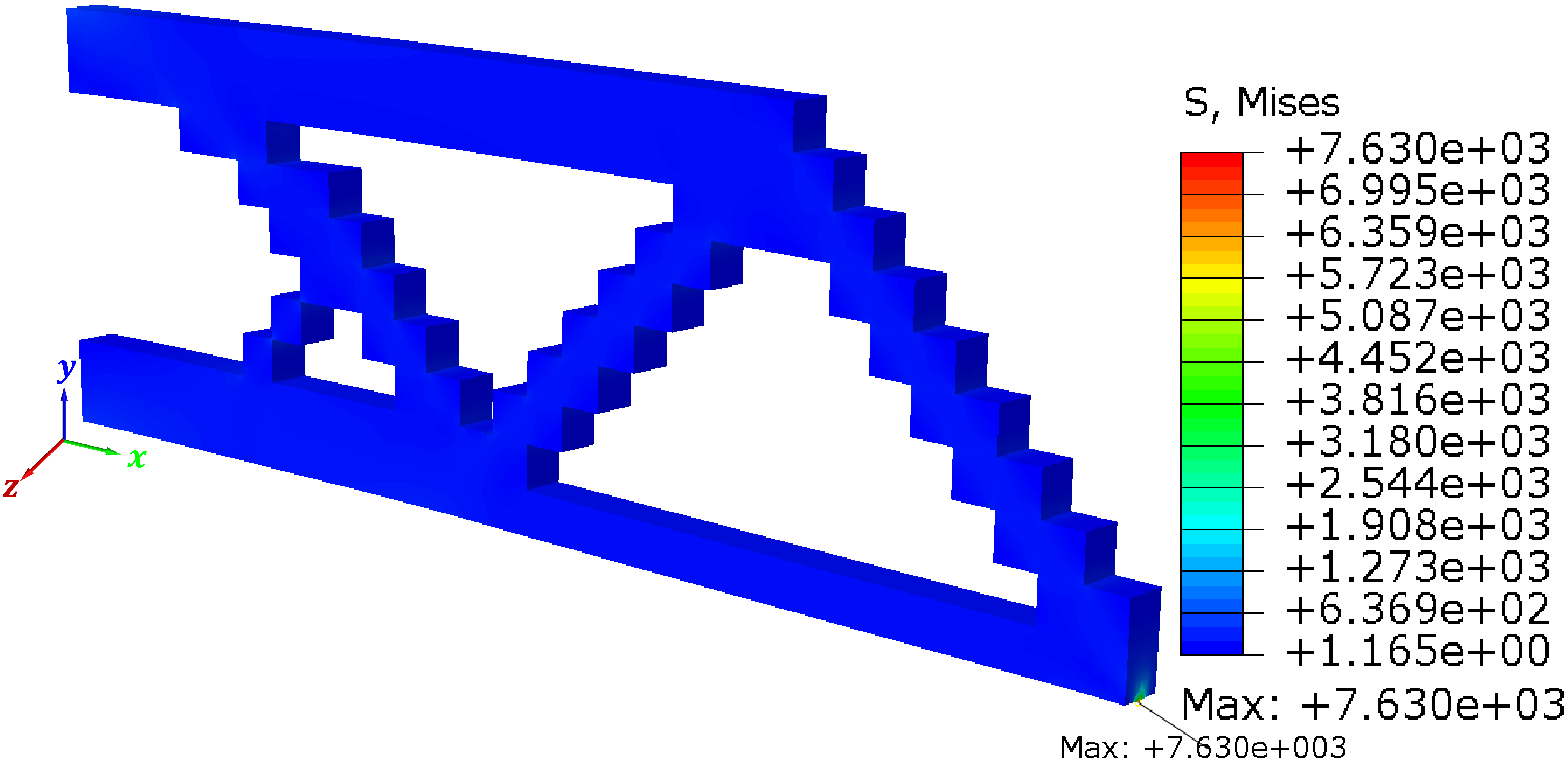}
\caption{}
\end{subfigure}%
\begin{subfigure}{0.5\linewidth}
\centering
\includegraphics[width=0.73\linewidth]{rebuttal3_figure/IH-GAN_stress_v45_r3.pdf}
\caption{}
\end{subfigure}
\caption{FE simulated displacement (\textit{first row:} (a) and (b)) and stress (\textit{second row:} (c) and (d)) contours of the beam redesigned for minimum compliance. \textit{Left:} Topologically optimized solid unit cells using SIMP and \textit{Right:} IH-GAN-based multi-type porous unit cells.}
\vspace{0pt}
\label{figureFEA_SIMP}
\end{figure}

Although the compliance minimization problem is a typical and popular case study that has been commonly used in the functionally graded porous structural designs \cite{li2019design, li2018optimal, panesar2018strategies, montoya2019density}, it has also been acknowledged that porous design does not bring extra benefits in terms of compliance minimization \cite{sivapuram2016simultaneous}. Given that our design ends up with porous unit cells, we further compare our IH-GAN design to the canonical design consisting of solid unit cells (\eg, topologically optimized solid isotropic materials using SIMP constrained by the same volume fraction). As illustrated in Figure~\ref{figureFEA_SIMP}, compared to the IH-GAN generated porous structure, the canonical design with solid isotropic materials arrives at a lower maximum displacement (0.1326 $mm$ {\it vs.} 0.1630 $mm$) and a lower compliance (7.1227 $N\cdot mm$ {\it vs.} 7.3876 $N\cdot mm$) in compliance minimization. The solid unit cells also lower the maximum concentrated stress ($\sigma_{max}$) compared to the IH-GAN porous structure (7,630.4 MPa {\it vs.} 10,302.0 MPa). As the concentrated stress is sensitive to the mesh quality for the FE simulation, to enable a fair comparison, we have standardized the mesh quality for each solid and porous unit cell such that each type of design (\ie, the IH-GAN porous design and the canonical solid design) have the same resolution in both macro- and micro-scales (\ie, a macro-scale resolution of $30\times10\times1$, an implicit surface resolution of $20\times20\times20$, and a volumetric mesh element size of 0.02). Even though the IH-GAN porous design achieves slightly higher compliance than the canonical design with solid isotropic materials, it shows decent performance in the minimum compliance problem. Considering other potential benefits (\eg, design diversity and high-resolution representation) porous structures can offer, porous designs are still a promising alternative candidate in structural optimization in terms of compliance minimization.

\subsection{All-space filling structures}

Adding the option to remove mass from regions that are not essential for load-bearing is definitely an interesting direction. Constrained by the envelope (Figure~\ref{fig:training_data}) of our relatively limited dataset of porous unit cells, our approach has a limitation of generating all-space filling porous structures under the present settings. 
To overcome this limitation, we need to expand the dataset to a more diverse one (as we will discuss in the next section) that also covers the origin in the property space. In that way, our approach can enable the option of mass removal without the need for additional steps. Nonetheless, for cases where we prefer structures to have a better heat dissipation (\eg, heat sink with load-carrying capability), self-supporting performance, or damage-resistant designs \cite{li2019design,yu2020three}, porous structures with unit cells placed everywhere can be a better option. They can achieve more energy dissipation (due to larger contact interface areas), be self-supporting, and be less damage-sensitive compared to solid structures \cite{li2019design}. Although we did not evaluate the thermal performance or add heat dissipation as an additional objective in this paper, we will study such scenarios in our future work. 

\subsection{Dataset diversity}

As we indicated above, by expanding the diversity of our dataset, our proposed IH-GAN approach can be used for broader applications. However, diversity still remains an open question. For almost every data-driven or ML-driven research, one would ask questions like ``how diverse is diverse?", ``where are the boundaries of the diverse data?" and ``when would the data-driven model fail as the data diversity increases?". To this end, topics like database expansion, Pareto frontier discovery (boundary expansion), and optimal diverse coverage could be covered. Our current database covers a specific region of the property space by using three classes of baseline unit cells and varying their weights and level set values. To expand the database with more diverse properties, we can add more classes of basic unit cells in the combination (\eg, including the anisotropic unit cells). However, before doing this, we need to find an appropriate way to assess or quantify the diversity. Specifically, we need to know if the diversity of a particular dataset is diverse enough for specific applications. For example, our current dataset has been diverse enough for our two case studies. However, the current property space could become limited when more diverse properties are desired for some other applications. (\eg, thermal, optical, and acoustic or combinations). 
With relatively limited data, a GAN model can be trained to learn a regularized design latent space that discovers the Pareto frontier of the real property space \cite{chen2021mo}. By pushing or exploring the learned Pareto frontier, one could expand the diversity coverage of the property space.
Additionally, for a massive number of diverse data samples, one would not want 10 million options. Rather, we want the fewest designs that maximize the diverse coverage. While certainly interesting and ultimately valuable, it would be unrealistic for us to complete all these in a feasible time range and be beyond the scope of a single paper. We believe it will be worthwhile answering such questions using a separate paper that could explore them more rigorously.

\section{Conclusion} \label{conclusion}

We proposed a GAN model to learn the IH mapping from properties to unit cell shapes that can be used to optimize functionally graded cellular structures. The cubic symmetric TPMS surfaces (P, D, and F-RD) were chosen and combined to create the dataset of isotropic cellular structures and represent each cellular unit cell as a six-dimensional shape vector (\ie, $\alpha_1, \alpha_2, \alpha_3, t_1, t_2, t_3$). The unit cell property space consists of effective Young's modulus, Poisson's ratio, and relative density (\ie, $E^H$, $\nu^H$, and $\rho$) in which $E^H$ and $\nu^H$ were computed using a voxel-based numerical homogenization method. Our proposed IH-GAN's one-to-many IH mapping was learned on the six-dimensional shape parameter space with the three-dimensional property space as the input conditions.

Our approach offers an end-to-end generative model that automatically learns the IH mapping from data without assuming a bijective polynomial relationship. Except for the unit cell density (volume fraction), we also consider unit cell types when building the mapping.
By including an auxiliary regressor in the IH-GAN, we can accurately generate the cellular unit cells that possess the desired properties ($R^2$-scores between target properties and properties of generated unit cells $>98\%$).

We also demonstrate the IH-GAN model's efficacy by implementing it on modified structural optimization problems to construct multi-scale functionally graded cellular structures (\eg, a cantilever beam in this paper) using multiple types of cellular unit cells. Our approach addresses the connectivity issue between different types of unit cells simultaneously without a need for further compatibility optimization. 
By performing FE simulations, we validate that the beam constructed by IH-GAN improves the functional performance (\eg, $79.7\%$ reduction in concentrated stress) compared to the conventional variable-density single-type beam structure. Moreover, the beam redesigned by IH-GAN can also achieve the target deformations with decent performance.

{\it Limitations and future work:} As a data-driven model, the IH-GAN cannot faithfully generate unit cell shapes given properties outside the training data distribution. This brings the need for constraining the solution space when performing structural optimization, which in turn may limit the performance of the optimized cellular structure. Future work could explore how to combine both data-driven and physics models to allow faithful data extrapolation beyond the training dataset. Another promising extension is to focus on the diversity exploration of the property space such that the IH-GAN model can be used for more diverse applications. To this end, topics like database expansion, Pareto frontier discovery (boundary expansion), and optimal diverse coverage could be covered. While in this paper, we use isotropic material properties to demonstrate the effectiveness of the proposed IH-GAN model, the model itself is agnostic to the type of properties. We can replace the isotropic material properties with anisotropic stiffness or other physics properties (\eg, thermal, optical, and acoustic or combinations). In our future work, we will perform new case studies to demonstrate and validate our proposed method in other application domains.



%
% ...

\bibliographystyle{elsarticle-num}

\bibliography{references}



\end{document}